\documentclass[reqno,11pt]{article}
\pdfoutput=1
\usepackage{jheppub}
\graphicspath{ {./graphs1/} }
\usepackage{epsfig}
\usepackage{amssymb}
\usepackage{verbatim}
\usepackage{amsmath}
\usepackage{hyperref}
\usepackage{dsfont}
\usepackage{slashed}
\usepackage[dvipsnames]{xcolor}
\usepackage{epstopdf}
\usepackage{graphicx}
\usepackage{color}

\usepackage{bbold}

\def\o{\omega}

\def\L{\Lambda}

\def\s{\sigma}

\def\a{\alpha}
\def\b{\beta}

\newcommand{\bc}{\begin{center}}
\newcommand{\ec}{\end{center}}

\newcommand{\ba}{\begin{eqnarray}}
\newcommand{\ea}{\end{eqnarray}}

\title{One-loop Parke-Taylor factors for quadratic propagators from massless scattering equations}

\author{Humberto Gomez,${}^{a,b}$ Cristhiam Lopez-Arcos${}^{b}$ and Pedro Talavera${}^{c}$}

\affiliation{$\,^{a}$Niels Bohr International Academy and Discovery Center, University of Copenhagen\\
Blegdamsvej 17, DK-2100 Copenhagen , Denmark,\\
${}^{b}$Universidad Santiago de Cali, Facultad de Ciencias Basicas,\\
Campus Pampalinda, Calle 5 No. 62-00, C\'{o}digo postal 76001, Santiago de Cali, Colombia}

\affiliation{${}^{c}$Institut de Ciencies del Cosmos, Universitat de Barcelona (IEEC-UB),\\ Marti i Franques 1, Barcelona 08028, Spain}

\emailAdd{humgomzu@gmail.com, crismalo@gmail.com, peterbretons@gmail.com}

\abstract{
In this paper we reconsider the Cachazo-He-Yuan construction (CHY) of the so called scattering amplitudes at one-loop, in order to obtain quadratic propagators.  In 
theories with colour ordering the key ingredient is the redefinition of the Parke-Taylor factors. After classifying all the possible one-loop CHY-integrands we conjecture a new one-loop amplitude for the massless Bi-adjoint $\Phi^3$ theory. The prescription directly reproduces the quadratic propagators of the {\sl traditional} Feynman approach.
}

\begin{document}

\maketitle


\section{Introduction} \label{sect:intro}

Our most accurate knowledge on quantum field theory in Minkowski $(3+1)$-d is almost entirely based on perturbation methods. 
At leading order, quantities such as scattering amplitudes are computed by adding tree-like diagrams. 
Even this elementary manipulation becomes a formidable task  at rather low multiplicity kinematics and become only technically feasible by using the Weyl-van der Waerden spinor calculus. 
  
The use of on-shell methods for the calculation of scattering amplitudes has come into attention since the last decade, following Witten's seminal work \cite{Witten:2003nn}. A remarkable development following these ideas is the approach by Cachazo-He-Yuan (CHY) \cite{Cachazo:2013gna,Cachazo:2013hca}, it has the great advantages of being applicable to several dimensions and also to a large array of theories \cite{Cachazo:2013iea,Cachazo:2014xea,Cachazo:2014nsa,Cachazo:2016njl}, even beyond field theory \cite{Mizera:2017sen,Mizera:2017cqs}. The main ingredient for this approach are the tree-level scattering equations \cite{Cachazo:2013gna}
\ba  
E_a := \sum_{b \ne a}\frac{k_a\cdot k_b}{\s_{ab}}=0,  \quad  \s_{ab}:=\s_a - \s_b , \quad  a=1,2,\ldots, n, 
\ea
where the $\s_a$'s denote punctures on the sphere. The tree-level S-matrix can be written in terms of contour integrals localized over solutions of these equations on the moduli space of n-punctured Riemann spheres 
\ba
{\cal A}_n = \int_{\Gamma} \,d\mu_n^{\rm tree}\,\, \cal{I}_{\rm tree}^{\rm CHY}(\s), 
\ea
where the integration measure, $d\mu_n^{\rm tree}$, is given by
\begin{equation}
d\mu_n^{\rm tree} =\frac{\prod_{a=1}^n d\s_a}{{\rm Vol}\,({\rm PSL}(2,\mathbb{C}))} \times \frac{(\s_{ij} \s_{jk} \s_{ki})}{    \prod_{b\neq i,j,k}^{n} E_b  }
\end{equation}
and the contour $\Gamma$ is defined by the $n-3$ independent scattering equations
\begin{equation}
E_b=0, \quad b\neq i,j,k\,.
\end{equation}
The integrand, $\cal{I}_{\rm tree}^{\rm CHY}$, depends on the described theory. There are other approaches that use the same moduli space \cite{Witten:2003nn,Roiban:2004yf,Cachazo:2012da,Cachazo:2012kg}, but restricted to four dimensions. 

There have been developed several methods to evaluate the integrals, from different perspectives. Some approaches  study the solutions to the scattering equations for particular kinematics and/or dimensions \cite{Cachazo:2013iaa,Cachazo:2013iea,Kalousios:2013eca,Lam:2014tga,Cachazo:2016sdc,He:2016vfi, Cachazo:2015nwa,Cachazo:2016ror}, others work with a polynomial form \cite{Kalousios:2015fya,Dolan:2014ega,Dolan:2015iln,Huang:2015yka,Cardona:2015ouc,Cardona:2015eba, Sogaard:2015dba,Bosma:2016ttj,Zlotnikov:2016wtk,Mafra:2016ltu}, or formulating sets of integration rules \cite{Baadsgaard:2015ifa,Baadsgaard:2015voa,Huang:2016zzb,Cardona:2016gon,Zhou:2017mfj}. A different approach was proposed in \cite{Gomez:2016bmv}, taking the double covered version of the sphere, the so called $\Lambda$-algorithm, which we will employ in this work. 

A generalization for loop level of the CHY formalism has been made. The ambitwistor and pure spinor ambitwistor worldsheet \cite{Mason:2013sva,Berkovits:2013xba} provided a prescription for a generalization to higher genus Riemann surfaces \cite{Geyer:2015bja,Geyer:2015jch,Adamo:2015hoa}. A different approach was also developed in \cite{Cachazo:2015aol,He:2015yua,Baadsgaard:2015hia}, where the forward limit with two more massive particles, playing the role of the loop momenta, were introduced. The scattering equations for massive particles were already studied in \cite{Dolan:2013isa,Naculich:2014naa}. Another alternative approach using an elliptic curve was developed in \cite{Cardona:2016wcr, Cardona:2016bpi}. 

The previous prescriptions give a new representation of the Feynman integrals with propagators linear in loop momenta. In order to find the equivalence with the usual Feynman propagators, $(\ell + K)^{-2}$, two additional steps must be taken: the first one is  the use of partial fractions, and the second one is the shifting of loop momenta \cite{Geyer:2015jch,Casali:2014hfa}. 

Recently, one of the authors \cite{Gomez:2017lhy} proposed a different approach to obtain the quadratic Feynman propagators directly from the CHY-integrands for the scalar $\Phi^3$ theory\footnote{There are some overlapping ideas with the recent paper published by Farrow and Lipstein  \cite{Farrow:2017eol}.}. The motivation came by analysing a Riemann surface of genus two after an unitary cut, which look exactly like a tree level diagram before the forward limit, but instead of the two massive particles associated to the loop momenta there are four massless particles. This new approach allows to work again with the scattering equations for massless particles, but at the expenses of increasing the number to $n+4$. In addition there is also the need to introduce a new measure of integrations that guarantees the cut and then take the forward limit.

In the present work, we follow the line of thought of \cite{Gomez:2017lhy} and propose a reformulation for the one-loop Parke-Taylor factors. Splitting the massive loop momenta $(\ell^+,\ell^-)$ into the four massless ones $((a_1,b_1),(b_2,a_2))$, we will have one-loop Parke-Taylor factors that will enter into CHY-integrands to lead directly to the usual Feynman propagators. The CHY-integrands in question are the ones for the Bi-adjoint $\Phi^3$ scalar theory.

\subsection*{Outline}

This paper is organized as follows. In section \ref{Parke-Taylor} we present our reformulation of the one-loop Parke-Taylor factors (PT). The expression is written in terms of the generalized holomorphic one-form on the Torus, $\omega^{a:b}_{i:j}$.  By exploiting algebraic identities we formulate the Theorem 1: each term in the PT factors can be decomposed into terms containing at least two $\omega^{a:b}_{i:j}$ factors, i.e.  the PT factors are rearranged in an expansion {\sl manifestly tadpole-free}\footnote{As it will be explained, the number of $\omega^{a:b}_{i:j}$  is related with the polygon of the loop, for example, two $\omega^{a:b}_{i:j}$ in the left integrand can only  generate a bubble or a triangle.} . 

In section \ref{classification} we write, classify and match with their Feynman integrands counterparts,  some general type of CHY-integrands that can appear at one-loop level. Since we are working with $n+4$ massless particles, the contour integrals can be calculated using any of the existing methods of integration. As we have mentioned already we employ the so-called $\L$-algorithm, with the choice of a new gauge fixing, to solve them. This allows to analytically evaluate arbitrary CHY-integrals using simple graphical rules. The classification is made tracing the structures defined in section \ref{Parke-Taylor}. As will become clear each element inside the partial amplitude have an unambigous correspondence with the elements of the CHY-graphs, starting with the n-gon, then following with the ones with tree level structures attached to their corners.

Section \ref{biadjoint}  shows our  proposal for the partial amplitude of the Bi-adjoint $\Phi^3$ theory at one-loop with quadratic propagators: first we give a simple review at tree level and one-loop with linear propagators, then we propose our formula using our definition for the PT factors.

In order to support our proposition in section \ref{Sexamples} we perform explicitly the calculation for the partial amplitudes of the three and four-point functions. We make an extensive use of the results of previous sections. In particular we emphasize the direct interpretation of the CHY-integrals
in terms of Feynman diagrams. 
This mapping is codified in the following equality at the integrand level 
\begin{center}
\begin{tabular}{| l |}
 \cline{1-1}  
 \\
 $\frac{1}{2^{N+1}} \int d\Omega \times s_{a_1 b_1} \times
 \int d\mu^{\rm tree}_{N+4}  $
 \parbox[c]{10em}{\includegraphics[scale=0.28]{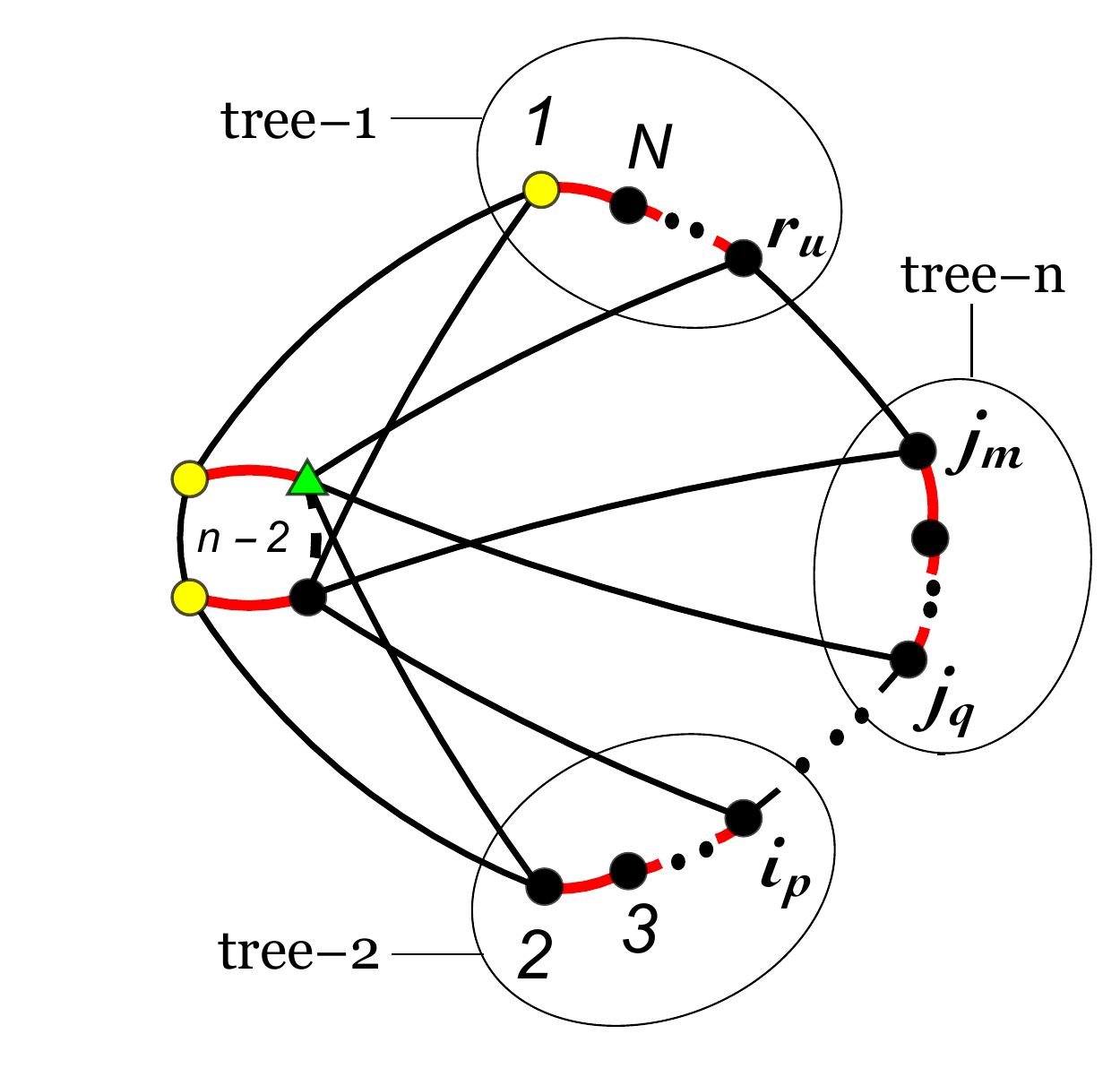}}
 =
 \hspace{-0.15cm}
  \parbox[c]{10em}{\includegraphics[scale=0.28]{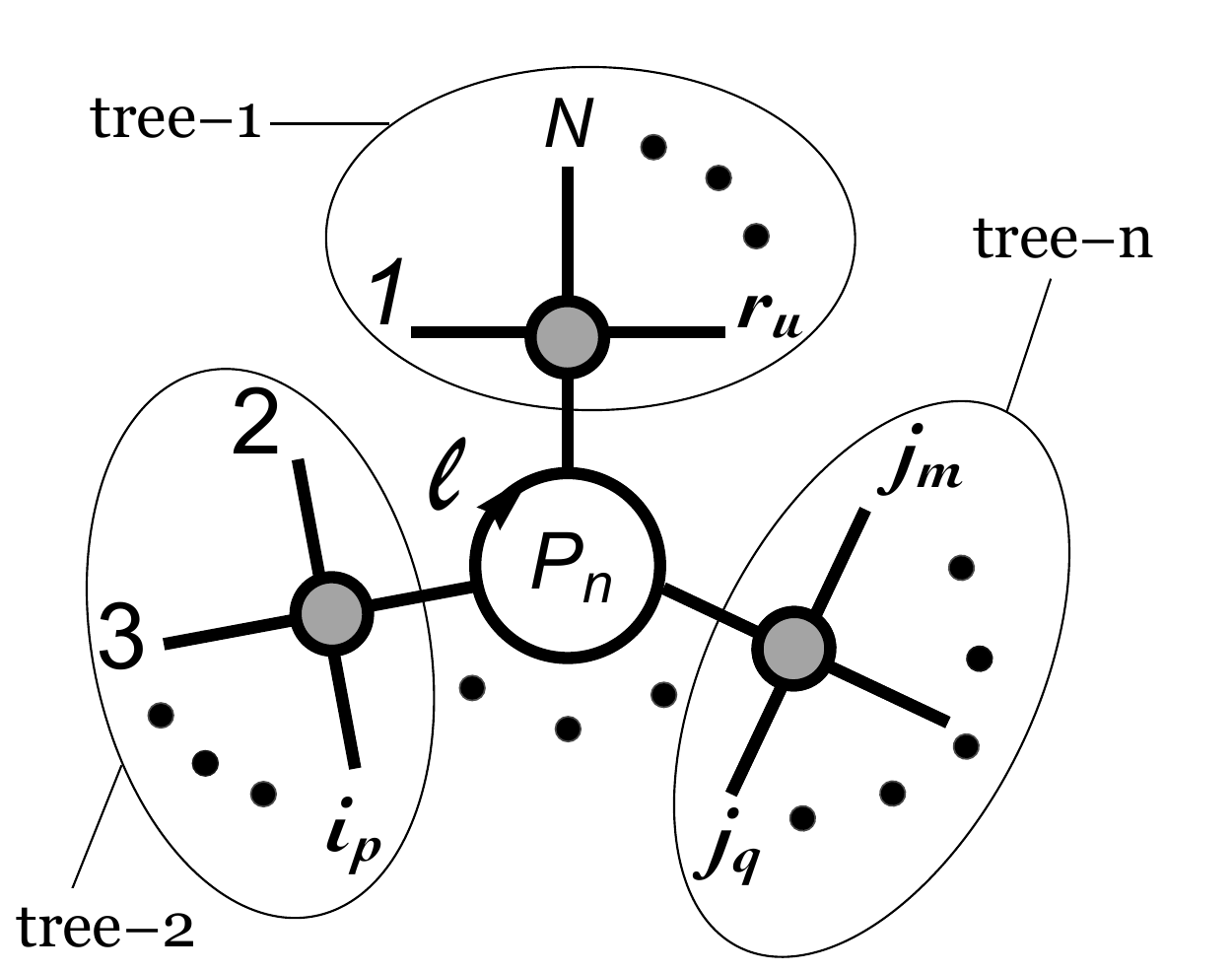}} 
\\
\\
\cline{1-1}
\end{tabular}
\end{center}
that constitutes one of the most important results of this work and it will be explained in detail during the course of this paper.

In section \ref{sectionEB} we comment on the issue of the external-leg bubble contributions. Diagrams involved are singular and need to be regularized. 
Next section, \ref{sectionepsilon}, is for illustrative purposes and is devoted to the $i \epsilon$ prescription and how to directly obtain it by dimension reduction.
Finally, in section \ref{discussions} we conclude by summarizing our findings.

For not disrupting the line of the paper more technical discussions have been gathered in some appendices: In Appendix \ref{chy-integrands} we explicitly show the sufficient form of the measure (\ref{dOmega})
to tackle the one-loop CHY-integral prescription. In particular how the momenta combination
it contains arrises. 
Proof of Theorem 1 is casted in Appendix \ref{Aproof} where it is discussed at length. Appendix \ref{Aconjeture} collects the relation between some techniques developed across the paper and the linear propagator prescription. We conclude by probing an statement of \cite{Cardona:2016bpi}.

Before beginning section \ref{Parke-Taylor}, we define the notation that is going to be used in the paper.

\subsection*{Notation}

For convenience, in this paper we use the following notation
\begin{align}\label{notations}
\s_{ij}  := \s_i - \s_j,  \qquad\qquad
\o_{i:j}^{a:b}  := \frac{\s_{ab}}{\s_{i a}\,\, \s_{j b}}.
\end{align}
Note that $\o_{i:j}^{a:b}$ are the generalization of the $(1,0)-$forms used in \cite{Gomez:2016cqb} to write the CHY-integrands at two-loop.
In addition, we define the $\s_{ab}$'s and  $\o_{i:j}^{a:b}$'s  chains as
\begin{align}\label{notations-2}
&(i_1,i_2,\ldots, i_p)  := \s_{i_1i_2} \cdots \s_{i_{p-1}i_p}\s_{i_p i_1},    \\
&(i_1,i_2,\ldots, i_p)^{a:b}_\o  := \o^{a:b}_{i_1:i_2} \cdots \o^{a:b}_{i_{p-1} : i_p}\o_{i_p : i_1}^{a:b}  
 \nonumber\\
& \qquad\qquad\qquad\quad
=  \o^{a:b}_{i_1:i_1} \cdots \o^{a:b}_{i_{p-1} : i_{p-1}}\o_{i_p : i_p}^{a:b},\nonumber\\ 
& (i_1^\s ,i^\s_2,\ldots, i_m^\o, i_{m+1}^\s,\ldots, i_n^\o,\ldots i_p^\s)^{a:b}  := \s_{i_1i_2}\s_{i_2i_3}\cdots \o^{a:b}_{i_{m}i_{m+1}}\s_{i_{m+1} i_{m+2}}  \cdots \o^{a:b}_{i_{n}i_{n+1}}\cdots \s_{i_{p-1}i_p}\s_{i_p i_1}.\nonumber
\end{align}

\begin{figure}[!h]
	\centering
	\includegraphics[width=5.0in]{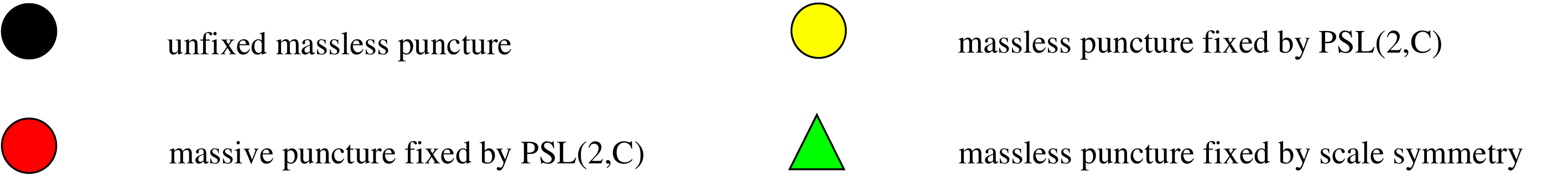}
	\caption{Vertex Color code in CHY-graphs for the $\L-$algorithm.}\label{color_codV}
\end{figure}

\begin{figure}[!h]
	\centering
	\includegraphics[width=4.5in]{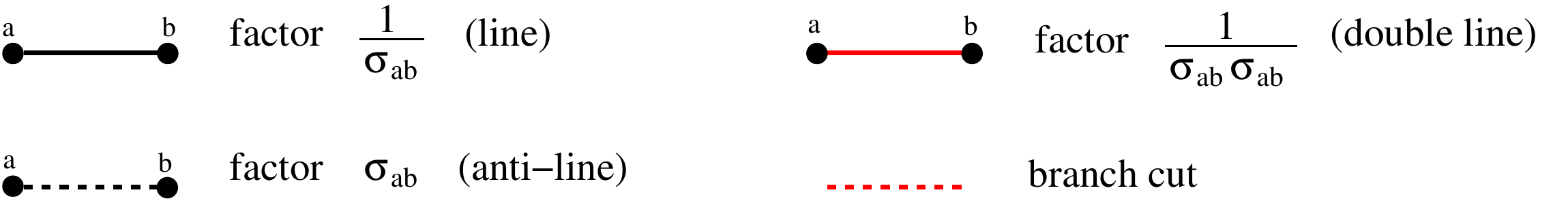}
	\caption{ Edges Color code in CHY-graphs for the $\L-$algorithm.}\label{color_codE}
\end{figure}

In order to have a graphical description for the CHY-integrands on a Riemann sphere (CHY-graphs), it is useful to represent each $\s_a$ puncture as a vertex, the factor ${1 \over \s_{ab}}$ as a line and the factor $\s_{ab}$ as a dashed line that we call the {\it anti-line}. Additionally, since we often use the $\L-$algorithm\footnote{It is useful to recall that the $\L-$algorithm fixes four punctures, three of them by the ${\rm PSL}(2,\mathbb{C})$ symmetry and the last one by the scale invariance.} \cite{Gomez:2016bmv}  we introduce the color code  given in Fig. \ref{color_codV} and \ref{color_codE}
for a  mnemonic understanding.
\noindent

\noindent

Finally, we introduce the momenta notation
\begin{equation}
k_{\{ a_1,\ldots, a_m\}}= [a_1,\ldots, a_m]:=\sum_{i=1}^m k_{a_i}, \quad
s_{a_1\ldots a_m}:=k_{\{ a_1,\ldots, a_m\}}^2, 
\quad k_{a_1\ldots a_m}:=\sum_{a_i<a_j}^m k_{a_i}\cdot k_{a_j}. \nonumber
\end{equation}

\section{Parke-Taylor at One-Loop}\label{Parke-Taylor}

In this section we define the PT factor at one-loop in the CHY prescription, which is totally similar to ones given in \cite{Geyer:2015jch, He:2015yua,Baadsgaard:2015hia}. After that, we carry out some manipulation in order to write algebraic identities that will be very convenient to perform computations using the $\L-$algorithm.

Before defining the PT factor at one-loop, it is useful to remind that, at tree level  in the CHY approach, it is given by the expression 
\begin{equation}
{\rm PT}_{\rm tree}[\pi]= \frac{1}{(\pi_1,\pi_2,\ldots , \pi_n)},
\end{equation}
where $\pi$ is a generic ordering and $n$ is the total number of particles.

Following the ideas presented in \cite{He:2015yua,Baadsgaard:2015hia},  we formulate:

\vspace{6mm}

\noindent {\bf Definition:} {\sl We define the Parke-Taylor factor at one-loop with ordering $\pi$ as}\,
\begin{align}
 {\bf PT}^{\, a_1:a_2}_{\rm 1-loop} [\pi]:=& 
 \sum_{\a\in {\rm cyc (\pi)}}
\frac{1}{\s_{\a_1\a_2}\s_{\a_2\a_3}\cdots \s_{\a_{n-1}\a_n}} \o^{a_1:a_2}_{\a_n:\a_1} \label{ptdef}\\
=&\,
{\rm PT}_{\rm tree}[\pi]\left( 
\s_{\pi_1\pi_2}\, \o_{\pi_1:\pi_2}^{a_1:a_2} + \s_{\pi_2\pi_3}\, \o_{\pi_2:\pi_3}^{a_1:a_2}
+\cdots + \s_{\pi_n\pi_1}\, \o_{\pi_n:\pi_1}^{a_1:a_2} \nonumber
\right).
\end{align}

\vspace{6mm}

As it will be discussed later,  the task of performing CHY-integrals using the PT defined in \eqref{ptdef} is not simple. The difficulty of these computations resides in the number of $\o^{a_1:a_2}_{i:j}$'s, more  $\o^{a_1:a_2}_{i:j}$'s  imply that the singular solutions of the scattering equations do not contribute. In fact, the minimum number of $\o^{a_1:a_2}_{i:j}$'s so the CHY-integrals become simpler is two, as we are going to explain in section \ref{classification}.  Nevertheless, it is always possible to manipulate algebraically ${\bf PT}^{\, a_1:a_2}_{\rm 1-loop} [\pi]$ and to decompose it as a linear combination of terms that contain at least two $\o^{a_1:a_2}_{i:j}$'s. 

 \vspace{6mm}

\noindent {\bf Theorem 1:} {\sl The ${\bf PT}^{\, a_1:a_2}_{\rm 1-loop} [\pi]$ factor, which was defined in \eqref{ptdef}, admits a power expansion in $\o^{a_1:a_2}_{i:j}$ with two as its lowest power}\,.


\vspace{6mm}

We shall only sketch some examples in order to illustrate this theorem, leaving the complete, technical proof, together with the precise construction using the Schouten-like identity for the $\s_{ij}$'s to the appendix \ref{Aproof}. 

The previous theorem allows to clarify, that the cancellations of the tadpoles in the bi-adjoint $\Phi^3$ theory can follow directly from an algebraic property of the one-loop PT factors and not necessarily from the anti-symmetry of the structure constant in the cubic vertex. 

Before proceeding we shall introduce the following definitions:
\begin{align}\label{pttype0}
 \mathbf{D}^{\, a_1:a_2}_{\rm type-0} [1,\ldots,n]^{ p\,\o} : = &(1,\dots,p)\, \o^{a_1:a_2}_{1:1}\cdots \o^{a_1:a_2}_{p:p} + (2,\dots,p+1)\, \o^{a_1:a_2}_{2:2}\cdots \o^{a_1:a_2}_{p+1:p+1}\nonumber \\
&+ \cdots +  (n,1,\dots,p-1)\, \o^{a_1:a_2}_{n:n} \o^{a_1:a_2}_{1:1}\cdots \o^{a_1:a_2}_{p-1:p-1},\\
 &{\rm with}\,\,\,\, 1< p\leq n,  \nonumber
\end{align}
\begin{align}\label{pttype1}
 \mathbf{D}^{\, a_1:a_2}_{\rm type-I} [1,\ldots,n]^{ p\,\o} :& =  (1^\s,\dots,p^\s,(p+1)^\o)^{a_1:a_2} \o^{a_1:a_2}_{2:2}\cdots \o^{a_1:a_2}_{p:p}\\
& 
  + (2^\s,\dots,(p+1)^\s,(p+2)^\o)^{a_1:a_2} \o^{a_1:a_2}_{3:3}\cdots \o^{a_1:a_2}_{p+1:p+1}\nonumber \\
&+ \cdots +  (n^\s,1^\s,\dots,(p-1)^\s,p^\o)^{a_1:a_2} \o^{a_1:a_2}_{1:1} \o^{a_1:a_2}_{2:2}\cdots \o^{a_1:a_2}_{p-1:p-1},\nonumber\\
 & {\rm with}\,\,\,\, 1<p\leq n-1, \nonumber  
\end{align}
\begin{align}\label{pttype2}
 \mathbf{D}^{\, a_1:a_2}_{\rm type-II} [1,\ldots,n]^{ p\,\o} :& =  (1^\s,2^\o,3^\s,4^\o\dots,p^\o)^{a_1:a_2}+(2^\s,3^\o,4^\s,5^\o\dots,(p+1)^\o)^{a_1:a_2}\\
& +\cdots + 
(n^\s,1^\o,2^\s,3^\o\dots,(p-1)^\o)^{a_1:a_2}\,, \nonumber\\
&{\rm with}\,\,\,\, p\in\{2,4,6,\ldots ,\frac{n}{2}\}. \nonumber  
\end{align}

Notice, that we have defined  the factors, $ \mathbf{D}^{\, a_1:a_2}_{\rm type-0} $, $ \mathbf{D}^{\, a_1:a_2}_{\rm type-I} $ and $ \mathbf{D}^{\, a_1:a_2}_{\rm type-II} $, with the  particular ordering $\{1,2,\ldots, n \}$, nevertheless, their definitions for another ordering are straightforward. These terms also carry the cyclic permutation invariance from the PT factor.

\subsection{Examples}

In this section we give some non-trivial examples  in order to illustrate the above proposition. The following identities are purely algebraic and they can be proven after a, somehow, direct computation.

\begin{itemize}

\item {\bf Two-Point}\\
\begin{align}
{\bf PT}^{\, a_1:a_2}_{\rm 1-loop} [1,2]=
\frac{1}{\s_{12}} \o^{a_1:a_2}_{2:1}+\frac{1}{\s_{21}} \o^{a_1:a_2}_{1:2}
= {\rm PT}_{\rm tree} [1,2] \,  \mathbf{D}^{\, a_1:a_2}_{\rm type-0} [1,2]^{ 2\o} , \label{PT-2pts}
\end{align}
where let us remind,  $ \mathbf{D}^{\, a_1:a_2}_{\rm type-0} [1,2]^{ 2\o} = (1,2)\, \o^{a_1:a_2}_{1:1}\o^{a_1:a_2}_{2:2}$.

\item{\bf Three-Point}\\
\begin{align}
 {\bf PT}^{\, a_1:a_2}_{\rm 1-loop} [1,2,3] &=
\frac{1}{\s_{12}\,\s_{23}} \o^{a_1:a_2}_{3:1}+\frac{1}{\s_{23}\,\s_{31}} \o^{a_1:a_2}_{1:2} +\frac{1}{\s_{31}\,\s_{12}} \o^{a_1:a_2}_{2:3} \nonumber\\
&=
{\rm PT}_{\rm tree} [1,2,3] \left[\, 2\mathbf{D}^{\, a_1:a_2}_{\rm type-0} [1,2,3]^{ 3\o} +  \mathbf{D}^{\, a_1:a_2}_{\rm type-I} [1,2,3]^{ 2\o} 
\right] \nonumber\\
&=
{\rm PT}_{\rm tree} [1,2,3] \left[\, \mathbf{D}^{\, a_1:a_2}_{\rm type-0} [3,2,1]^{ 3\o} +  \mathbf{D}^{\, a_1:a_2}_{\rm type-I} [3,2,1]^{ 2\o} 
\right] . 
\label{PT-3pts_2}
\end{align}

\item{\bf Four-Point}\\
\begin{align}
 \frac{ {\bf PT}^{\, a_1:a_2}_{\rm 1-loop} [1,2,3,4]}{ {\rm PT}_{\rm tree} [1,2,3,4]}&=
 \s_{12}\, \o^{a_1:a_2}_{1:2}
+\s_{23} \, \o^{a_1:a_2}_{2:3}
+\s_{34} \o^{a_1:a_2}_{3:4}
+\s_{41} \, \o^{a_1:a_2}_{4:1}
\nonumber \\
&=
3\mathbf{D}^{\, a_1:a_2}_{\rm type-0} [1,2,3,4]^{ 4\o} + \, 2\mathbf{D}^{\, a_1:a_2}_{\rm type-I} [1,2,3,4]^{ 3\o} 
\nonumber\\
& 
\hspace{.4cm}+  \mathbf{D}^{\, a_1:a_2}_{\rm type-I} [1,2,3,4]^{ 2\o}  
+ \frac{1}{2}\,\mathbf{D}^{\, a_1:a_2}_{\rm type-II} [1,2,3,4]^{ 2\o} \nonumber\\
&=
\mathbf{D}^{\, a_1:a_2}_{\rm type-0} [4,3,2,1]^{ 4\o} + \, \mathbf{D}^{\, a_1:a_2}_{\rm type-I} [4,3,2,1]^{ 3\o} 
\nonumber\\
& 
\hspace{.4cm}+  \mathbf{D}^{\, a_1:a_2}_{\rm type-I} [4,3,2,1]^{ 2\o}  
+ \frac{1}{2}\,\mathbf{D}^{\, a_1:a_2}_{\rm type-II} [4,3,2,1]^{ 2\o} .
\label{PT-4pts_2}
\end{align}

Note that the $(1/2)$ factor in $\mathbf{D}^{\, a_1:a_2}_{\rm type-II} [4,3,2,1]^{ 2\o}$ comes from the fact there is a double counting, i.e. 
\begin{align}
\mathbf{D}^{\, a_1:a_2}_{\rm type-II}[4,3,2,1]^{ 2\o}&=(4^\s,3^\o,2^\s,1^\o)^{a_1:a_2}+(3^\s,2^\o,1^\s,4^\o)^{a_1:a_2}\nonumber\\
&\,
+(2^\s,1^\o,4^\s,3^\o)^{a_1:a_2}+(1^\s,4^\o,3^\s,2^\o)^{a_1:a_2}\nonumber\\
&=2\left[\,(4^\s,3^\o,2^\s,1^\o)^{a_1:a_2}+(3^\s,2^\o,1^\s,4^\o)^{a_1:a_2}\,\right].
\end{align}
\end{itemize}

\section{One-Loop Feynman Integrands Classification from CHY Approach}\label{classification}

The main goal of this section, following \cite{Gomez:2017lhy}, is to classify some CHY-integrals at 
one-loop.  Notice that the one-loop CHY-integral prescription given  in  \cite{Gomez:2017lhy} has the particular structure\footnote{For more details see appendix \ref{chy-integrands}.}
\begin{align}
\mathbf{I}_n=&\int \frac{d^D\ell}{(2\pi)^D} \,\, \mathfrak{I}_n \,\, ,\nonumber\\
\mathfrak{I}_n:=&\frac{1}{2^{n+1}}\int d\Omega \,\,\, s_{a_1b_1}\, \int d\mu_{n+4}^{\rm tree} \,
   \left[ \frac{{\cal I}_{L}^{\, a_1:a_2} (\s)}{(a_1,b_1,b_2,a_2)} \right]  \times  \left[ \frac{{\cal I}_{R}^{\, b_1:b_2} (\s)}{(a_1,b_1,b_2,a_2)} \right]
\label{prop1}
\end{align}
where $n$ is the number of massless external particles and  the  $d\mu_{n+4}^{\rm tree}$ is the tree level measure defined in \cite{Cachazo:2013hca} 
\begin{align}
d\mu_{n+4}^{\rm tree}:=& \,
 \frac{   \prod_{A=1}^{n+4}\,d\s_A          }{{\rm Vol}\,\,({\rm PSL}(2,\mathbb{C}))} \times \frac{(\s_{1b_1}\,\s_{b_1b_2}\, \s_{b_21})}{ \prod_{ A\neq 1, b_1, b_2}^{n+4}  E_A}\nonumber\\
 &   ^{\underrightarrow{\,\quad{\rm fixing\,\, PSL}(2,\mathbb{C})\,\quad }}\,\,\,
\frac{d\s_{a_1}}{E_{a_1}} \times
\frac{d\s_{a_2}}{E_{a_2}} \times
 \prod_{i=2}^{n} \frac{d\s_{i}}{E_{i}}\times (\s_{1b_1}\,\s_{b_1b_2}\, \s_{b_21}  )^2 , \label{dmu}\\
E_A := &\sum_{B=1\atop B\ne A }^{n+4}\frac{k_A\cdot k_B}{\s_{AB}}=0,  \qquad A=1,2,\ldots, n+4, \label{Sequations}
\end{align}
with the identification
\begin{align}\label{identification}
&\{ k_{n+1} , k_{n+2},k_{n+3},k_{n+4}\}:=\{ k_{a_1} , k_{a_2},k_{b_1},k_{b_2}\}\nonumber,\\
&\{ \s_{n+1} , \s_{n+2},\s_{n+3},\s_{n+4}\}:=\{ \s_{a_1} , \s_{a_2},\s_{b_1},\s_{b_2}\},\\
&\{ E_{n+1} , E_{n+2},E_{n+3},E_{n+4}\}:=\{ E_{a_1} , E_{a_2},E_{b_1},E_{b_2}\}\nonumber .
\end{align}
Notice that in \eqref{dmu}, without loss of generality,  we have fixed $\{\s_{1},\s_{b_1},\s_{b_2} \}$
and  $\{ E_{1},E_{b_1},E_{b_2} \}$. Additionally, let us remind  the  measure, $d\Omega$, it is given by the expression\footnote{In this paper we are considering that the $D-$dimensional momentum space is real, i.e. $k_i\in \mathbb{R}^{D-1,1}$. Therefore, the Dirac delta functions in \eqref{dOmega} are well defined.}
\begin{align}\label{dOmega}
d\Omega :=& \, d^D(k_{a_1}+k_{b_1})\, \delta^{(D)}(k_{a_1}+k_{b_1} - \ell)\, d^{D}k_ {a_2}\, d^{D}k_{ b_2}\,\delta^{(D)} (k_{a_2}+k_{a_1})\,\,\delta^{(D)} (k_{b_2}+k_{b_1}).
\end{align}
This measure is introduced to take the forward limit of the four on-shell loop momenta. The motivation comes because momenta will appear combined in a particular way after the integration of the $\sigma_A$'s. In Appendix \ref{chy-integrands} we give a more detailed explanation of how this particular combination appears.

It is useful to recall that the integration over the $\sigma_A$'s variables is a contour integral, which is localized over the solution of the scattering equations, i.e. $E_A=0$. In addition, in this paper we are just interested to focus at the integrand level, in other words in $\mathfrak{I}_n $, therefore we do not worry to write the measure $\frac{d^D\ell}{(2\pi)^D}$.  

The classification is going to be made by taking into account the CHY-integrands appearing in the ${\bf D}^{\, a:b}_{\rm type- \,*}$ definitions given in \eqref{pttype0},\eqref{pttype1} and \eqref{pttype2}. {\sl i.e.}
\begin{align*}
 {\cal I}_{L}^{\, a_1:a_2} (\s)  &=  {\rm PT_{tree}}[1,2,\ldots,n]\times {\bf D}^{\, a_1:a_2}_{\rm type- \,*} [n,n-1,\ldots, 1] , \\
  {\cal I}_{R}^{\, b_1:b_2} (\s)  &= {\rm PT_{tree}}[1,2,\ldots,n] \times \left(  \s_{12}\, \o ^{b_1:b_2}_{1:2} \right) \nonumber .
\end{align*}

\begin{figure}[!h]
	\centering
	\includegraphics[width=2.0in]{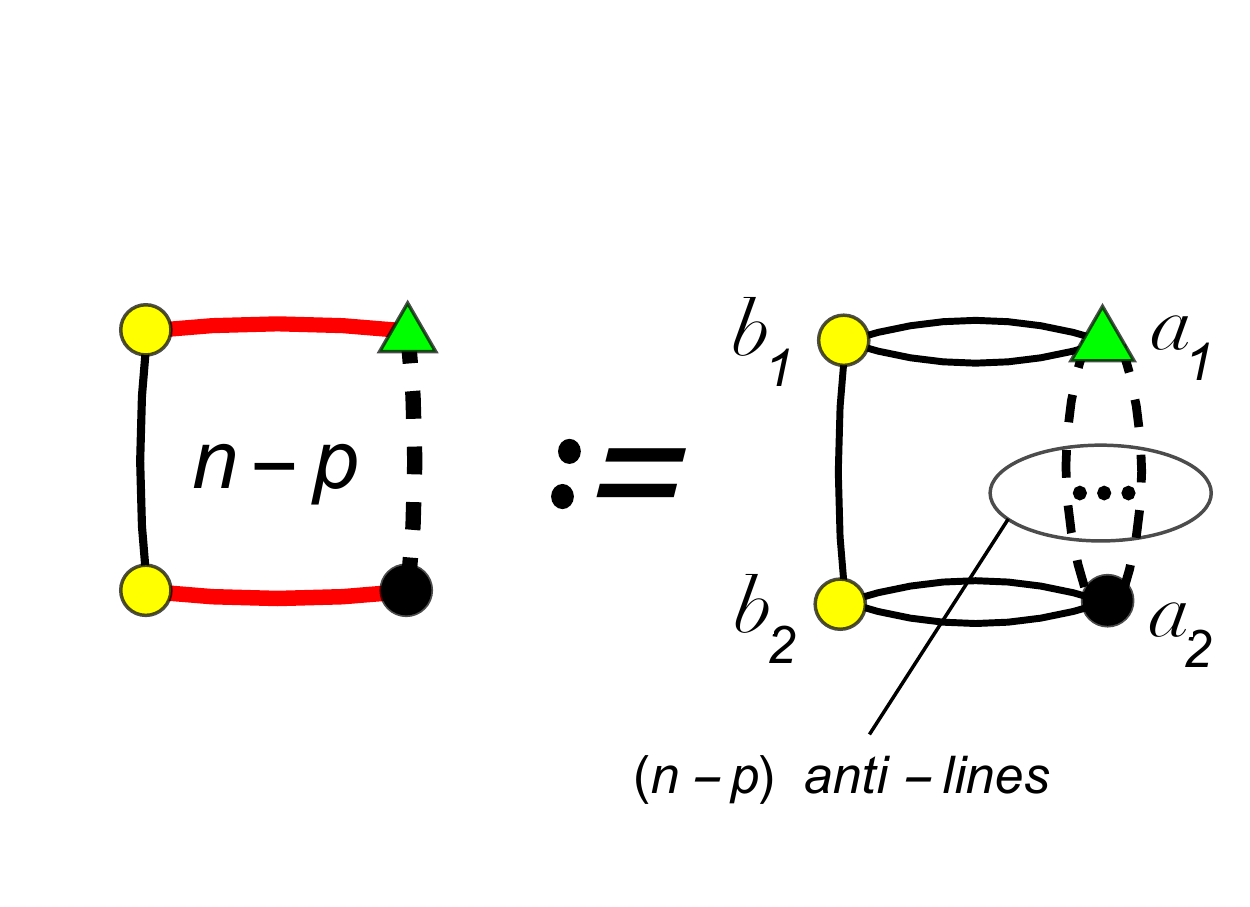}
	\hspace{1.6cm}
		\includegraphics[width=1.8in]{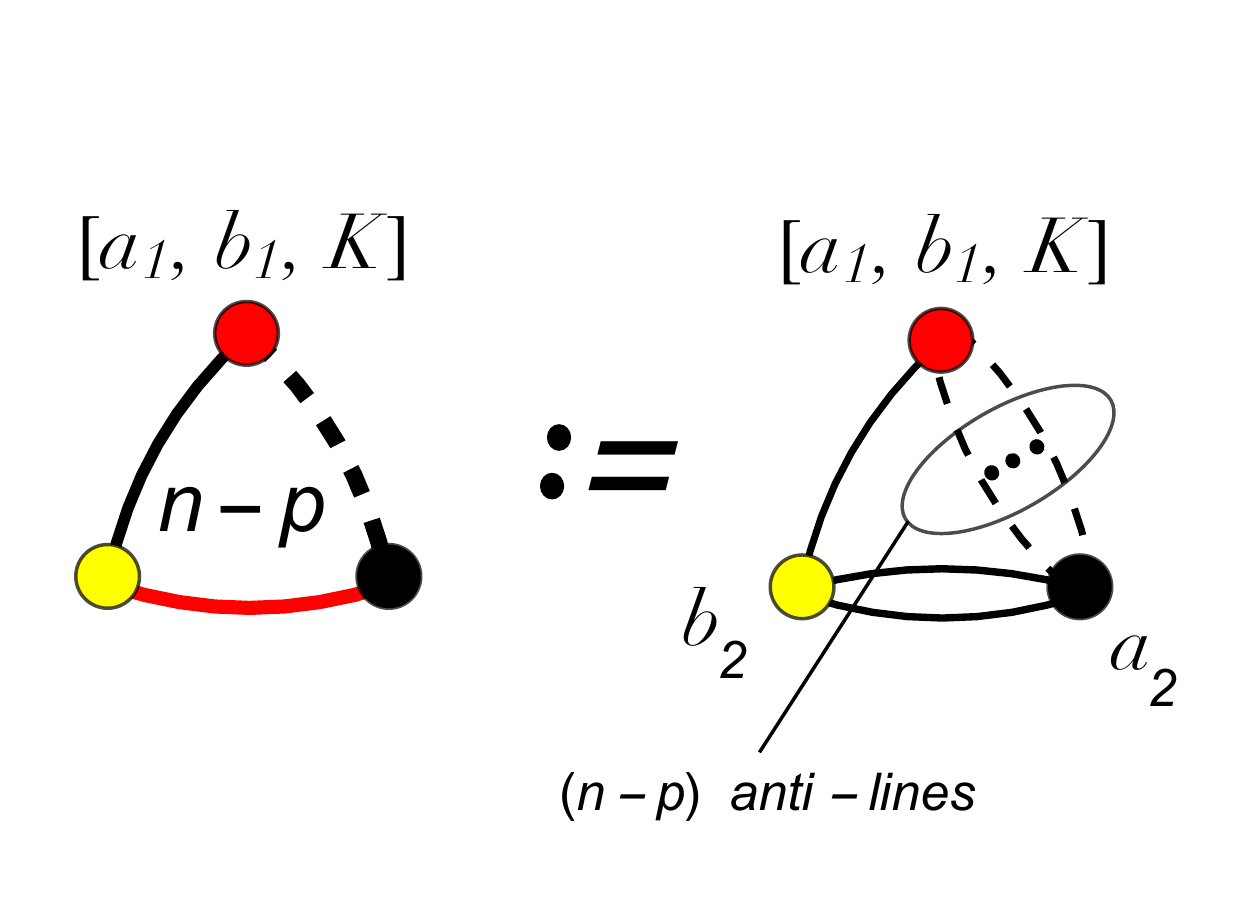}
	\caption{Loop box and triangle definition .}\label{loopS}
\end{figure}

In order to not saturate the notation on the CHY-graphs, we  introduce the definition given in Fig. \ref{loopS}.
\noindent

\subsection{One Loop Integrands Classification}\label{Sclassification}

\begin{figure}[!h]
	\centering
	\includegraphics[width=2.0in]{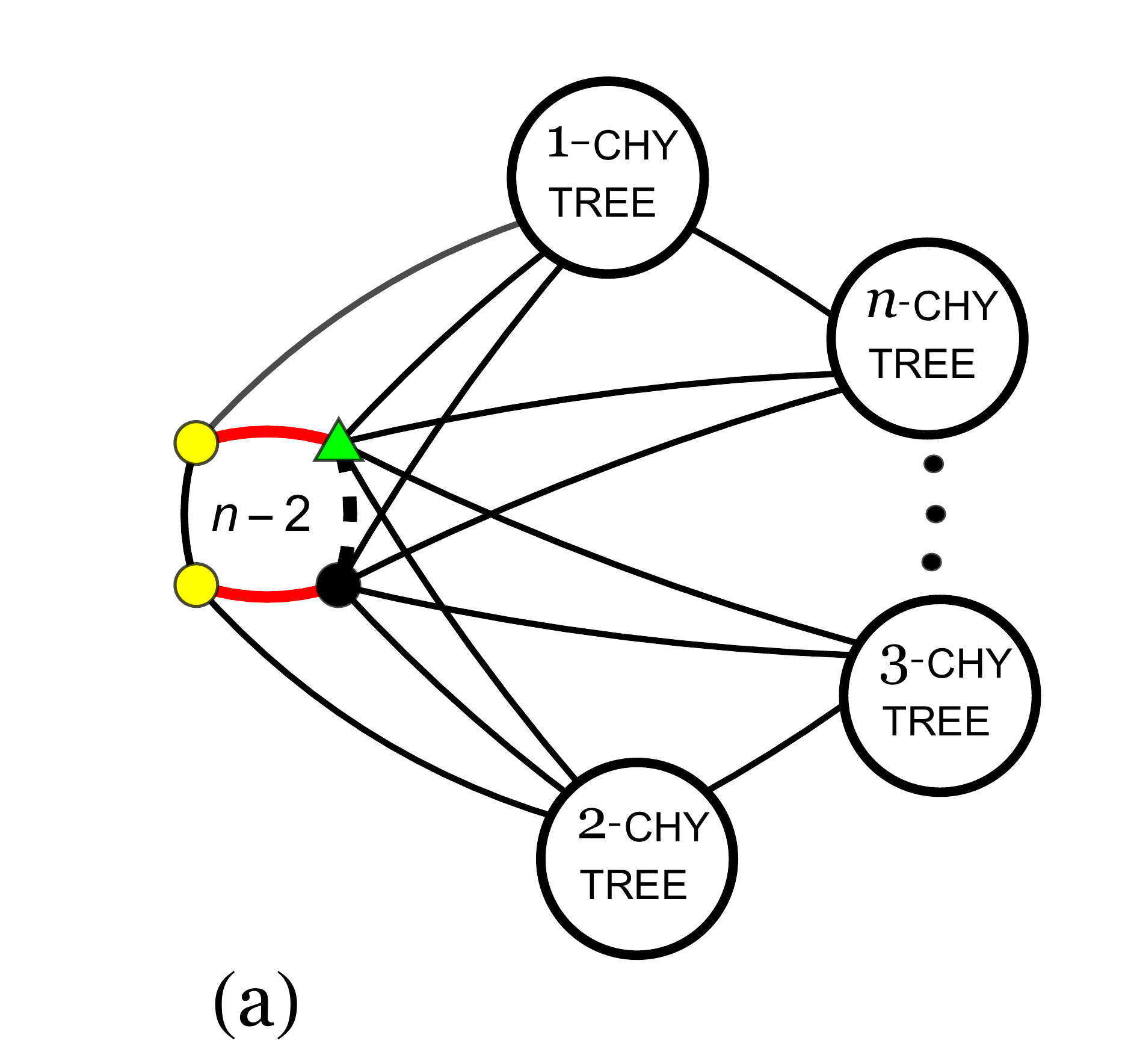}
		\includegraphics[width=2.0in]{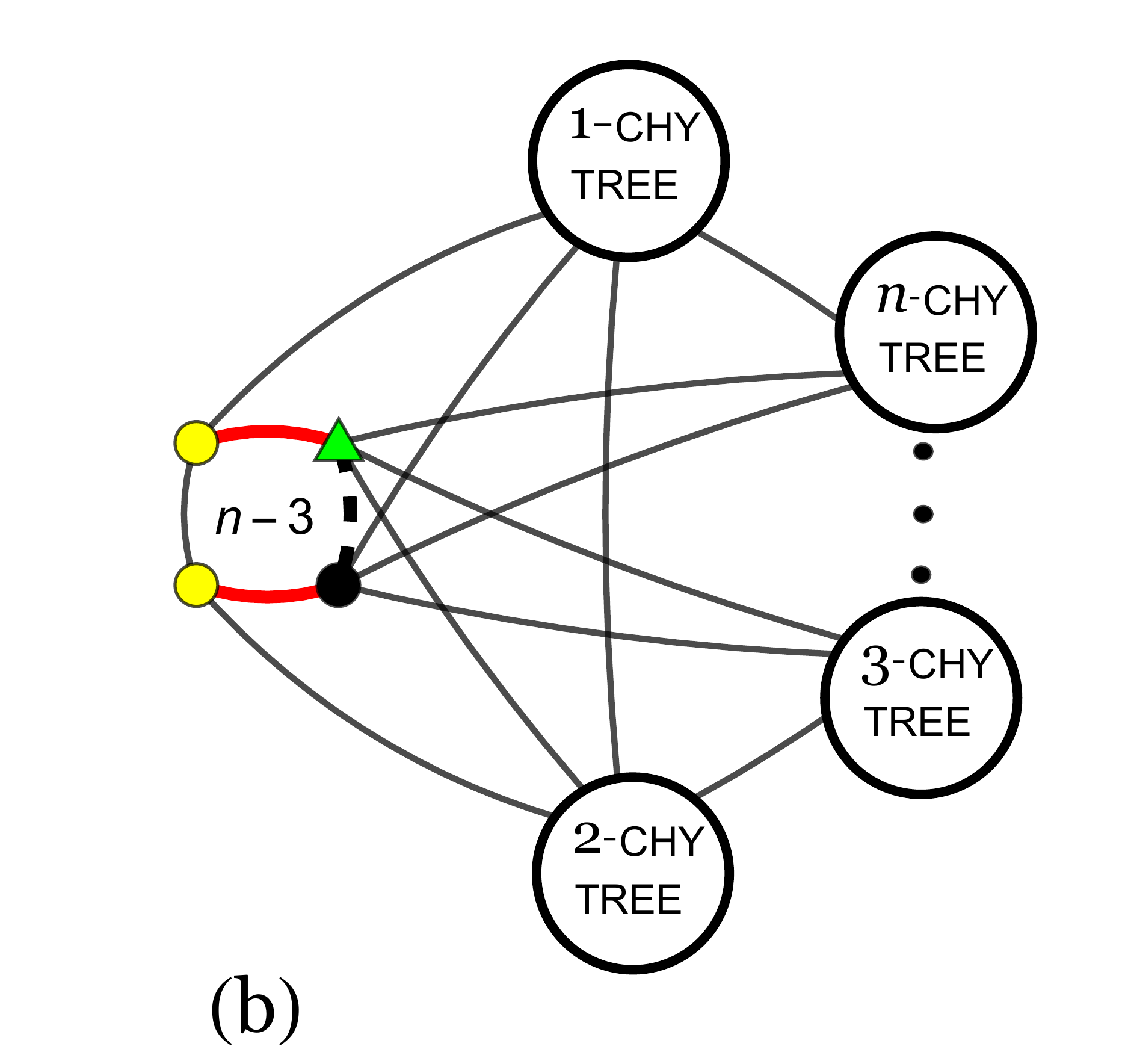}
			\includegraphics[width=2.0in]{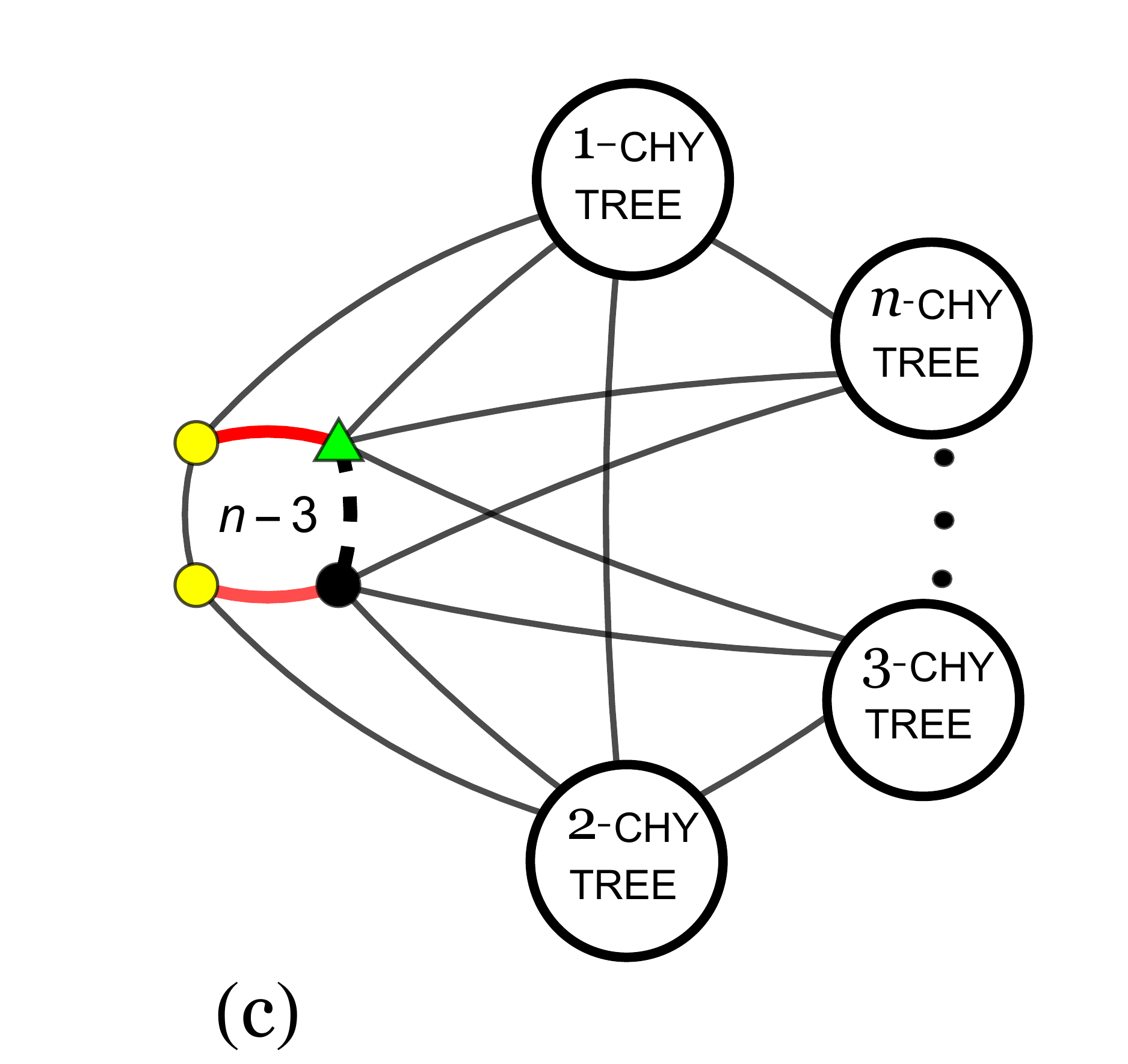}
	\caption{Fundamental CHY-Integrands at one-loop.}\label{3-CHY-graph}
\end{figure}

\begin{figure}[!h]
	\centering
	\includegraphics[width=2.0in]{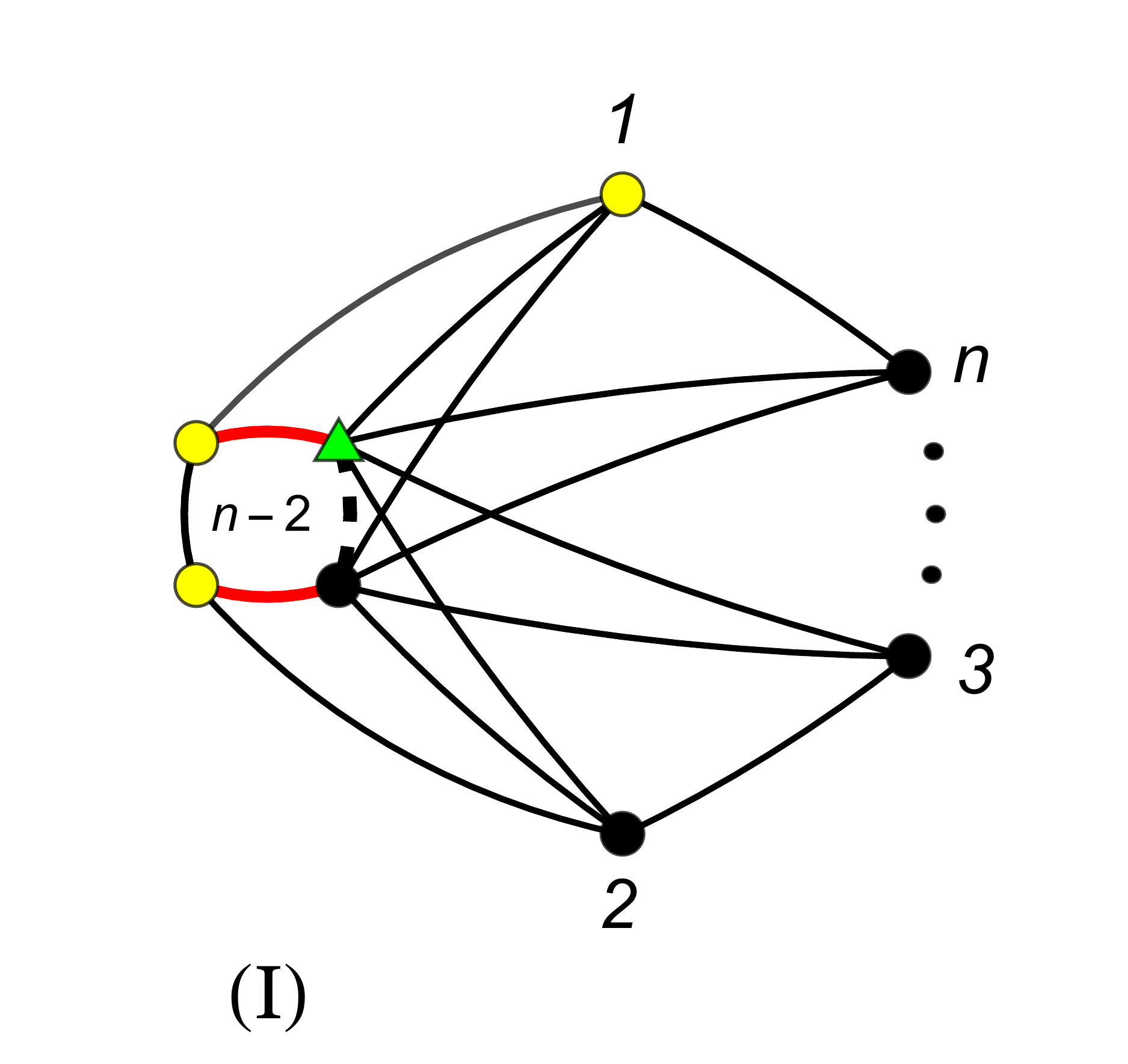}
		\includegraphics[width=2.0in]{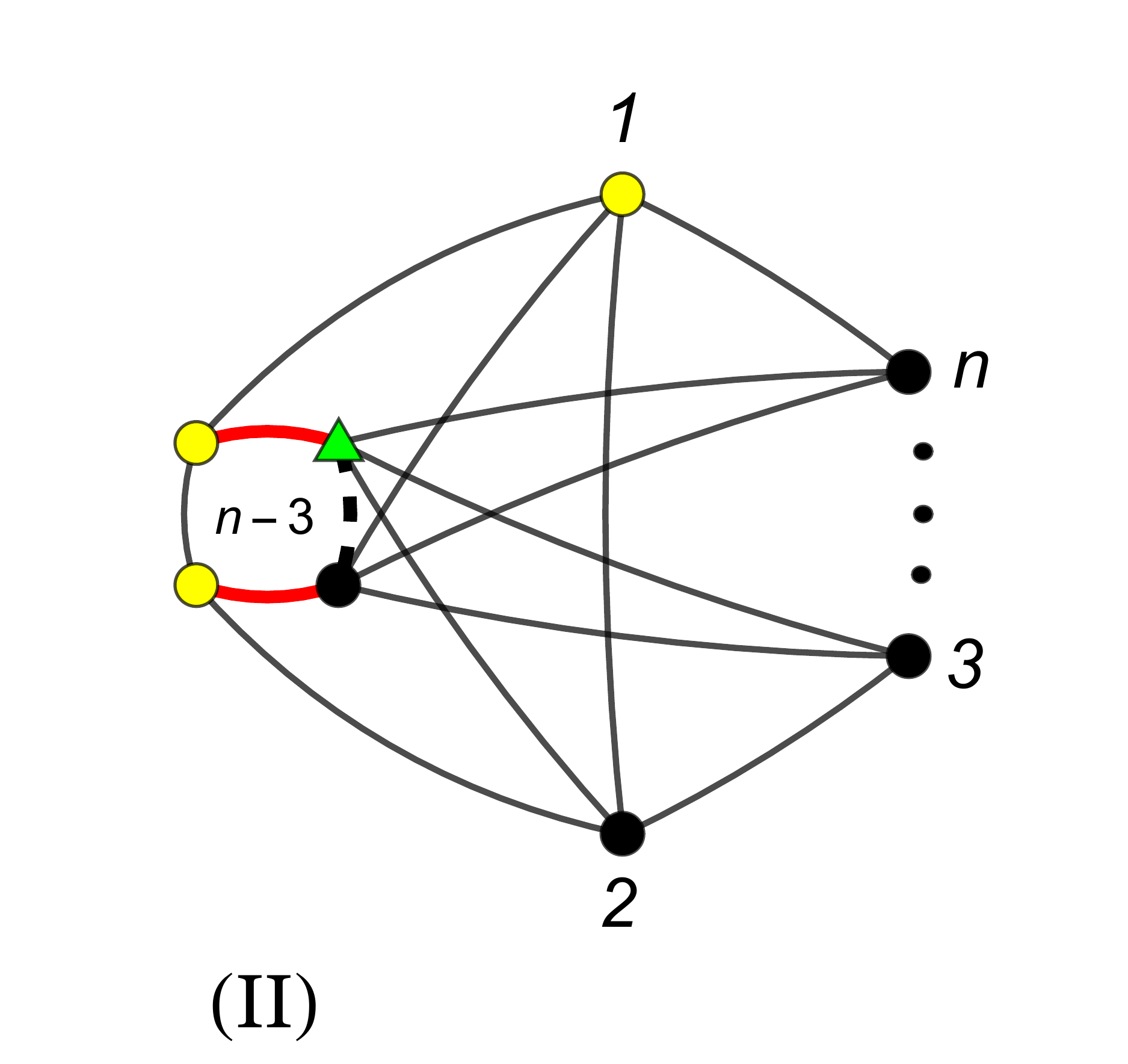}
			\includegraphics[width=2.0in]{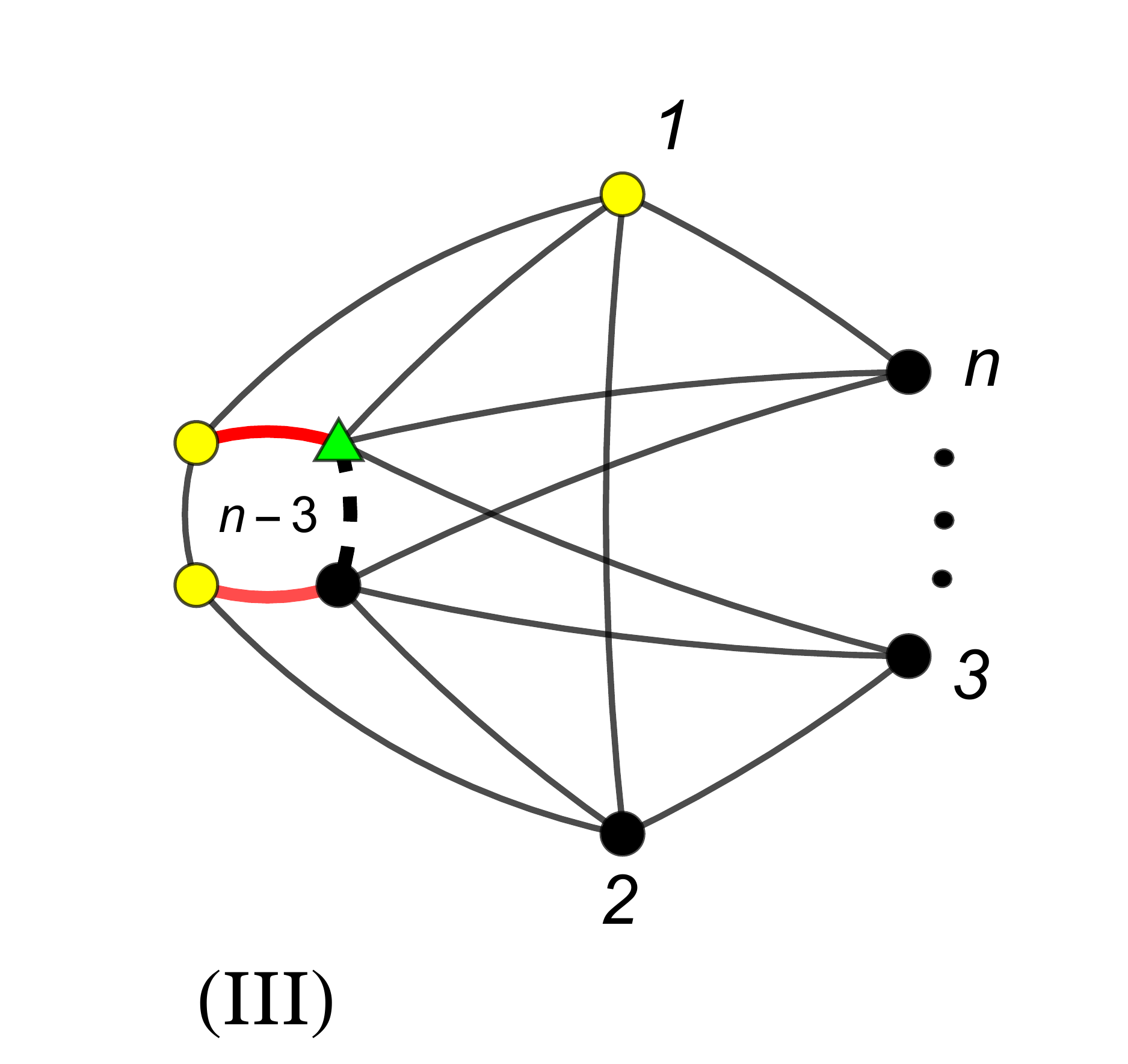}
	\caption{Simplest cases of CHY-graphs from Fig. \ref{3-CHY-graph}.}\label{G-CHY-graph}
\end{figure}

As it is going to be checked in section \ref{Sexamples},  the terms ${\bf D}^{\, a:b}_{\rm type- \,*}$'s
have  a physical meaning, as there is a relation between them and Feynman diagrams. Basically, the number of 
$\o^{a_1:a_2}_{i:j}$'s in 
$$
\mathcal{I}_L^{a_1:a_2}= {\rm PT_{tree}}[1,2,\ldots,n]\times {\bf D}^{\, a_1:a_2}_{\rm type- \,*} [n,n-1,\ldots, 1]
$$
corresponds to the number of legs attached to the loop in the Feynman diagram, and the term in 
$$
\mathcal{I}_R^{b_1:b_2}={\rm PT_{tree}}[1,2,\ldots,n] \times \left(  \s_{ij}\, \o ^{b_1:b_2}_{i:j} \right)
$$
with a single $\o^{b_1:b_2}_{i:j}$ will be in charge of its ordering.  In fact, the CHY-integrand, $\frac{\mathcal{I}_L^{a_1:a_2}\times \mathcal{I}_R^{b_1:b_2}}{(a_1,b_1,b_2,a_2)^2}$,  can be represented as a linear combination of the three CHY-graphs given in Fig. \ref{3-CHY-graph}.
\noindent
In particular, in Fig. \ref{G-CHY-graph} we consider the simplest cases. 
\noindent

In order to apply the $\L$-algorithm to integrate the CHY-graphs in Fig. \ref{G-CHY-graph},
we have fixed the punctures $\{\s_{1},\s_{b_1},\s_{b_2}\}$ by the ${\rm PSL}(2,\mathbb{C})$ symmetry and the puncture $\{ \s_{a_1} \}$ by the scale symmetry  \cite{Gomez:2016bmv}. Clearly, 
we are using a different gauge than the one introduced in \cite{Gomez:2017lhy}. This new gauge will allows us to work \`{a} la Feynman, i.e. each cut on the CHY-graph becomes a Feynman propagator, {\sl quadratic in momenta}.

The CHY-graphs in Fig. \ref{G-CHY-graph} come from the following set of integrands 
\ba\label{nomeg}
{\rm (I)}&&=\frac{{\rm PT}_{\rm tree}[1,\ldots, n]\,\mathbf{D}^{\, a_1:a_2}_{\rm type-0} [n,\ldots, 1]^{ n\o}\times{\rm PT}_{\rm tree}[1,\ldots,n]\,\s_{12}\,\o^{b_1:b_2}_{1:2}}{(a_1,b_1,b_2,a_2)^2} \\
&& =\frac{(n,n-1,\ldots ,1 )\times \o_{1:1}^{a_1:a_2} \o_{2:2}^{a_1:a_2} \cdots \o_{n:n}^{a_1:a_2}  \times \s_{12}\,\o^{b_1:b_2}_{1:2}}{(a_1,b_1,b_2,a_2)^2\times (1,2,\ldots ,n )\times (1,2,\ldots ,n )}
:=\mathbf{I}^{\rm CHY}_{{\rm (I)}}[1,\ldots ,n]^{1:2} +\cdots  \nonumber ,\\
&&\nonumber\\
{\rm (II)}&&= \frac{{\rm PT}_{\rm tree}[1,\ldots, n]\,\mathbf{D}^{\, a_1:a_2}_{\rm type-I} [n,\ldots, 1]^{ (n-1)\o}\times{\rm PT}_{\rm tree}[1,\ldots,n]\,\s_{12}\,\o^{b_1:b_2}_{1:2}}{(a_1,b_1,b_2,a_2)^2} 
\nonumber\\
&& =\cdots + \frac{(1^\s,n^\s\ldots ,3^\s,2^\o )^{a_1:a_2}\times \o_{3:3}^{a_1:a_2}  \cdots \o_{n:n}^{a_1:a_2}  \times \s_{12}\,\o^{b_1:b_2}_{1:2}}{(a_1,b_1,b_2,a_2)^2\times (1,2,\ldots ,n )\times (1,2,\ldots ,n )}
:= \mathbf{I}^{\rm CHY}_{{\rm (II)}}[1,\ldots ,n]^{1:2} +\cdots  \nonumber ,\\
&&\nonumber\\
{\rm (III)}&& =\frac{{\rm PT}_{\rm tree}[n,\ldots, 1]\,\mathbf{D}^{\, a_1:a_2}_{\rm type-I} [1,\ldots, n]^{ (n-1)\o}\times{\rm PT}_{\rm tree}[1,\ldots,n]\,\s_{12}\,\o^{b_1:b_2}_{1:2}}{(a_1,b_1,b_2,a_2)^2} 
\nonumber\\
&& = \cdots + \frac{(2^\s,\ldots, n^\s ,1^\o )^{a_1:a_2}\times \o_{3:3}^{a_1:a_2}  \cdots \o_{n:n}^{a_1:a_2}  \times \s_{12}\,\o^{b_1:b_2}_{1:2}}{(a_1,b_1,b_2,a_2)^2\times (1,2,\ldots ,n )\times (n,n-1,\ldots ,1 )}+\cdots
:= \mathbf{I}^{\rm CHY}_{{\rm (III)}}[1,\ldots ,n]^{1:2}+\cdots  \, ,
\nonumber
\ea
where ellipsis stand for additional terms with the same structure\footnote{Sometimes, it is necessary to manipulate the integrand to obtain the CHY-graphs in Fig. \ref{3-CHY-graph}.} as those given in the graphs of Fig. \ref{3-CHY-graph}.

Bearing in mind that the $\L-$algorithm is able to solve the CHY-graphs up to an overall sign the first point to tackle is how to fix this ambiguity\footnote{Notice that the overall sign can be fixed by the technique developed in \cite{Cardona:2016gon}, where the anti-lines have been considered.}.
Let us consider the integrands, $\mathbf{I}^{\rm CHY}_{{\rm (I)}}[1,\ldots ,n]^{1:2} , \mathbf{I}^{\rm CHY}_{{\rm (II)}}[1,\ldots ,n]^{1:2}$ and $\mathbf{I}^{\rm CHY}_{{\rm (III)}}[1,\ldots ,n]^{1:2} $ above. Pulling out the common factors they become
\ba\label{nomegOr}
\mathbf{I}^{\rm CHY}_{{\rm (I)}}[1,\ldots ,n]^{1:2}
&=& (-1)\left[(-1)^{n+1}\,\,\frac{\s_{a_1 a_2}^{n-2}}{\s_{a_1 b_1}^2 \s_{a_2b_2}^2 \s_{b_1b_2}}\times\frac{1}{ \prod_{i=1}^n  \s_{ia_1}\s_{ia_2} }    \times \frac{1}{\s_{1b_1} \s_{2 b_2}\s_{23}\cdots \s_{n1}}\right] ,\\
\mathbf{I}^{\rm CHY}_{{\rm (II)}}[1,\ldots ,n]^{1:2}
&=& (-1)\left[(-1)^{n+1}\,\,\frac{\s_{a_1 a_2}^{n-3}}{\s_{a_1 b_1}^2 \s_{a_2b_2}^2 \s_{b_1b_2}}\times\frac{1}{\s_{1a_2}\s_{2a_1} \prod_{i=3}^n\s_{i a_1}\s_{i a_2}}    \times \frac{1}{\s_{1b_1} \s_{2 b_2}(1,\ldots,n)}\right]\nonumber ,\\
\mathbf{I}^{\rm CHY}_{{\rm (III)}}[1,\ldots ,n]^{1:2}
&=& (-1)\left[(-1)^{n+1}\,\,\frac{\s_{a_1 a_2}^{n-3}}{\s_{a_1 b_1}^2 \s_{a_2b_2}^2 \s_{b_1b_2}}\times\frac{1}{\s_{1a_1}\s_{2a_2} \prod_{i=3}^n\s_{i a_1}\s_{i a_2}}    \times \frac{1}{\s_{1b_1} \s_{2 b_2}(1,\ldots,n)}\right]\nonumber.
\ea
We claim that the terms in the square brackets have a direct representation in terms of the CHY-graphs in Fig. \ref{G-CHY-graph}.


In the following we make use of the $\L-$algorithm to perform the set of integrals
$\mathbf{I}^{\rm CHY}_{{\rm (I)}}[1,\ldots ,n]^{1:2}$, $\mathbf{I}^{\rm CHY}_{{\rm (II)}}[1,\ldots ,n]^{1:2}$ and $\mathbf{I}^{\rm CHY}_{{\rm (III)}}[1,\ldots ,n]^{1:2} $. The conjectured overall sign inside the brackets in (\ref{nomegOr}), $(-1)^{n+1}$,  is checked numerically afterwards\footnote{For the numerical checking we have fixed the punctures and scattering equations, $\{(\s_1, E_1), (\s_2,E_2), (\s_3,E_3)\}$, namely, the Faddeev-Popov determinant  becomes $(\s_{12}\s_{23}\s_{31})^2$.} .

\vspace{6mm}

\noindent {\bf Proposition 1:} {\sl There is an equality among the CHY-integral of
 ${\rm (I)}$, (\ref{nomegOr}), and the one-loop n-point Feynman integrand}
\begin{eqnarray}\label{computation-npts}
\mathfrak{I}^{\rm CHY}_{\rm (I)}[1,\ldots, n]^{1:2} :=\frac{1}{2^{n+1}}\int d\Omega \,\, s_{a_1b_1} \int d\mu_{n+4}^{\rm tree}\,\,
\mathbf{I}^{\rm CHY}_{{\rm (I)}}[1,\ldots ,n]^{1:2}\,\,
=
\hspace{-0.3cm}
\parbox[c]{10em}{\includegraphics[scale=0.2]{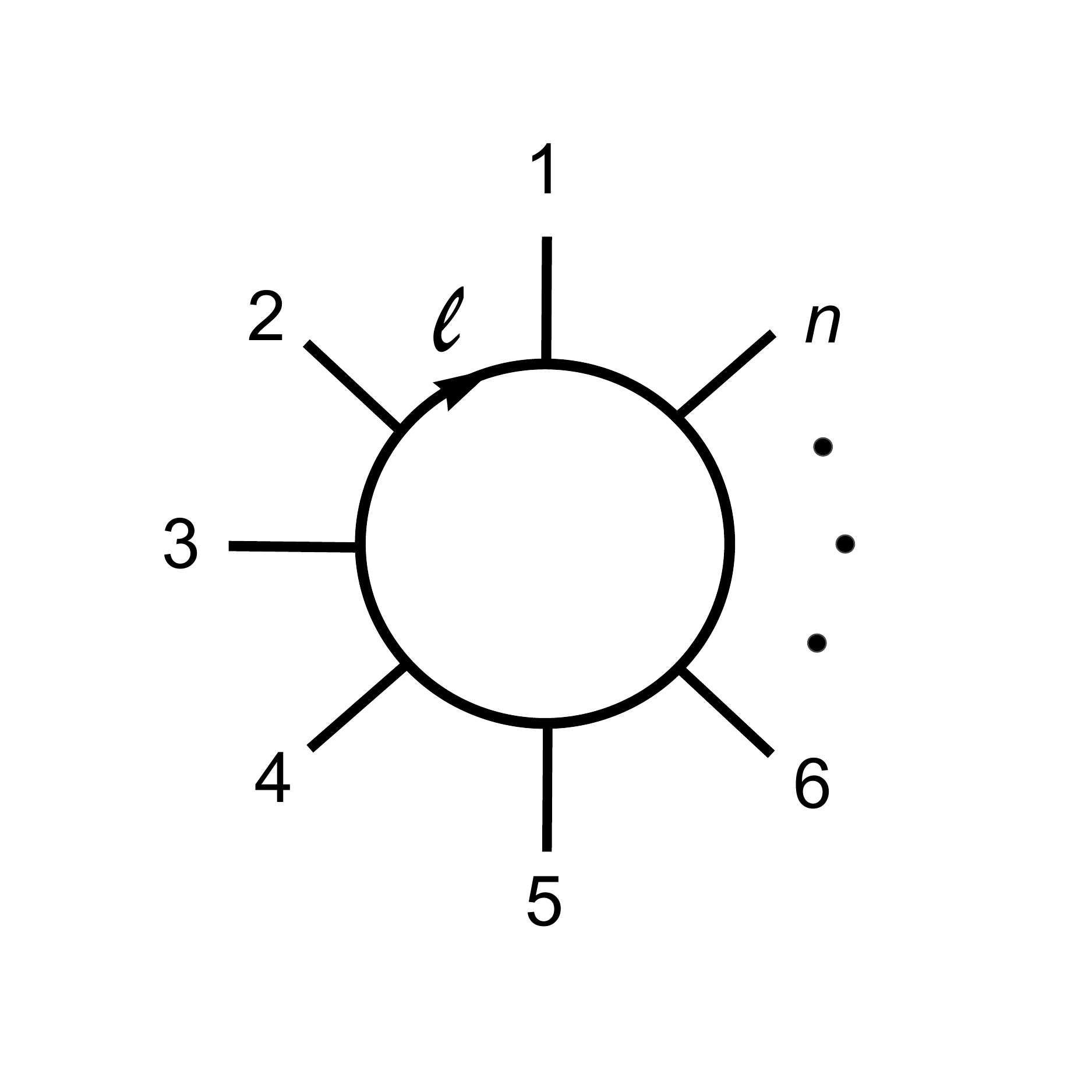}}\,\,
\label{new1}
\end{eqnarray}

\vspace{6mm}

\noindent {\bf Proof:}  The result follows from a direct calculation, 
\begin{eqnarray}
&&\mathfrak{I}^{\rm CHY}_{\rm (I)}[1,\ldots, n]^{1:2} = \frac{1}{2^{n+1}}\int d\Omega \times s_{a_1b_1}\times \int d\mu_{n+4}^{\rm tree} \,\,
\mathbf{I}^{\rm CHY}_{{\rm (I)}}[1,\ldots ,n]^{1:2}\nonumber \\ 
&&= \frac{(-1)} {2^{n+1}}
\int d\Omega \, s_{a_1b_1}\, \int d\mu_{n+4}^{\rm tree} 
\left[(-1)^{n+1}\,\,\frac{\s_{a_1 a_2}^{n-2}}{\s_{a_1 b_1}^2 \s_{a_2b_2}^2 \s_{b_1b_2}}\times\frac{1}{\prod_{i=1}^n  \s_{ia_1}\s_{ia_2} }    \times \frac{1}{\s_{1b_1} \s_{2 b_2}\s_{23}\cdots \s_{n1}}\right]\nonumber \\
&&= \frac{(-1)} {2^{n+1}}
\int d\Omega \, s_{a_1b_1}\, \int d\mu_{n+4}^{\rm tree} 
\parbox[c]{10em}{\includegraphics[scale=0.21]{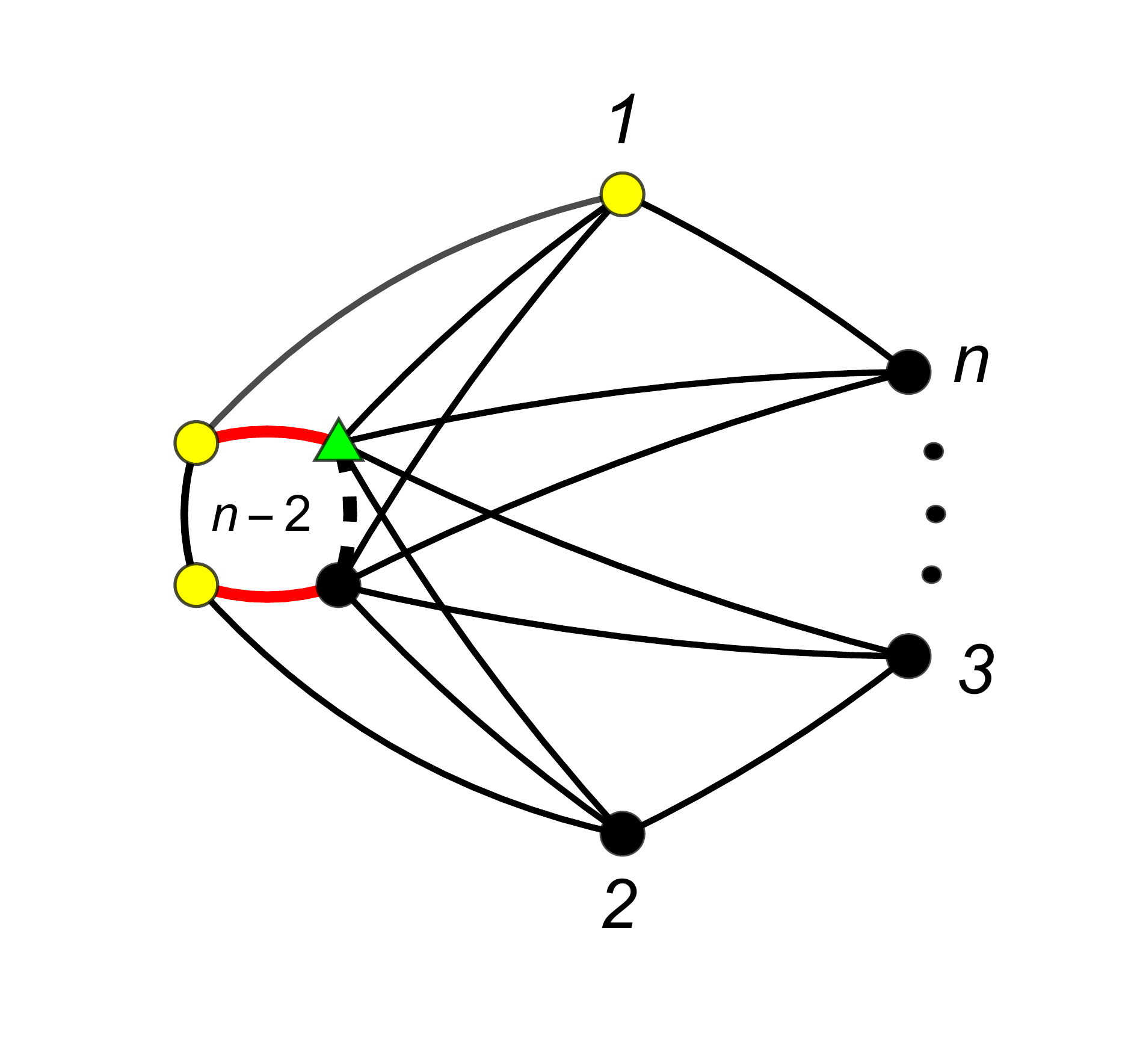}}\,\,,
\label{G-graph11}
\end{eqnarray}
where in the last step we have mapped each element in the expression inside the bracket to the CHY-graph:  The first factor is the box, namely the loop momenta sector,  the second one is in charge of connecting the loop momenta to all the external points, and the third factor is the one that gives the ordering. 
With these identifications all the integrands of this type will look alike but with the external points permuted.

In order to compute the $\int d\mu_{n+4}^{\rm tree}$ integral within the $\L$-algotithm \cite{Gomez:2016bmv}, we introduce the ``nearest neighbour'' gauge fixing.  One can show that it is necessary to perform a total of $(n+1)-$consecutive cuts, each one introducing a single propagator, like in the Feynman diagrams.  For instance,  using the $\L-$rules, there is only one non zero cut on the CHY-graph in \eqref{G-graph11}, as it is shown on left graph in \eqref{1loopnp}. 
\begin{eqnarray}
\centering
\parbox[c]{10em}{\includegraphics[scale=0.24]{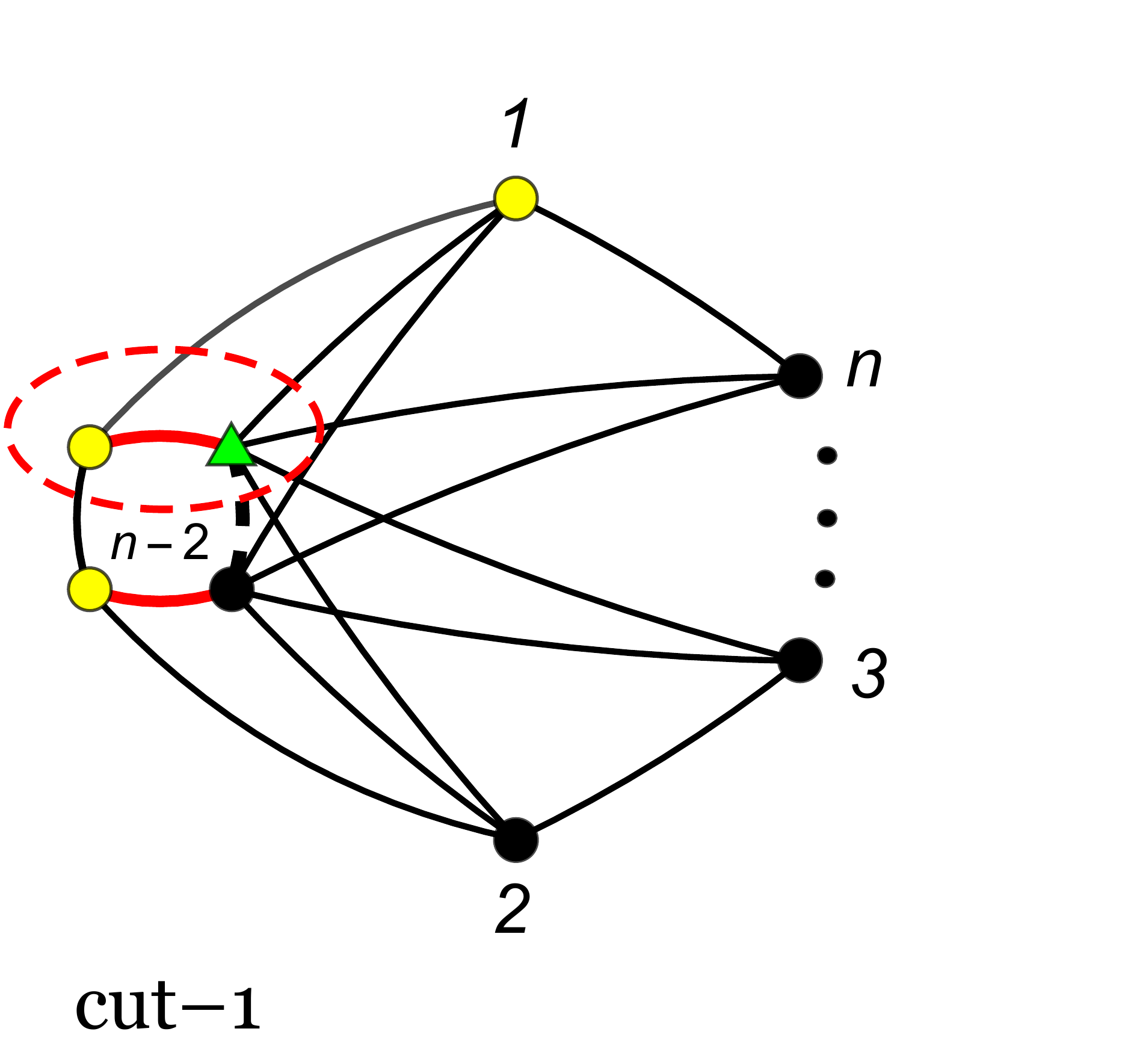}} 
=\,\,\,
\frac{1}{k_{a_1 b_1}} \times
\left(
\parbox[c]{12em}{\includegraphics[scale=0.24]{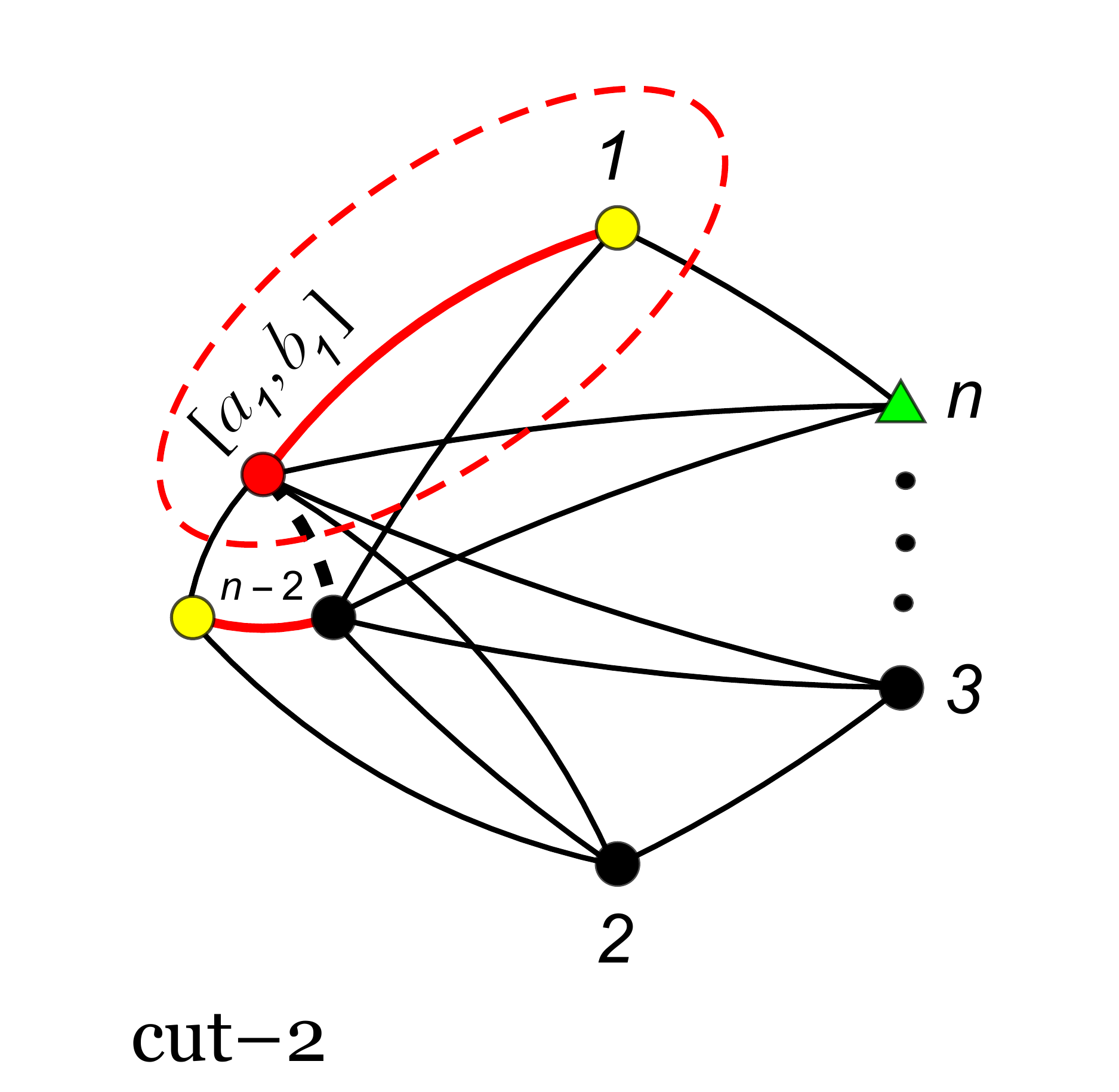}}
\right)\,\, .
\label{1loopnp}
\end{eqnarray}
This first cut gives the factor $\frac{1}{k_{a_1b_1}}$ and the resulting CHY-graph contains now  a massive puncture with momentum,  $k_{a_1}+k_{b_1}$, which is connected with a double line to the puncture $\s_1$. By scale symmetry we fix the nearest neighbour to $\s_1$ (next to its right), i.e.  $\s_n$ , such as it is shown on right graph in \eqref{1loopnp}, see also Fig. \ref{loopS}.

The resulting CHY-graph in \eqref{1loopnp} contains also only one non zero cut,  which we have dubbed {\bf cut-2}. This cut is simple to compute and its result is given by
\begin{eqnarray}
\centering
\parbox[c]{10em}{\includegraphics[scale=0.24]{G-cut2.pdf}} \,\,
=\,\,\,
\frac{1}{k_{234\cdots n a_2 b_2}} \times
\left(
\parbox[c]{12em}{\includegraphics[scale=0.24]{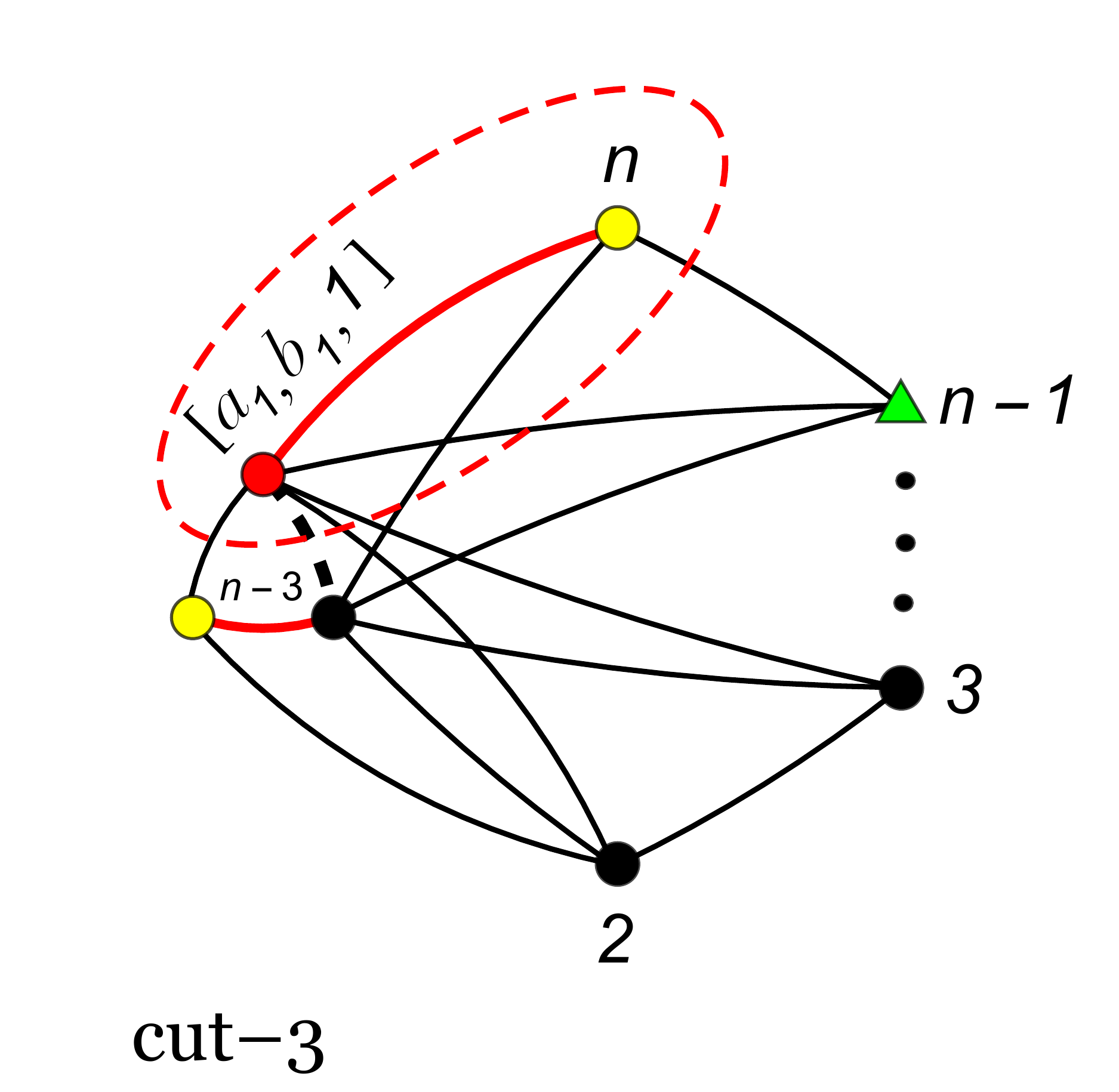}}
\right),
\label{G-cut_2}
\end{eqnarray}
where clearly the resulting graph keeps the same form but with one less puncture.  Iterating $(n-2)$ times this procedure we are led with 
\begin{eqnarray}
\centering
\parbox[c]{9em}{\includegraphics[scale=0.24]{G-cut1.pdf}} 
=
\frac{1}{k_{a_1 b_1} \times k_{234\cdots  n a_2 b_2}\times \cdots \times k_{2a_2 b_2}} \times
\left(
\parbox[c]{8em}{\includegraphics[scale=0.24]{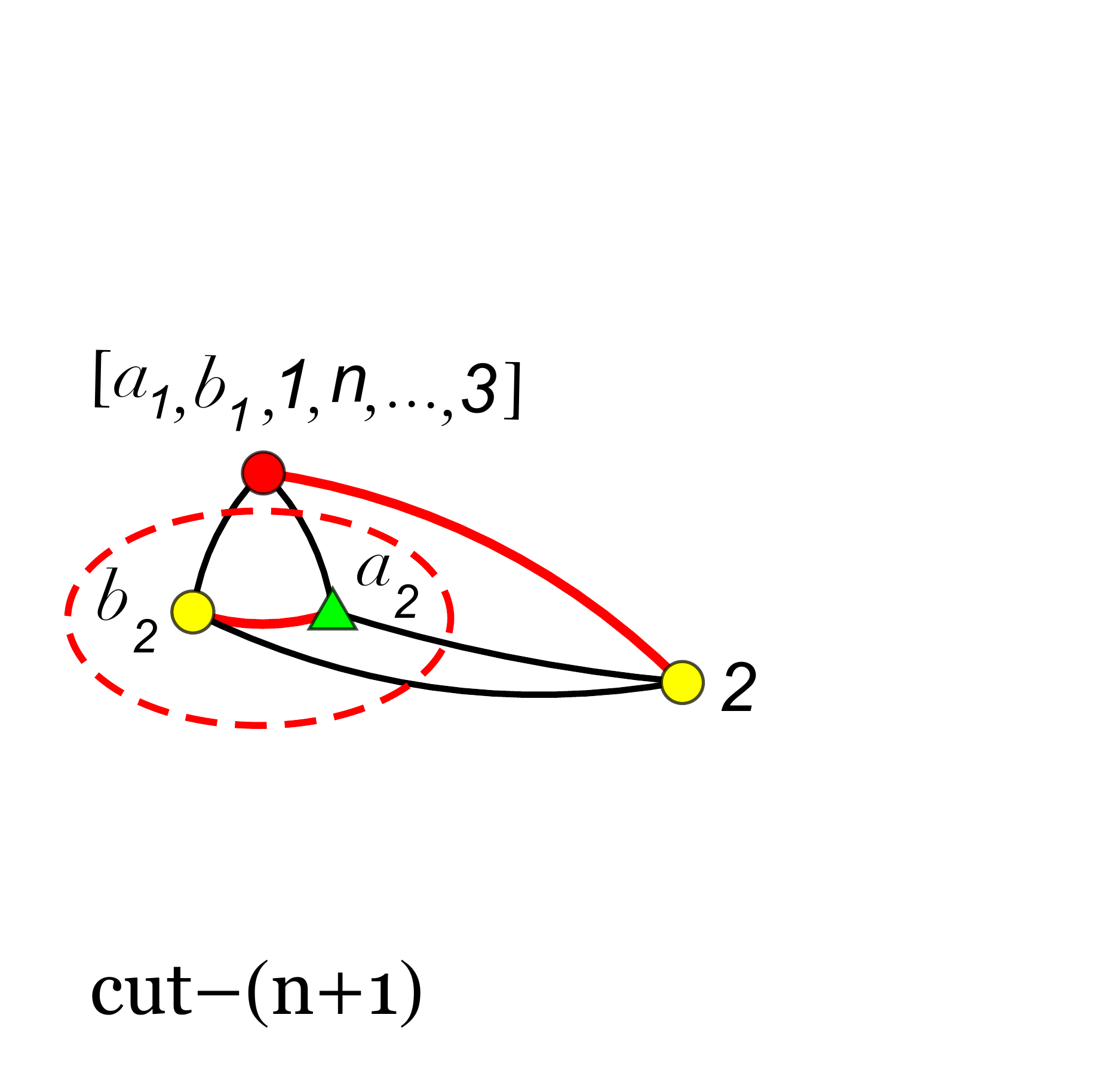}}
\right)
\label{G-cut_n}
\end{eqnarray}
This final resulting graph has just one non-zero cut,
as it is shown in \eqref{G-cut_n}, which we have called ``{\bf cut-(n+1)}" and its result by the $\L-$rules is  $\frac{1}{k_{a_2b_2}}$. 

Summarizing, so far we have proven the equality
\begin{eqnarray}
 \int d\mu_{n+4}^{\rm tree} 
\parbox[c]{9em}{\includegraphics[scale=0.21]{G-graph11.pdf}}
=
\frac{2^{n+1}}{s_{a_1 b_1} \times s_{1 a_1 b_1}\times  s_{1n a_1 b_1}\times \cdots\times  s_{1n (n-1) \cdots 3a_1 b_1} \times s_{a_2 b_2}} \,\, ,\nonumber
\end{eqnarray}
where the momentum conservation condition, $\sum_{i=1}^n k_i+k_{a_1}+k_{a_2}+k_{b_1}+k_{b_2}=0$, has been used.

Carrying out the last integration, $\int d\Omega\times s_{a_1 b_1}$, we are left with the following expression
\begin{eqnarray}
&&
\frac{1}{2^{n+1}}\int d\Omega\times s_{a_1 b_1}  \times \frac{2^{n+1}}{s_{a_1 b_1} \times s_{1 a_1 b_1}\times  s_{1n a_1 b_1}\times \cdots\times  s_{1n (n-1) \cdots 3a_1 b_1} \times s_{a_2 b_2}} \nonumber\\
&&
=
\frac{1}{\ell^2(\ell+k_1)^2(\ell+k_1+k_n)^2(\ell+k_1+k_n+k_{n-1})^2\cdots(\ell+\sum_{i,i\neq 2}^nk_i)^2} \,\,=
\parbox[c]{9em}{\includegraphics[scale=0.2]{feyn-p.pdf}} \,\, ,
\nonumber
\end{eqnarray}
where we have finally identified the algebraic expression with the pictorical one-loop n-point Feynman integrand. This equality, which corresponds to our initial claim, has also been checked numerically.
$\quad\blacksquare$

As a final remark for this first proposition, we can generalize  
the previous calculation to the case of attaching CHY tree-level graphs  instead of points as in \eqref{G-graph11}. Schematically it is
\begin{eqnarray}
\frac{1}{2^{N+1}}\int d\Omega\times  s_{a_1 b_1} \int d\mu_{N+4}^{\rm tree} 
\parbox[c]{10em}{\includegraphics[scale=0.21]{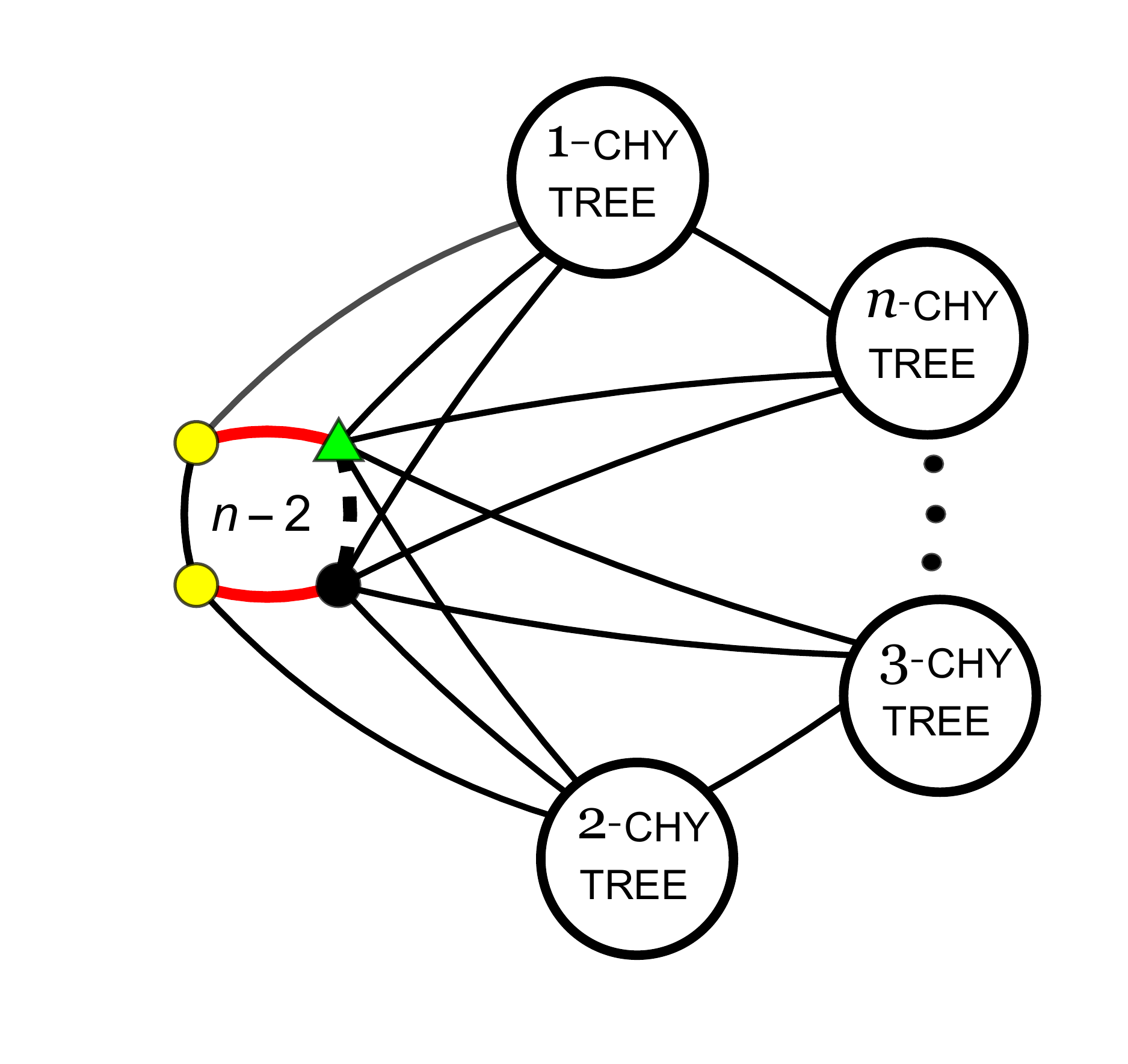}}
=
\parbox[c]{8em}{\includegraphics[scale=0.2]{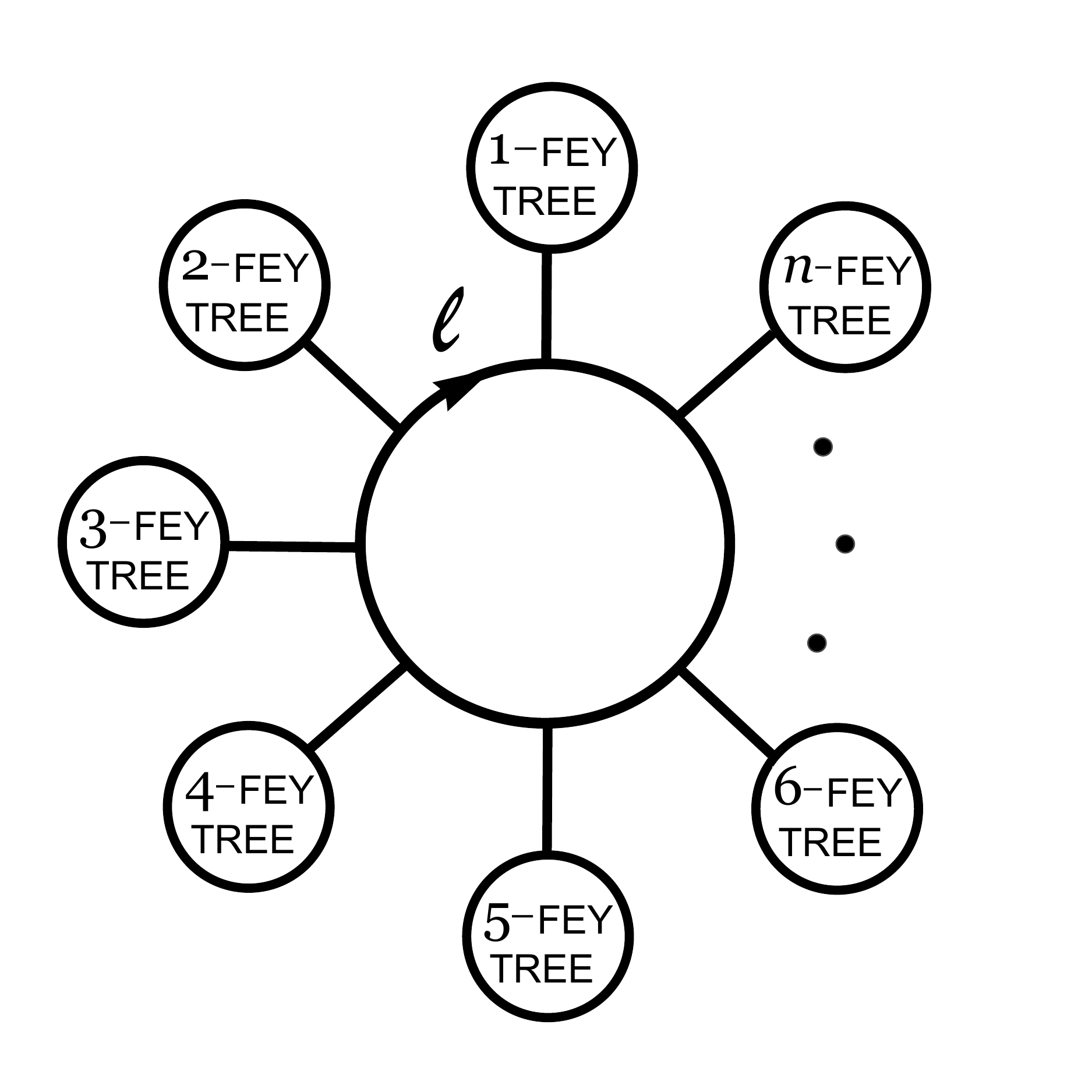}}\,\, ,
\label{gen-diag}
\end{eqnarray}
where $N$ is the total number of external particles. The gluing process to join the CHY tree-level graphs at the one-loop skeleton can be found in \cite{Gomez:2016cqb,Cardona:2016wcr}.
These kind of graphs can be solved in the same fashion as we did for the $n$-point case.

\vspace{6mm}

\noindent {\bf Proposition 2:} {\sl The CHY-integral of
 ${\rm (II)}$, (\ref{nomegOr}), identically vanishes}.
\begin{eqnarray}
&&\mathfrak{I}^{\rm CHY}_{\rm (II)}[1,\ldots, n]^{1:2} := \frac{1}{2^{n+1}}\int d\Omega \times s_{a_1b_1}\times \int d\mu_{n+4}^{\rm tree} \,\,
\mathbf{I}^{\rm CHY}_{{\rm (II)}}[1,\ldots ,n]^{1:2}= 0\, .
\label{new2}
\end{eqnarray}

\vspace{6mm}

\noindent {\bf Proof:}  As in the previous case, the proof is straightforward, but a little tedious.
\begin{eqnarray}
&&\mathfrak{I}^{\rm CHY}_{\rm (II)}[1,\ldots, n]^{1:2} = \frac{1}{2^{n+1}}\int d\Omega \times s_{a_1b_1}\times \int d\mu_{n+4}^{\rm tree} \,\,
\mathbf{I}^{\rm CHY}_{{\rm (II)}}[1,\ldots ,n]^{1:2}\nonumber \\ 
&&= -
\int \frac{d\Omega}{2^{n+1}} \, s_{a_1b_1}\, \int d\mu_{n+4}^{\rm tree} 
\left[(-1)^{n+1}\,\,\frac{\s_{a_1 a_2}^{n-3}}{\s_{a_1 b_1}^2 \s_{a_2b_2}^2 \s_{b_1b_2}}\times\frac{1}{\s_{1a_2}\s_{2a_1} \prod_{i=3}^n\s_{i a_1}\s_{i a_2}}    \times \frac{1}{\s_{1b_1} \s_{2 b_2}(1,\ldots,n)}\right]\nonumber \\
&&= \frac{(-1)} {2^{n+1}}
\int d\Omega \, s_{a_1b_1}\, \int d\mu_{n+4}^{\rm tree} 
\parbox[c]{10em}{\includegraphics[scale=0.21]{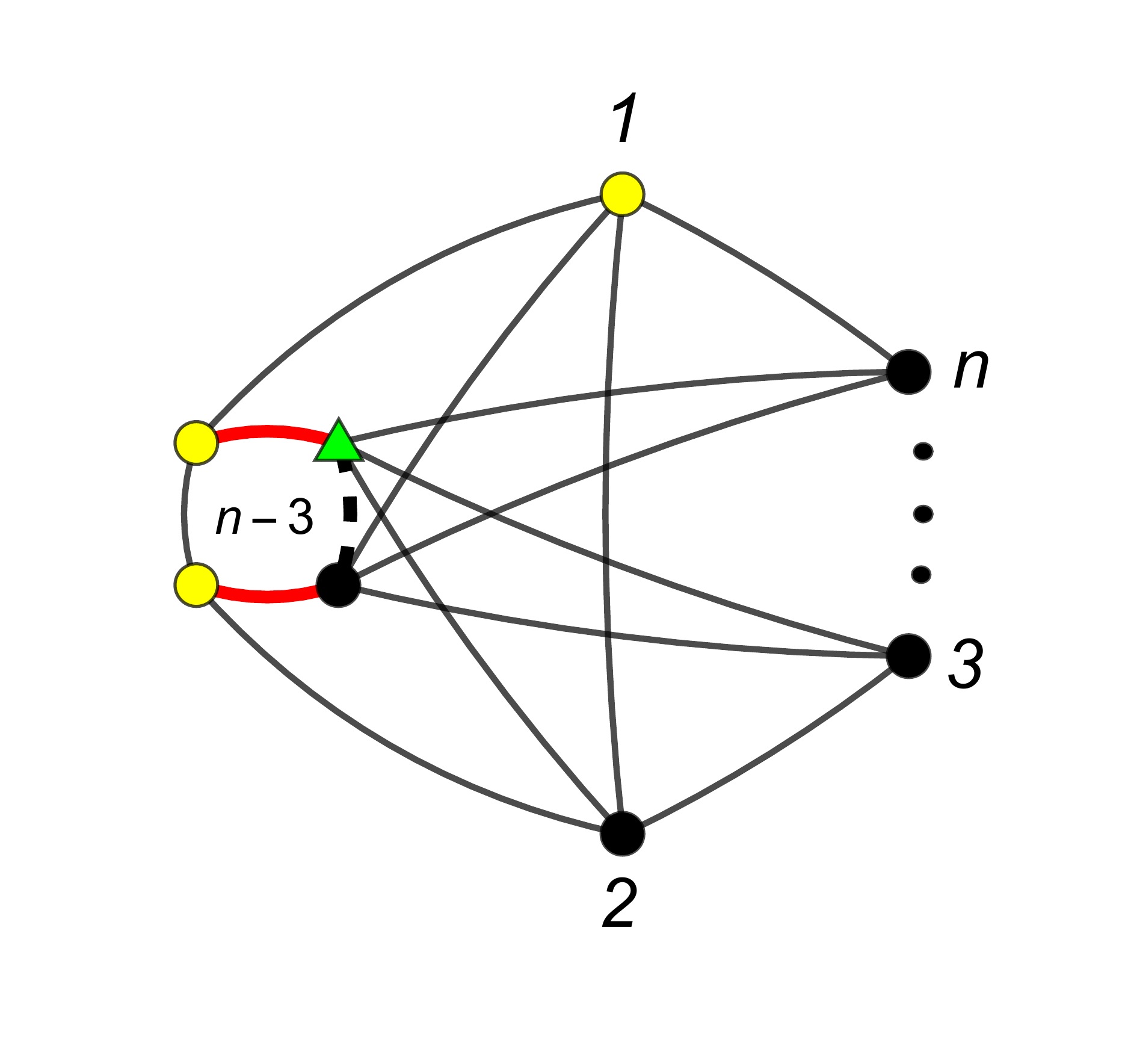}}\,\, .
\label{G-graph22}
\end{eqnarray}

Applying the $\L$-rules on the graph in \eqref{G-graph22} to compute the integral, $\int d\mu^{\rm tree}_{n+4}$, one obtains that there is just one non-zero cut
\begin{eqnarray}
\centering
\parbox[c]{10.3em}{\includegraphics[scale=0.24]{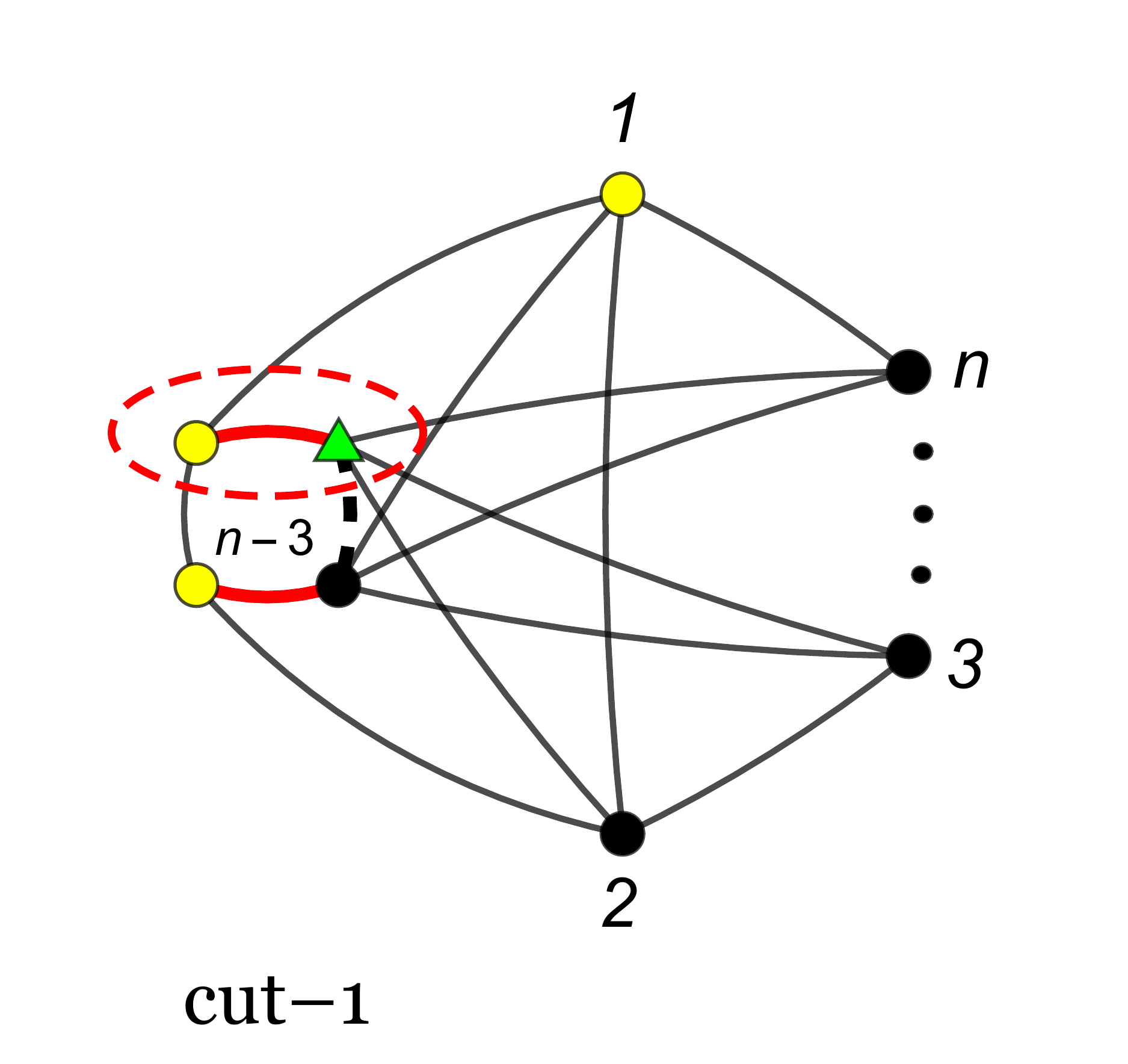}} 
=\,\,
\frac{1}{k_{a_1 b_1}} \times
\left(
\parbox[c]{12em}{\includegraphics[scale=0.24]{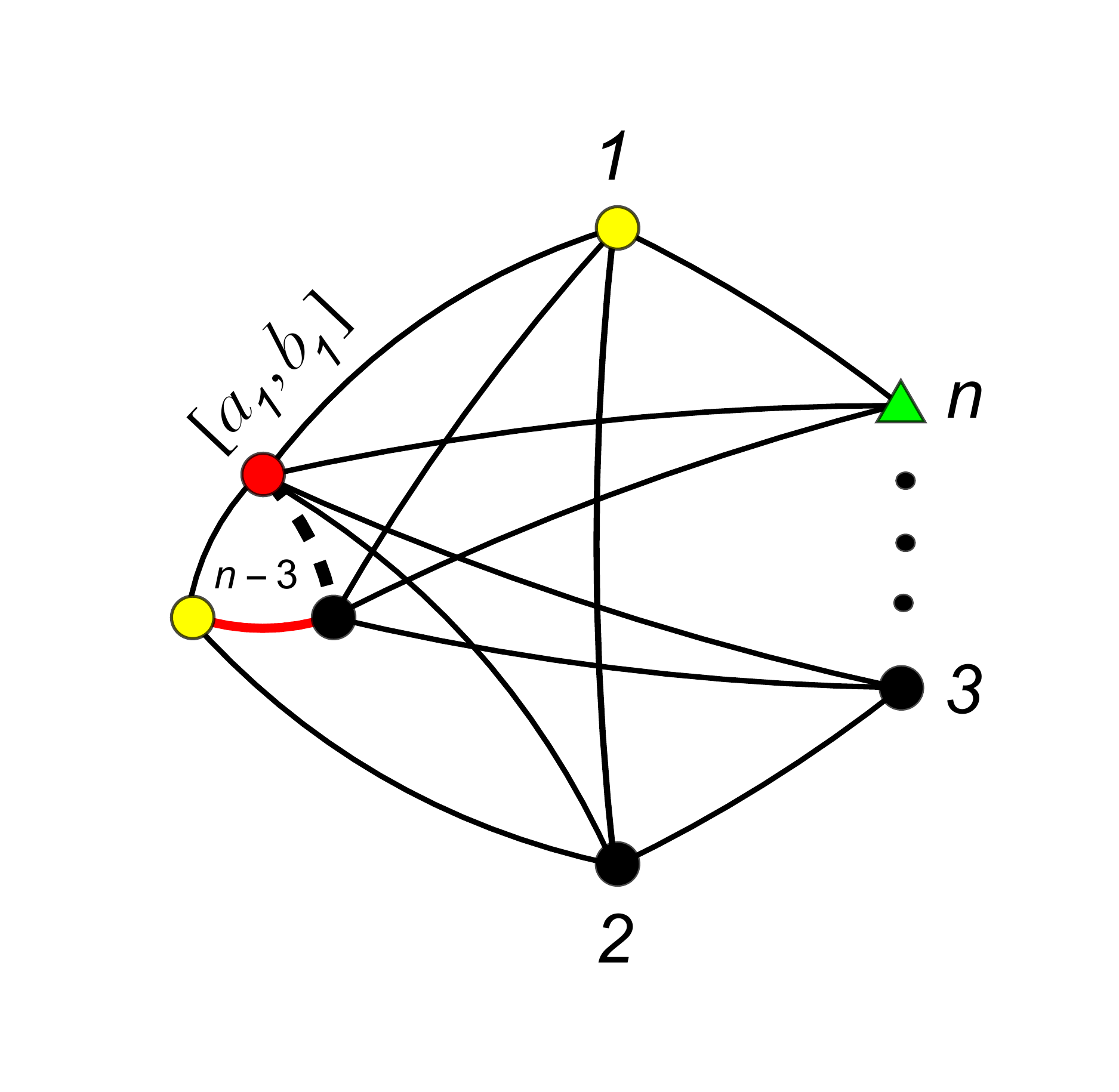}}
\right)\,\, .
\label{1loopnp2}
\end{eqnarray}

Furthermore all possible cuts on the resulting CHY-graph in \eqref{1loopnp2} are zero, therefore one can conclude that the integral, $\mathfrak{I}^{\rm CHY}_{\rm (II)}[1,\ldots, n]^{1:2}$, vanishes, i.e.  
\begin{eqnarray}
\mathfrak{I}^{\rm CHY}_{\rm (II)}[1,\ldots, n]^{1:2} 
= \frac{(-1)} {2^{n+1}}
\int d\Omega \, s_{a_1b_1}\, \int d\mu_{n+4}^{\rm tree} 
\parbox[c]{10em}{\includegraphics[scale=0.21]{G-graph22.pdf}}=0\,\, .\quad\blacksquare
\label{G-graph222}
\end{eqnarray}

This calculation can be generalized to the case of attaching CHY tree-level graphs  instead of points. Schematically it is
\begin{eqnarray}
\int d\Omega \, s_{a_1b_1}\, \int d\mu_{N+4}^{\rm tree} 
\parbox[c]{10em}{\includegraphics[scale=0.21]{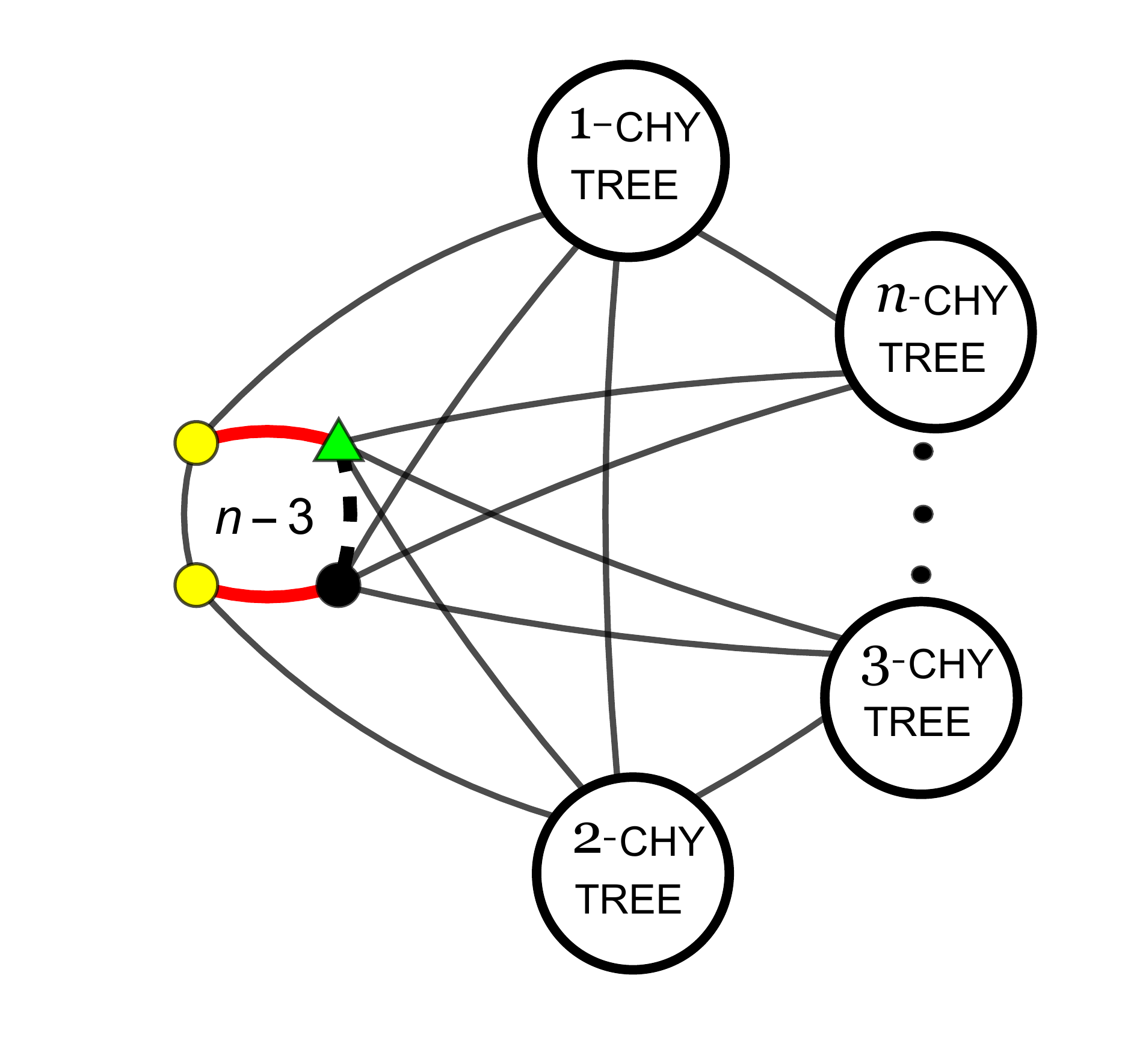}}=0\,\, ,
\label{Gen-graph2222}
\end{eqnarray}
where $N$ is the total number of external particles. These type of graphs, with tree-level CHY-graphs attached, can be solved in a similar way as we did for the case in \eqref{G-graph222}.


\vspace{6mm}

\noindent {\bf Proposition 3:} {\sl There is a  correspondence between the CHY-integral of
${\rm  (III)}$, (\ref{nomegOr}), and the one-loop n-point Feynman integrand}
\begin{eqnarray} \label{integral-3}
\mathfrak{I}^{\rm CHY}_{\rm (III)}[1,\ldots, n]^{1:2} := \frac{1}{2^{n+1}}\int d\Omega \,\, s_{a_1b_1} \int d\mu_{n+4}^{\rm tree} \,\,
\mathbf{I}^{\rm CHY}_{{\rm (III)}}[1,\ldots ,n]^{1:2}
= (-1)
\hspace{-0.3cm}
\parbox[c]{12em}{\includegraphics[scale=0.2]{feyn-p.pdf}}\,\,
\label{G-graph12}
\end{eqnarray}
\vspace{6mm}

\noindent {\bf Proof:}  The result follows from a direct calculation, 
\begin{eqnarray}
&&\mathfrak{I}^{\rm CHY}_{\rm (III)}[1,\ldots, n]^{1:2} = \frac{1}{2^{n+1}}\int d\Omega \times s_{a_1b_1}\times \int d\mu_{n+4}^{\rm tree} \,\,
\mathbf{I}^{\rm CHY}_{{\rm (III)}}[1,\ldots ,n]^{1:2}\nonumber \\ 
&&= \frac{(-1)} {2^{n+1}}
\int d\Omega \, s_{a_1b_1}\, \int d\mu_{n+4}^{\rm tree} 
\left[(-1)^{n+1}\,\frac{\s_{a_1 a_2}^{n-3}}{\s_{a_1 b_1}^2 \s_{a_2b_2}^2 \s_{b_1b_2}}\times\frac{1}{\s_{1a_1}\s_{2a_2} \prod_{i=3}^n\s_{i a_1}\s_{i a_2}}    \times \frac{1}{\s_{1b_1} \s_{2 b_2}(1,\ldots,n)}\right]\nonumber \\
&&= \frac{(-1)} {2^{n+1}}
\int d\Omega \, s_{a_1b_1}\, \int d\mu_{n+4}^{\rm tree} 
\parbox[c]{10em}{\includegraphics[scale=0.21]{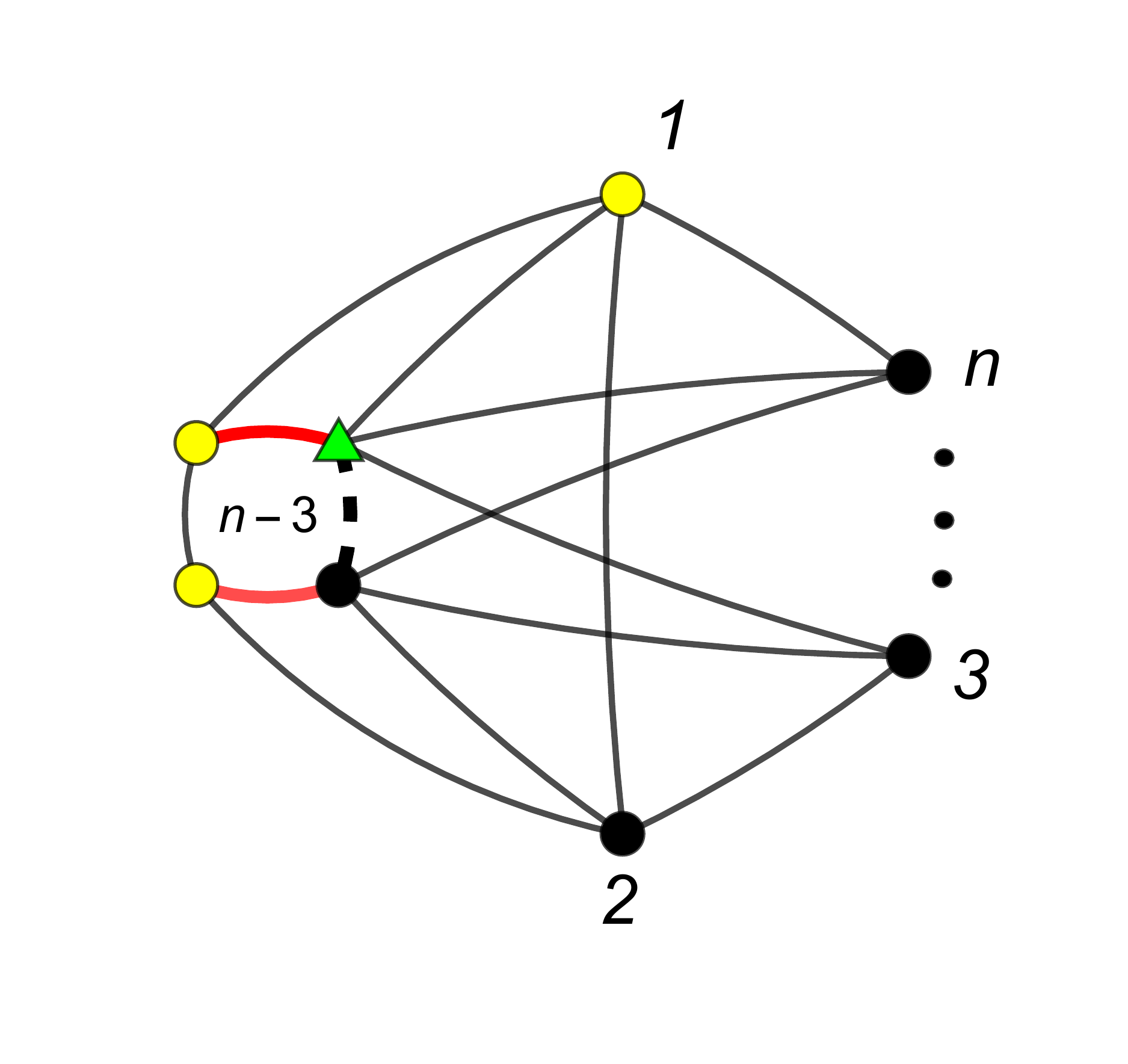}}\,\, .
\label{G-graph33}
\end{eqnarray}
To compute this integral we use the previous results together with the cross-ratio identity
\ba
\mathbb{1}=-\frac{\s_{12}\,\s_{a_1a_2}}{\s_{1a_2}\s_{2a_1}}+\frac{\s_{1a_1}\s_{2a_2}}{\s_{1a_2}\s_{2a_1}}\,.
\ea
It is straightforward to check that by multiplying the CHY-integrand (or graph) in \eqref{G-graph33}  times the above identity  and using Proposition 2, the integral,  $\mathfrak{I}^{\rm CHY}_{\rm (III)}[1,\ldots, n]^{1:2}$, becomes
\begin{align}\label{i3-i1}
\mathfrak{I}^{\rm CHY}_{\rm (III)}[1,\ldots, n]^{1:2} &= \frac{1}{2^{n+1}}\int d\Omega \times s_{a_1b_1}\times \int d\mu_{n+4}^{\rm tree} \,\,
\mathbf{I}^{\rm CHY}_{{\rm (III)}}[1,\ldots ,n]^{1:2} \times \mathbb{1} \nonumber\\
&=-\mathfrak{I}^{\rm CHY}_{\rm (I)}[1,\ldots, n]^{1:2}+\mathfrak{I}^{\rm CHY}_{\rm (II)}[1,\ldots, n]^{1:2}\nonumber \\
&=-\mathfrak{I}^{\rm CHY}_{\rm (I)}[1,\ldots, n]^{1:2} \, ,
\end{align}
and consequently
\begin{eqnarray}
\frac{1}{2^{n+1}}\int d\Omega\times  s_{a_1 b_1} \int d\mu_{n+4}^{\rm tree} 
\parbox[c]{9.5em}{\includegraphics[scale=0.21]{G-graph33.pdf}}
= \,\,\,(-1)
\parbox[c]{9em}{\includegraphics[scale=0.2]{feyn-p.pdf}} \,\,\blacksquare
\label{menos-n-gon-ordering}
\end{eqnarray}

This result can also be generalized for external CHY tree graphs. Schematically it is
\begin{eqnarray}
\frac{1}{2^{N+1}}\int d\Omega\times  s_{a_1 b_1} \int d\mu_{N+4}^{\rm tree} 
\parbox[c]{10em}{\includegraphics[scale=0.21]{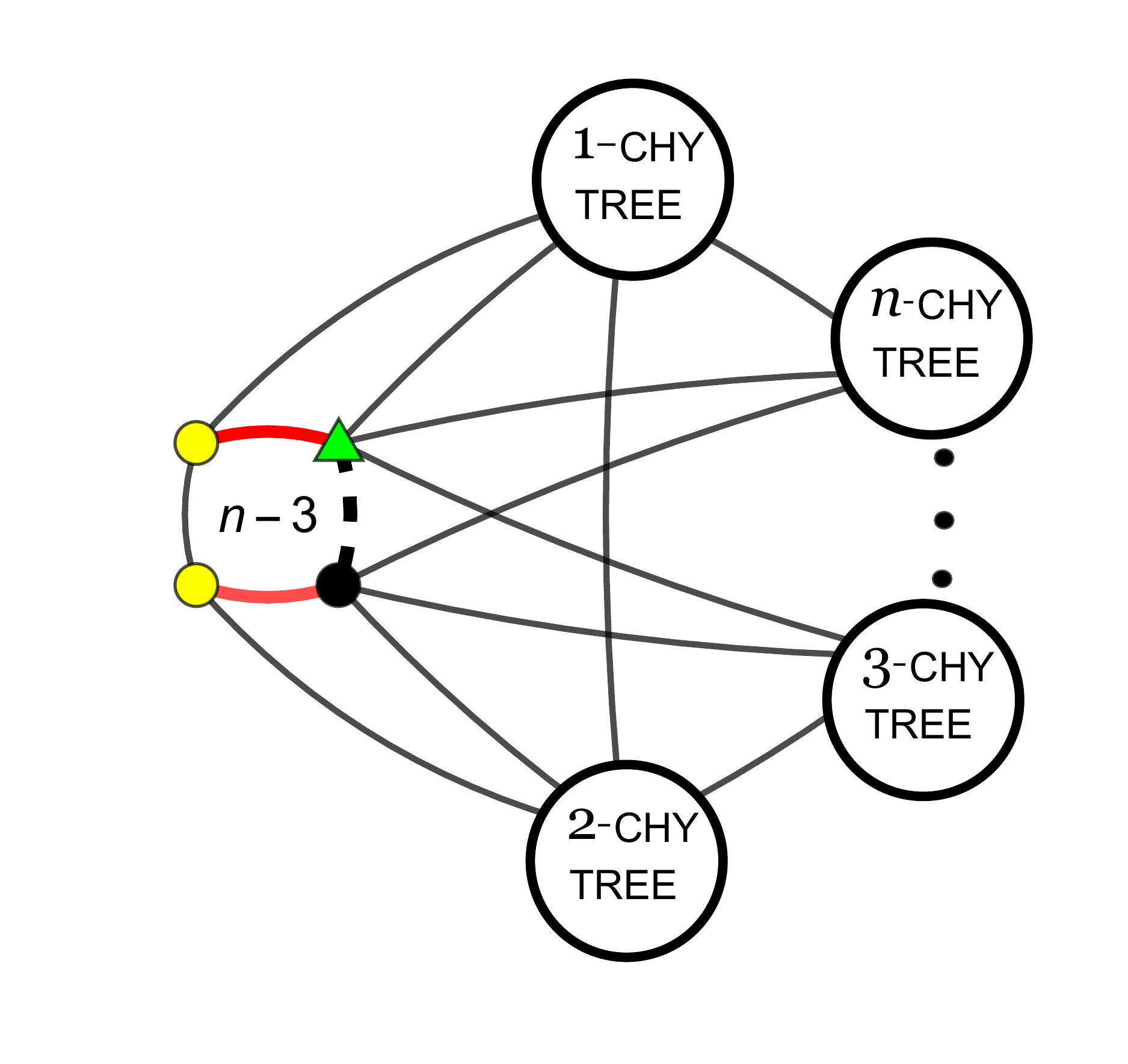}}
= \,\,\,(-1)
\parbox[c]{9em}{\includegraphics[scale=0.2]{G-Fey-type_a.pdf}}\,\, ,
\end{eqnarray}
where $N$ is the total number of external particles and again we can solve these type of CHY-graphs like we have done in the previous cases.

Now that we have a classification for the type of integrands that will appear in our calculations we can employ the results in the computation of the integrands for the Bi-adjoint $\Phi^3$ scalar theory.

\section{The bi-adjoint $\Phi^3$ scalar theory}\label{biadjoint}

The scattering amplitudes at tree level for the massless bi-adjoint $\Phi^3$ scalar theory are given by the elements \cite{Cachazo:2013iea}
\begin{align}
m_n^{\rm tree}[\pi|\rho]:= {1 \over 2^{n-3}}\int \,\, d\mu_n^{\rm tree} \,\,
\cal{I}_{\rm t}^{\rm CHY}[\pi|\rho]
\label{mtree}
\end{align}
where 
\begin{equation}
\cal{I}_{\rm tree}^{\rm CHY}[\pi |\rho] := {\rm PT}_{\rm tree} [\pi]\times{\rm PT}_{\rm tree} [\rho]
\end{equation}
and the measure, $d\mu_n^{\rm tree}$, is given in \eqref{dmu}.

In \cite{Cachazo:2013iea}, it was shown that the integral, $m_n^{\rm tree}[\pi|\rho]$, is composed by the sum over all the trivalent Feynman diagrams containing
 two planar embeddings, consistent with the $\pi$ ($\rho$) ordering respectively\footnote{There are many techniques to compute $m_n^{\rm tree}[\pi|\rho]$, such as ones given in \cite{Gomez:2016cqb,Chen:2017edo,  Cardona:2016gon,Huang:2016zzb,Cardona:2015ouc,Kalousios:2015fya,Mafra:2016ltu,Cachazo:2013iea,Baadsgaard:2015voa,Cachazo:2015nwa,Cardona:2015eba}.}. Specifically $m_n^{\rm tree}[\pi|\rho]$ reduces to the sum over all elements contained in the intersection among these two planar ordering. Schematically it can be written  as \cite{Cachazo:2013iea}
\begin{align}
m_n^{\rm tree}[\pi|\rho]= (-1)^{n-3+n_{\pi |\rho}}m_n^{\rm tree}[\pi|\pi]\cap m_n^{\rm tree}[\rho|\rho]\,,
\label{ord_intersc}
\end{align}
where $n_{\pi|\rho}$ is determined by the number of flips between two permutations $\pi, \rho$ \cite{Cachazo:2013iea}. 
For example, let us consider the element, $m_4^{\rm tree}[1234|1243]$, it reads
\begin{eqnarray}
m_4^{\rm tree}[1234|1243]=m_4^{\rm tree}[1234|1234]\cap m_4^{\rm tree}[1243|1243] =
\parbox[c]{10em}{\includegraphics[scale=0.38]{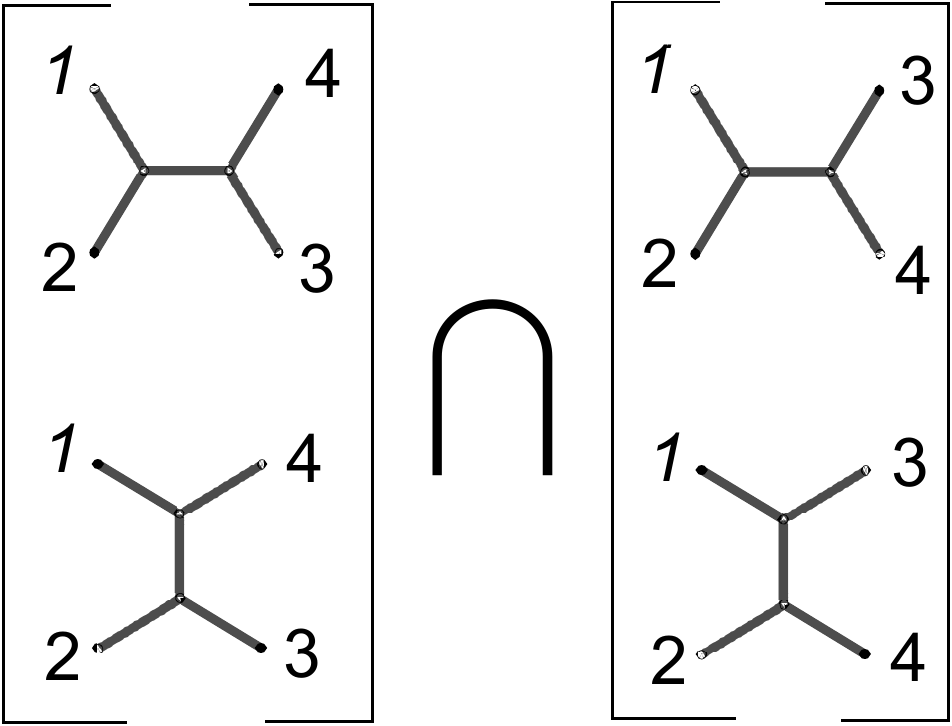}}=
\parbox[c]{6em}{\includegraphics[scale=0.41]{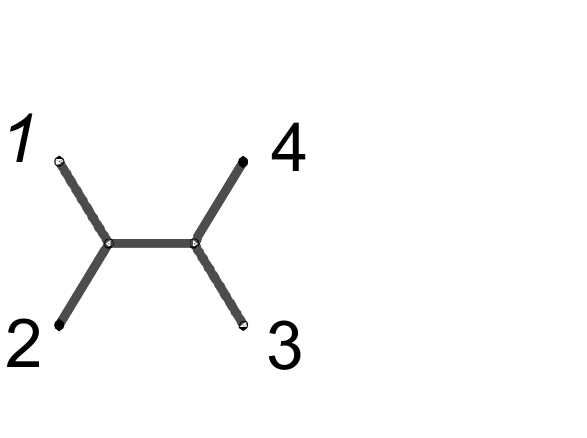}}
\end{eqnarray} 
which is the right answer. In the previous example one can appreciate the advantage of \eqref{ord_intersc}, since the calculation for both orderings is exactly the same but with external legs permuted to the given ordering. Therefore, we only have to do the full calculation in the canonical ordering (i.e. (1,2,...,n)). This will prove extremely useful for the one-loop calculations.

\subsection{Bi-adjoint $\Phi^3$ at One-Loop}\label{oldP}

The CHY prescription to obtain the scattering integrands at one-loop, respecting the planar orderings ($\pi$, $\rho$), is given by the elements  
\begin{align}\label{oldproposal}
m^{\rm 1-loop}_n[\pi|\rho] = {1 \over 2^{n-1}\,\, \ell^2}\int d\mu^{\rm 1-loop}_{n+2} \,\,
\mathcal{I}_{\rm 1-loop}^{\rm CHY}[\pi|\rho],
\end{align}
where 
\begin{equation}
\mathcal{I}_{\rm 1-loop}^{\rm CHY}[\pi|\rho]:= \frac{1}{(\ell^+,\ell^-)^2}\times {\bf PT}_{\rm 1-loop}^{\,\ell^+:\ell^-}[\pi]\times {\bf PT}_{\rm 1-loop}^{\,\ell^+:\ell^-}[\rho]
\end{equation}
and the measure $d\mu^{\rm 1-loop}_{n+2}$
\begin{align}\label{1LL-measure}
d\mu^{\rm1-loop}_{n+2} :=&\frac{d\s_{\ell^+}  d\s_{\ell^-}  \prod_{a=1}^n  d\s_{a}   } {{\rm Vol}\,\,({\rm PSL}(2,\mathbb{C}))}\times \frac{ (\s_{\ell^+\ell^-}\,\s_{\ell^- 1}\, \s_{1 \ell^+}  )   }{\prod_{a=2}^{n} E^{\rm 1-loop}_{a}},\nonumber\\
&   ^{\underrightarrow{\,\quad{\rm fixing\,\, PSL}(2,\mathbb{C})\,\quad }}\,\,\,\prod_{i=2}^{n} \frac{d\s_{i}}{E^{\rm 1-loop}_{i}}\times (\s_{\ell^+\ell^-}\,\s_{\ell^- 1}\, \s_{1 \ell^+}  )^2
\end{align}
with
\begin{align}\label{SE_1-loop}
&E_a^{\rm 1-loop}:= \sum^n_{b=1\atop b\neq a}\frac{k_a\cdot k_b}{\s_{ab}} +  \frac{k_a\cdot \ell^+}{\s_{a\ell^+}} + \frac{k_a\cdot \ell^-}{\s_{a\ell^-}},\quad a=1,\ldots ,n\\
& E_{\ell ^{\pm}}^{\rm 1-loop}:= \sum^{n}_{b=1}\frac{\ell^{\pm}\cdot k_b}{\s_{\ell^{\pm}b}}, 
\qquad  (\ell^+)^\mu=-(\ell^-)^\mu := \ell^\mu, \quad \ell^2\neq 0. \nonumber 
\end{align}
where, without loss of generality,  we have fixed $\{\s_{\ell^+},\s_{\ell^-},\s_{1} \}$
and  $\{ E^{\rm 1-loop}_{\ell^+},E^{\rm 1-loop}_{\ell^-},E^{\rm 1-loop}_{1} \}$.
 
As it has been shown in \cite{Geyer:2015jch, He:2015yua,Baadsgaard:2015hia},  the CHY-integral in \eqref{oldproposal} reproduces the linear propagators in the internal loop momentum $\ell^\mu$, i.e. in the Q-cut representation \cite{Baadsgaard:2015twa,Huang:2015cwh,Feng:2016msc}. Therefore, as it is very well known, these results match with the traditional Feynman propagators just after using the partial fraction identity and performing a shift in the loop momentum.

\subsection{A New Proposal}\label{NEWproposal}

In this section we propose a new CHY prescription  for the  S-matrix at one-loop of the bi-adjoint $\Phi^3$ scalar theory, that takes into account the planar orderings. In this proposal we will be able reproduce directly the quadratic propagators, in the same way as the traditional Feynman approach.


Borrowing the line of reasoning in  \cite{Gomez:2017lhy} and the prescription \eqref{oldproposal} we 
arrive at:

\vspace{6mm}

\noindent {\bf Definition:} {\sl The partial amplitude $\mathfrak{M}_n^{\rm 1-loop}[\pi|\rho]$ is given as}\,
\begin{align}\label{newproposal}
\mathfrak{M}_n^{\rm 1-loop}[\pi|\rho]:= {1 \over 2^{n+1}}\int d\Omega \times (k_{a_1}+k_{b_1})^2\times \int d\mu_{n+4}^{\rm tree} \,\,
\mathbf{I}_{\rm 1-loop}^{\rm CHY}[\pi|\rho],
\end{align}
with 
\begin{equation}\label{chyint_1-loop}
\mathbf{I}_{\rm 1-loop}^{\rm CHY}[\pi |\rho] :=  
 {\bf PT}^{\, a_1:a_2}_{\rm 1-loop} [\pi]  \times  \frac{1}{(a_1,b_1,b_2,a_2)^2}    \times     {\bf PT}^{\, b_1:b_2}_{\rm 1-loop} [\rho]  .
\end{equation}

\vspace{6mm}

We conjecture that the amplitude, $\mathfrak{M}_n^{\rm 1-loop}[\pi|\rho]$, reproduces the integrands of the S-matrix elements  at one-loop for the $\Phi^3$ bi-adjoint massless theory.  We are going to present several examples in order to support this statement.


Notice that despite the similarity among the prescriptions \eqref{oldproposal} and \eqref{newproposal}, there are significant differences between them: {\sl i)} The total number of punctures do not match, namely in \eqref{oldproposal}  there are $n+2$ punctures, out of which two are massive, while in \eqref{newproposal} there are $n+4$ massless punctures.  {\sl ii)} The scattering equations are neither the same,  although in \cite{Cachazo:2015aol} it was shown that the massive scattering equations in \eqref{SE_1-loop} can be obtained from \eqref{Sequations} after dimensional reduction.  Additionally, {\sl iii)} the final outcomes are different, as it is going to be shown later, \eqref{oldproposal} produces linear propagators in $\ell^\mu$, while \eqref{newproposal} is able to reproduce the quadratic propagators as the traditional Feynman approach.

After some manipulations $\mathbf{I}_{\rm 1-loop}^{\rm CHY}[\pi |\rho]$ in \eqref{chyint_1-loop} becomes
\begin{align}\label{PTn-PTn}
&\mathbf{I}_{\rm 1-loop}^{\rm CHY}[\pi |\rho]  := {\bf PT}^{\, a_1:a_2}_{\rm 1-loop} [\pi]  \times  \frac{1}{(a_1,b_1,b_2,a_2)^2}    \times     {\bf PT}^{\, b_1:b_2}_{\rm 1-loop} [\rho] \nonumber\\
&= \sum_{\a\in \rm cyc (\pi)} \sum_{\b\in \rm cyc (\rho)} {\rm PT_{tree}} (\a_1,\ldots , \a_n, a_1,b_1,b_2,a_2) \times
{\rm PT_{tree}} (\b_1,\ldots , \b_n, b_1,a_1,a_2,b_2).
\end{align}
As a consequence the integral, $\int d\mu_{n+4}^{\rm tree}\, \mathbf{I}_{\rm 1-loop}^{\rm CHY}[\pi |\rho]$, is just a sum over trivalent tree level planar Feynman diagrams. However, this is not a very efficient and useful way to proceed because there is a large number of singular Feynman diagrams that do not contribute to the partial Amplitude, i.e. these diagrams must cancel out among them. 

\begin{figure}[!h]
  \centering
  \qquad\qquad\qquad
 \includegraphics[scale=0.3]{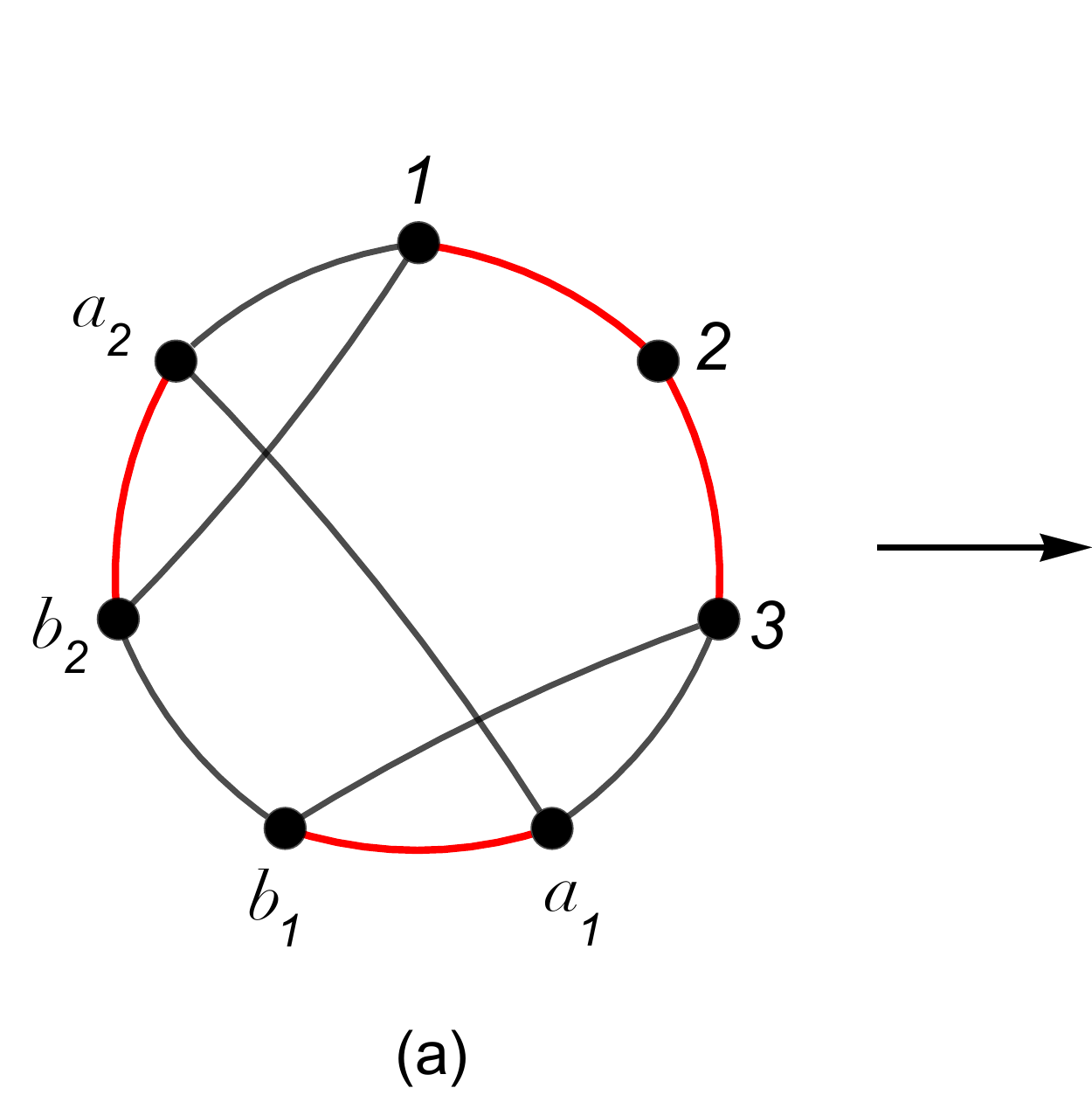}
 \includegraphics[scale=0.3]{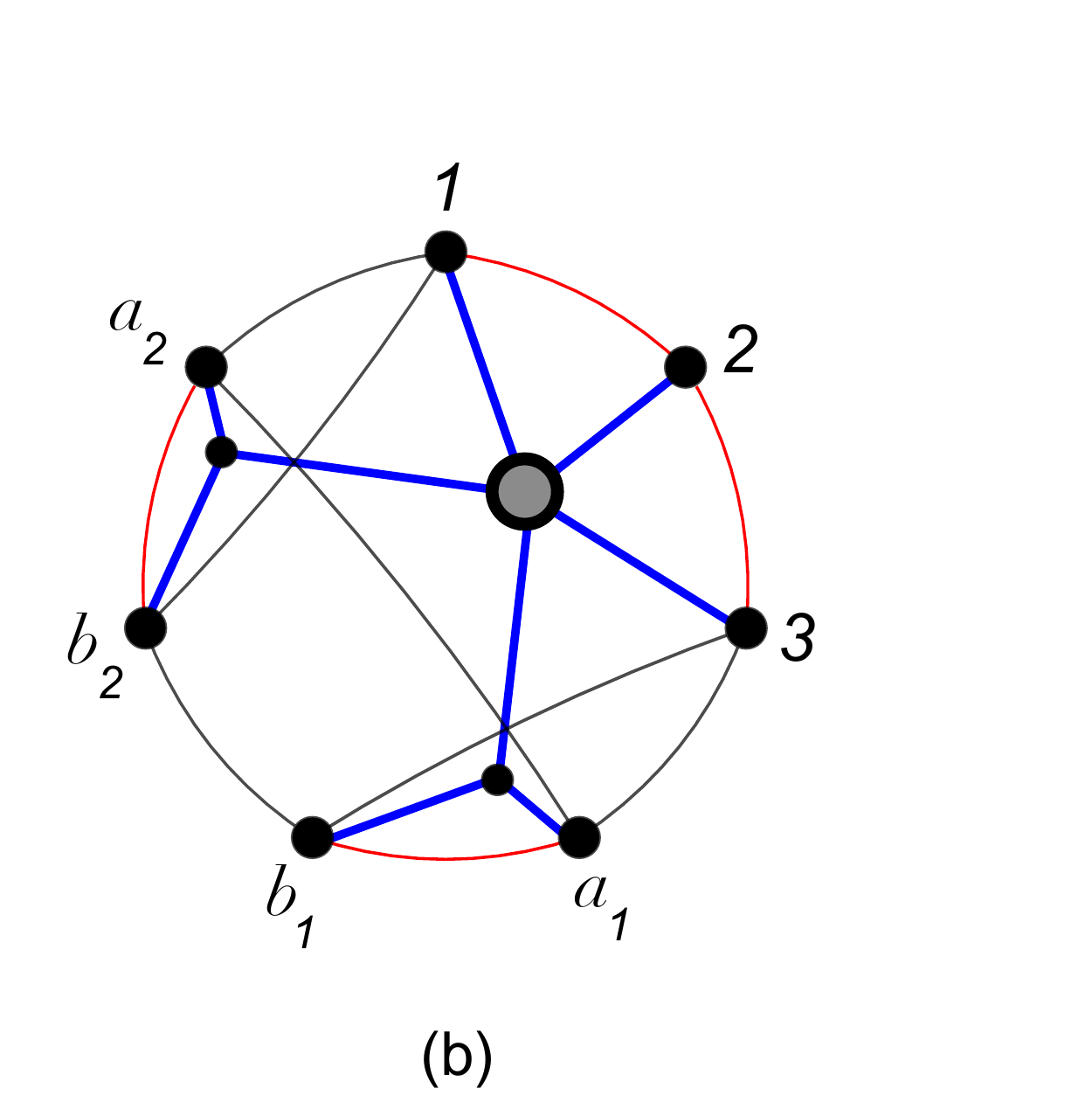} 
  \caption{(a) CHY-graph for the first term in \eqref{PT3-PT3}. (b) All trivalent Feynman diagrams associated to the CHY-graph (The grey circle means the sum over all possible trivalent vertices.). }\label{ex-chy-fey}
\end{figure}

For example, consider the CHY-integrand
\begin{equation}\label{PT3-PT3}
\mathbf{I}_{\rm 1-loop}^{\rm CHY}[1,2,3 |1,2,3] =  {\rm PT_{tree}} (1,2,3, a_1,b_1,b_2,a_2) \times
{\rm PT_{tree}} (1,2,3, b_1,a_1,a_2,b_2) + \cdots
\end{equation}
where ellipsis stand for terms obtained under cyclic permutations. The CHY-graph for the first term of the expansion in \eqref{PT3-PT3} is represented on the left hand side of Fig. \ref{ex-chy-fey}. 

\begin{figure}[!h]
  \centering
\includegraphics[scale=0.2]{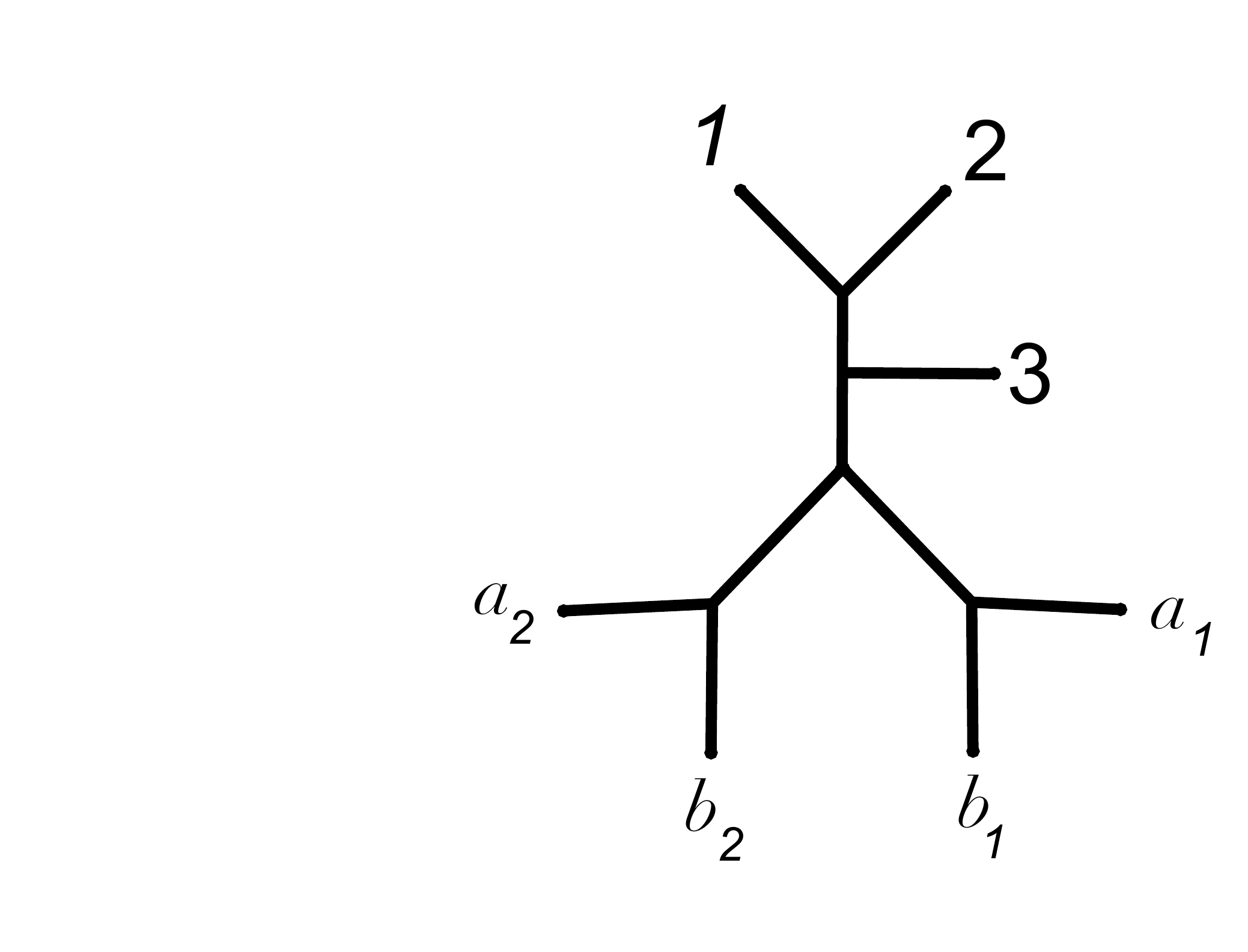}
\includegraphics[scale=0.2]{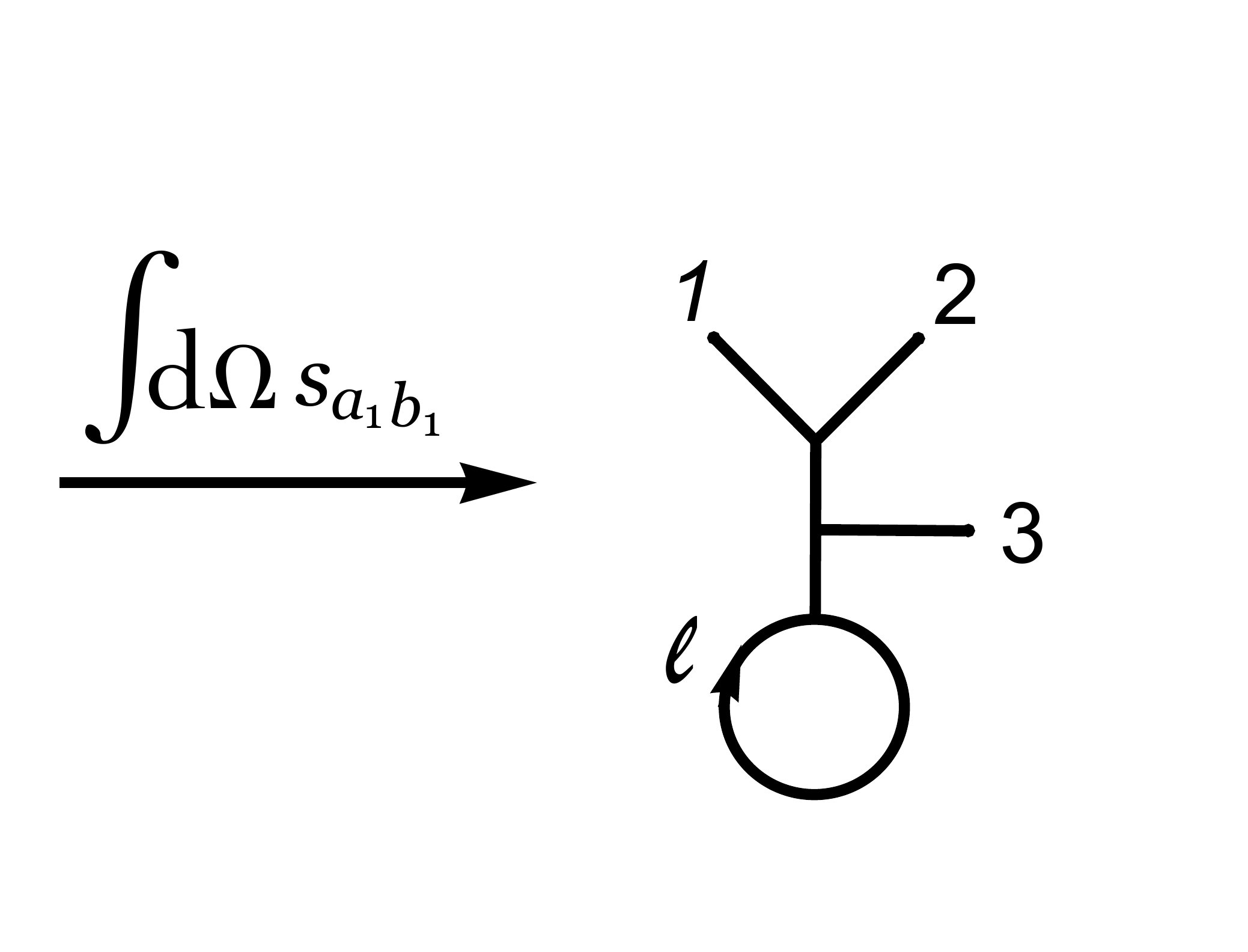} 
\caption{One of all possible Feynman diagrams given in Fig. \ref{ex-chy-fey}(b). Resulting tadpole after making the forward limit, i.e.  by integrating $\int d\Omega\,(k_{a_1}+k_{b_1})^2$.}\label{ex-fey-tree-loop}
\end{figure}

\noindent
The integral
$$\int d\mu_{3+4}^{\rm tree}\,{\rm PT_{tree}} (1,2,3, a_1,b_1,b_2,a_2) \times
{\rm PT_{tree}} (1,2,3, b_1,a_1,a_2,b_2)\,,$$  contains the sum over  all possible trivalent planar Feynman diagrams, \cite{Cachazo:2013iea,Mafra:2016ltu,Baadsgaard:2015voa,Gomez:2016bmv},  that have been depicted with a grey circle and blue lines  in Fig.  \ref{ex-chy-fey}(b). In particular, in Fig. \ref{ex-fey-tree-loop}(a) we give one example. Clearly, by performing the $\int d\Omega\, s_{a_1 b_1}$ integral of this diagram, one obtains a tadpole, such as it is shown in Fig. \ref{ex-fey-tree-loop}(b).  This singular diagram must not contribute to the amplitude and therefore it cancels out with another one which comes from the next contributions. This kind of analysis is tedious since the number of tree level diagrams in \eqref{PTn-PTn} is large.
\noindent

In order to handle this group of cancelling diagrams without going through the detailed analysis described above, we shall rely on the findings of sections \ref{Parke-Taylor} and \ref{classification} to classify the Feynman diagrams at one-loop from the CHY approach.  In the next section we are going to show several examples where this new technology is applied and the conjecture over $\mathfrak{M}_n^{\rm 1-loop}[\pi|\rho]$ will be checked.

Finally, note that the method used in \cite{He:2015yua, Baadsgaard:2015hia} could be applied here with a small variation:  the off-shell momenta, $\ell^\mu$ and $-\ell^\mu$, are split  into two on-shell momenta, $(k_{a_1}^\mu,k_{b_1}^\mu)$ and $(k_{a_2}^\mu,k_{b_2}^\mu)$, as is shown in Fig. \ref{ex-chy-fey}.

\section{Examples}\label{Sexamples}

In the following section we will use all the previous results to calculate the particular cases $n=3$ and $n=4$. Since $n=3$ is simpler, it contains only two possible orderings, we will show all the contributions by direct calculation of the integrands in \eqref{PT-3pts_2}.

Before giving the examples, it is useful to introduce the following notation
\begin{eqnarray}
\parbox[c]{8em}{\includegraphics[scale=0.18]{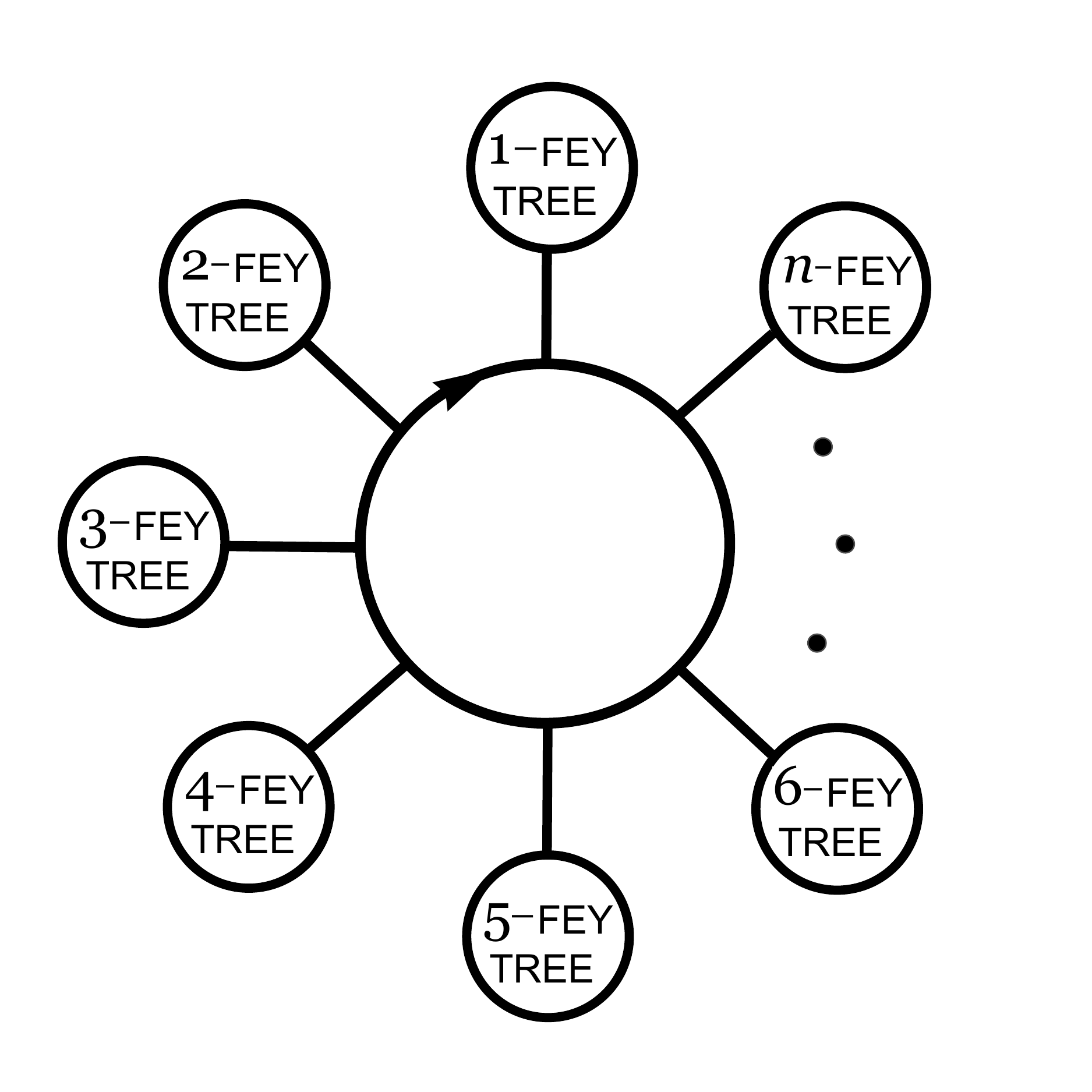}} :=
\parbox[c]{8em}{\includegraphics[scale=0.18]{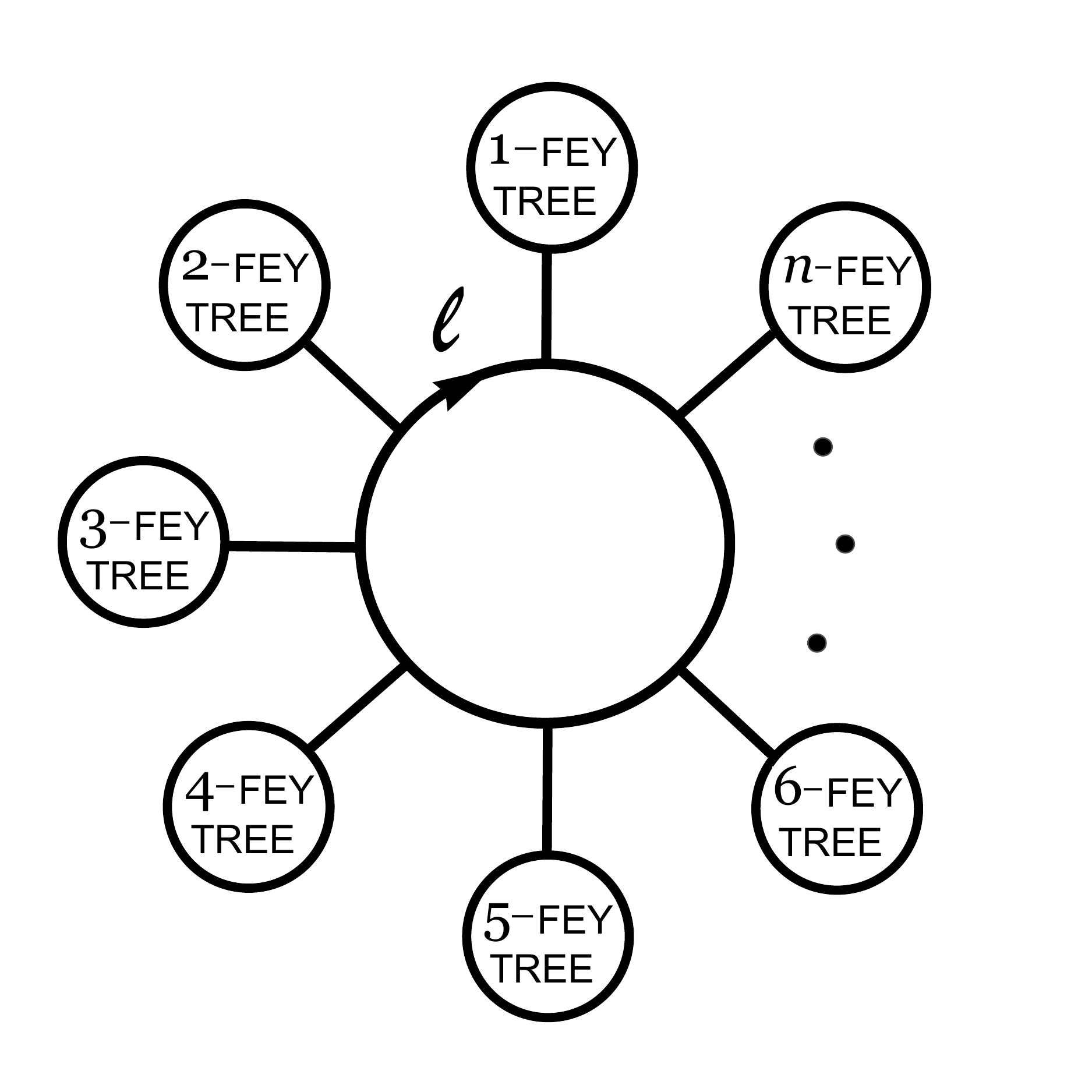}} +
\parbox[c]{8em}{\includegraphics[scale=0.18]{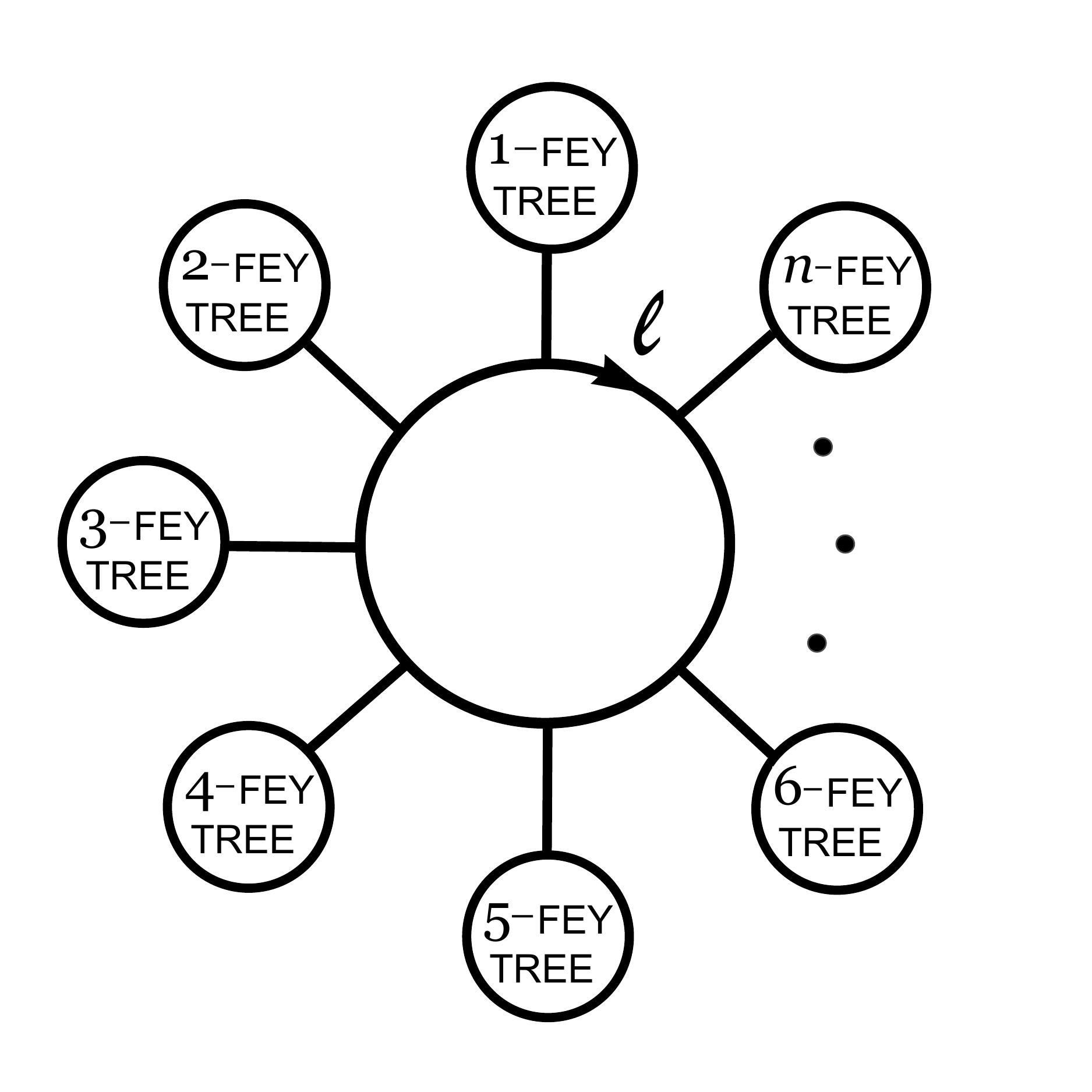}} +
\cdots +
\parbox[c]{10em}{\includegraphics[scale=0.18]{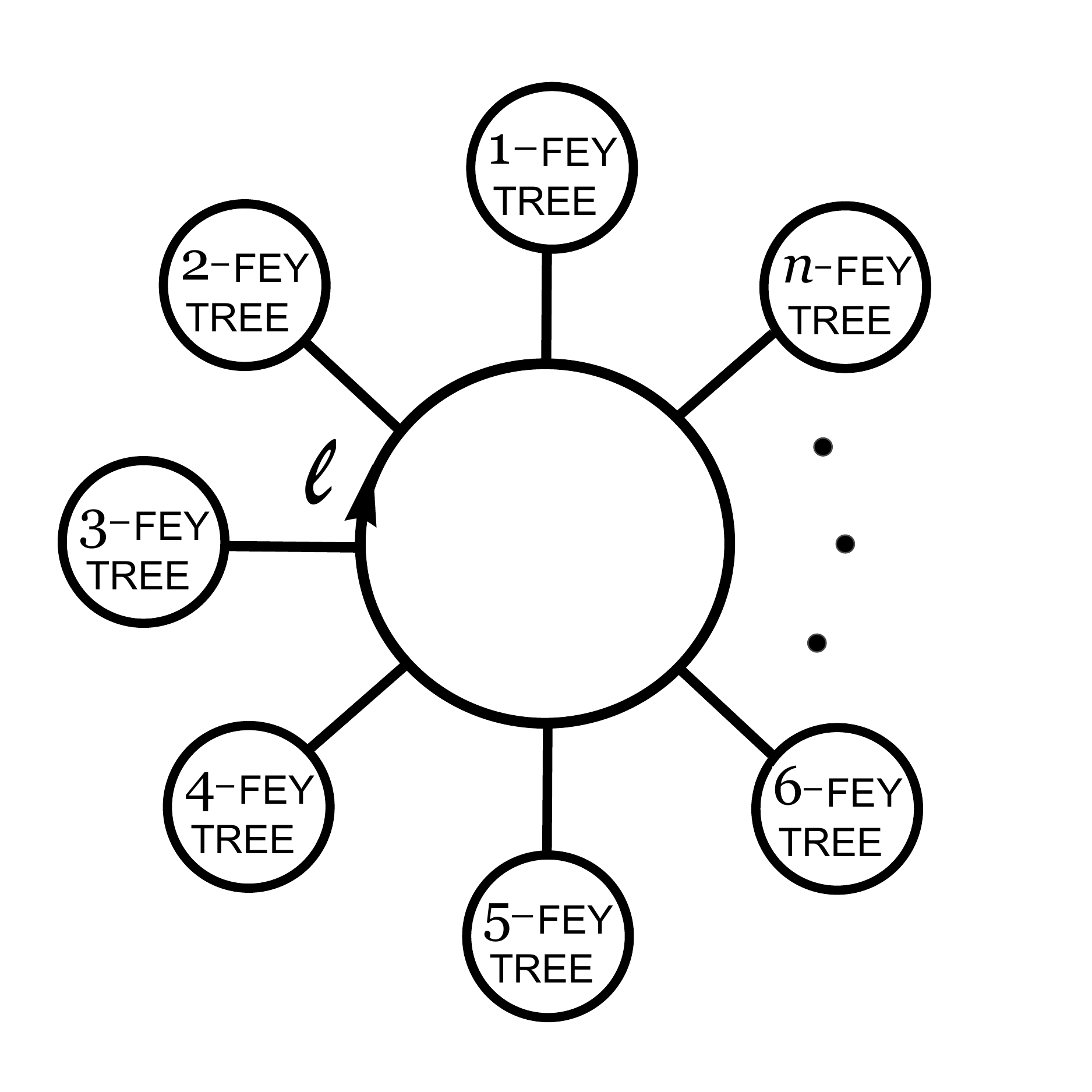}}.
\label{fey-1loop-T}
\end{eqnarray}

\subsection{Three-point}

In this case there are just two partial amplitudes, one coming from $\mathfrak{M}_3^{\rm 1-loop}[1,2,3|1,2,3]$ and another from $\mathfrak{M}_3^{\rm 1-loop}[1,2,3|3,2,1]$. In addition, as inferred from  \eqref{PT-3pts_2}, we have two expressions for ${\bf PT}^{\, a_1:a_2}_{\rm 1-loop} [1,2,3]$, one written in terms of the ${\bf D}^{\, a:b}_{\rm type- \,*}$'s from the ordering (1,2,3), and the other in terms of the ordering (3,2,1). In the following we disentangle each of them.

\subsubsection{Obtaining $\mathfrak{M}_3^{\rm 1-loop}[1,2,3|1,2,3] $} 
Let us start with the first CHY-integrand
\ba
\mathbf{I}_{\rm 1-loop}^{\rm CHY}[1,2,3|1,2,3] &=& \frac{{\rm PT}_{\rm tree}[1,2,3] \times {\rm PT}_{\rm tree}[1,2,3]}{(a_1,b_1,b_2,a_2)^2}\times
\nonumber
\\
&&
\left[\, 2\mathbf{D}^{\, a_1:a_2}_{\rm type-0} [1,2,3]^{ 3\o} +  \mathbf{D}^{\, a_1:a_2}_{\rm type-I} [1,2,3]^{ 2\o} \right]
\times\left[\s_{12}\, \o^{b_1:b_2}_{1:2} + \s_{23} \, \o^{b_1:b_2}_{2:3} + \s_{31} \o^{b_1:b_2}_{3:1} \right]\nonumber\\
&=&-\left[\,2\mathbf{I}^{\rm CHY}_{\rm (I)}[1,2,3]^{1:2} + \mathbf{I}^{\rm CHY}_{\rm (III)}[1,2,3]^{1:2} + \mathbf{I}^{\rm CHY}_{\rm (I)}[1,(2,3)]^{1:2} + \mathbf{I}^{\rm CHY}_{\rm (I)}[(3,1),2]^{1:2 }\right.\nonumber\\
&&
\left.\quad
+ {\rm cyc(1,2,3)}
\right]
\ea
where we have introduced the shorthand notation
\begin{align}
\mathbf{I}^{\rm CHY}_{{\rm (I)}}[i,(j,k)]^{i:j}
&:= (-1)\left[(-1)\times\frac{1}{\s_{a_1 b_1}^2 \s_{a_2b_2}^2 \s_{b_1b_2}}\times\frac{1}{  \s_{ia_1}\s_{ia_2} \times  \s_{ja_1}\s_{ka_2} }    \times \frac{1}{\s_{ib_1} \s_{j b_2} \s_{ki}\times (j,k)}\right] ,
\nonumber\\
\mathbf{I}^{\rm CHY}_{{\rm (I)}}[(k,i),j]^{i:j}
&:= (-1)\left[(-1)\times\frac{1}{\s_{a_1 b_1}^2 \s_{a_2b_2}^2 \s_{b_1b_2}}\times\frac{1}{\s_{ka_1}\s_{ia_2}\times  \s_{ja_1}\s_{ja_2} }    \times \frac{1}{\s_{ib_1} \s_{j b_2} \s_{jk}\times (k,i)}\right] .
\nonumber
\end{align}

In terms of the CHY-graphs one has
\begin{eqnarray}\label{CHY(123.123)}
&&\mathfrak{M}_3^{\rm 1-loop}[1,2,3|1,2,3]
= \frac{1}{2^4} \int d\Omega \times s_{a_1 b_1} \times \\
&&
\int d\mu_{3+4}^{\rm tree}\left\{ 2
\hspace{-0.3cm}
\parbox[c]{8em}{\includegraphics[scale=0.18]{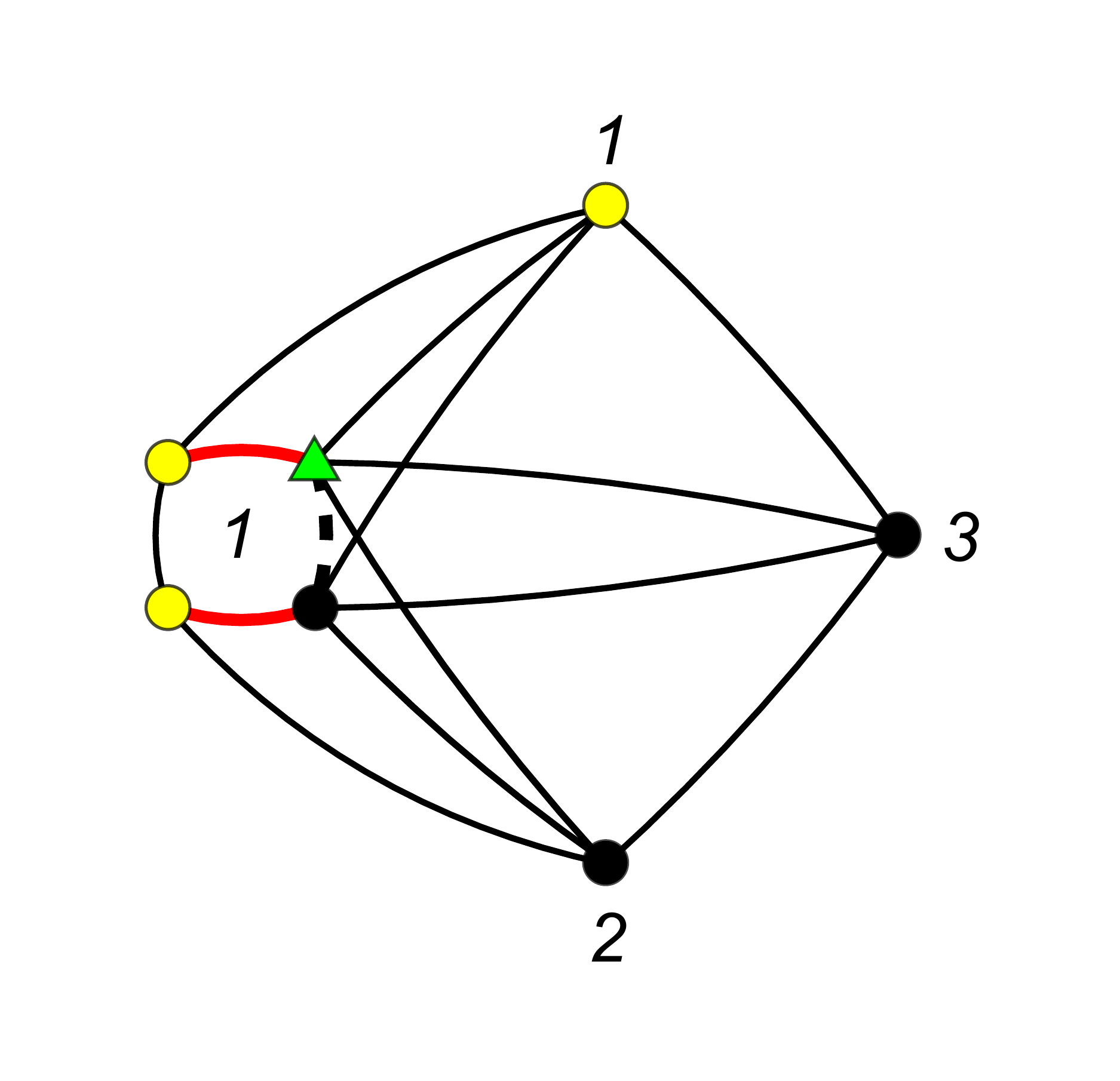}}\,+
\hspace{-0.3cm}
\parbox[c]{8em}{\includegraphics[scale=0.18]{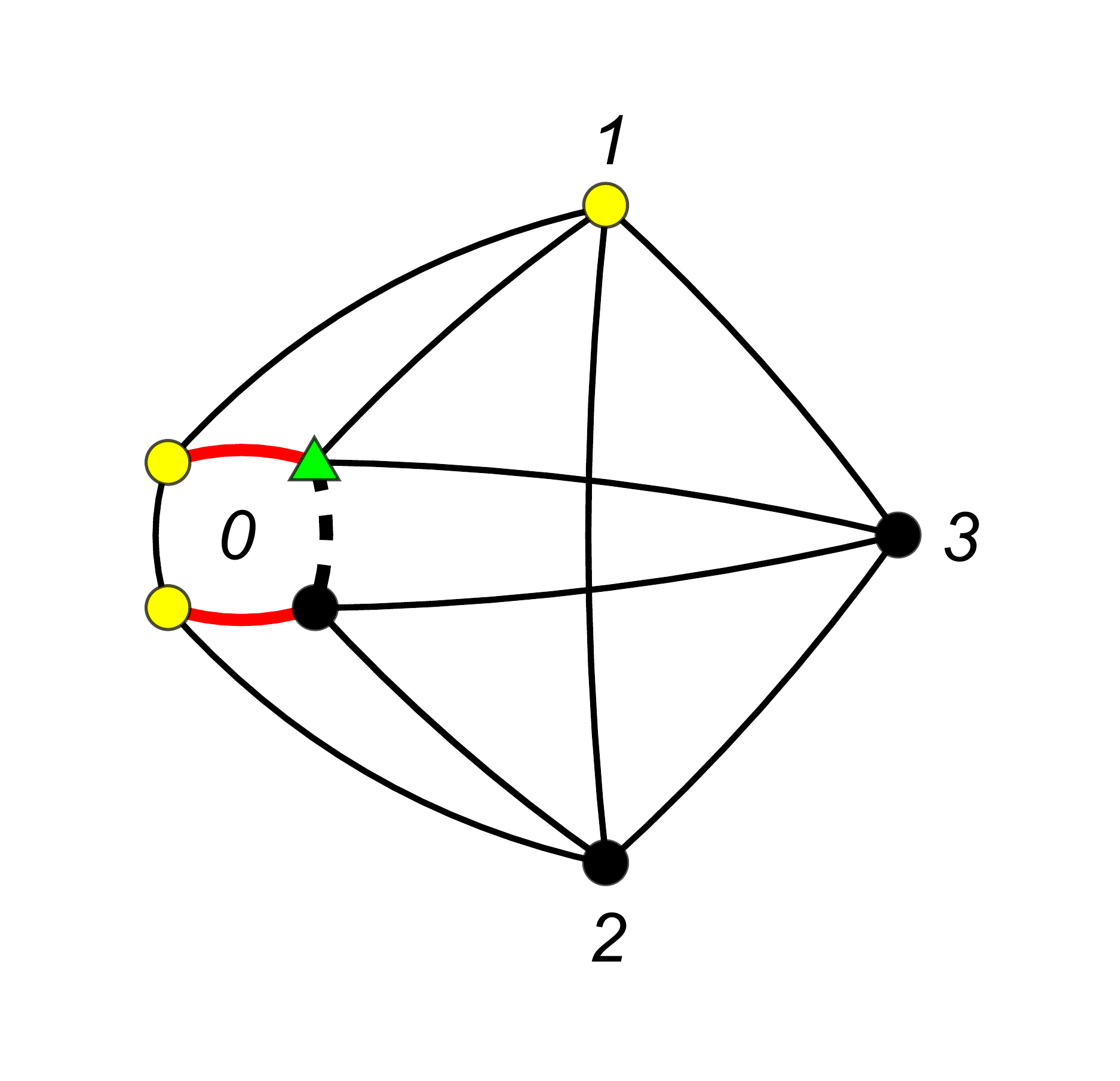}}\,+
\hspace{-0.4cm}
\parbox[c]{7em}{\includegraphics[scale=0.18]{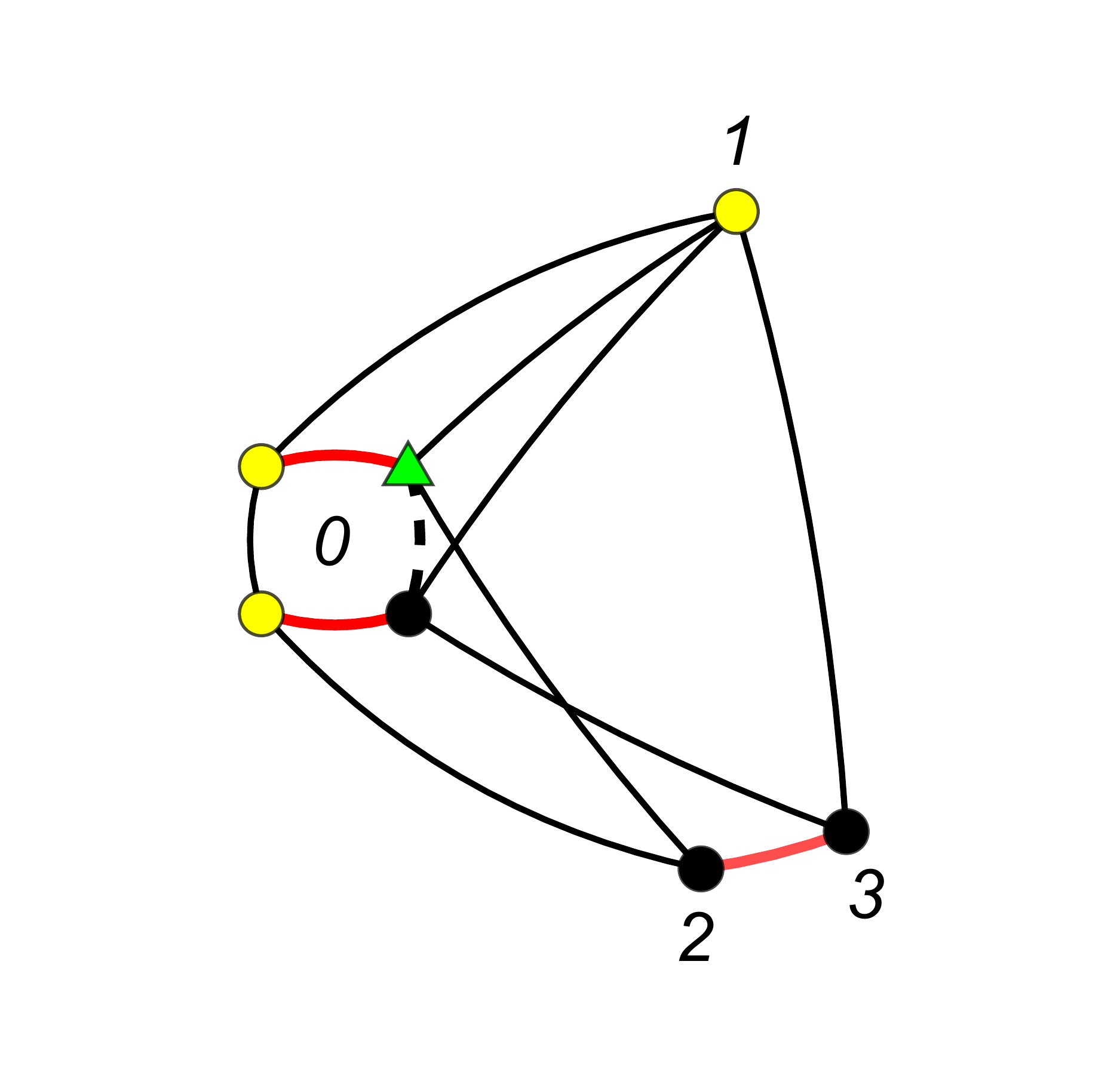}}\,+
\hspace{-0.4cm}
\parbox[c]{8em}{\includegraphics[scale=0.18]{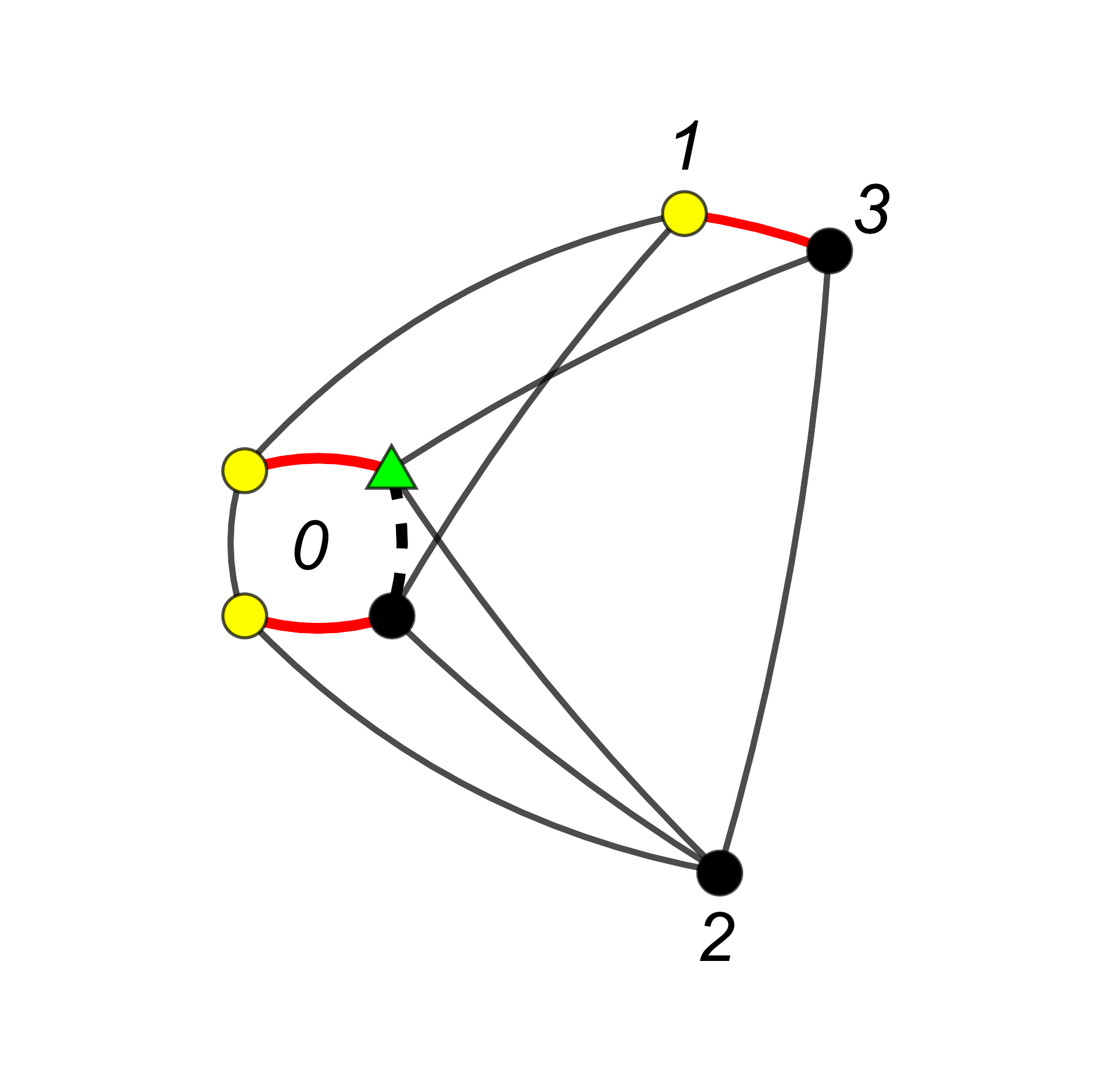}}
\hspace{-0.3cm}
+ {\rm cyc(1,2,3)}\,\right\}.\nonumber
\end{eqnarray}
That can be simplified further using  \eqref{i3-i1}

\begin{eqnarray}\label{CHY(123.123)2}
&&\mathfrak{M}_3^{\rm 1-loop}[1,2,3|1,2,3]
= \frac{1}{2^4} \int d\Omega \times s_{a_1 b_1} \times \\
&&
\int d\mu_{3+4}^{\rm tree}\left\{ 
\hspace{-0.3cm}
\parbox[c]{8em}{\includegraphics[scale=0.18]{1-l_3-p_c1.pdf}}\,+
\hspace{-0.4cm}
\parbox[c]{7em}{\includegraphics[scale=0.18]{1-l_3-p_c3.pdf}}\,+
\hspace{-0.4cm}
\parbox[c]{8em}{\includegraphics[scale=0.18]{1-l_3-p_c4.pdf}}
\hspace{-0.3cm}
+ {\rm cyc(1,2,3)}\,\right\}.\nonumber
\end{eqnarray}
From the general result in section \ref{Sclassification}  the first integral reduces to
\begin{eqnarray}
\frac{1}{2^4} \int d\Omega \, s_{a_1 b_1}  \int d\mu_{3+4}^{\rm tree}\left\{
\hspace{-0.3cm}
\parbox[c]{8em}{\includegraphics[scale=0.18]{1-l_3-p_c1.pdf}}+
 {\rm cyc} (1,2,3)\right\}
=
\hspace{-1.2cm}
\parbox[c]{10em}{\includegraphics[scale=0.23]{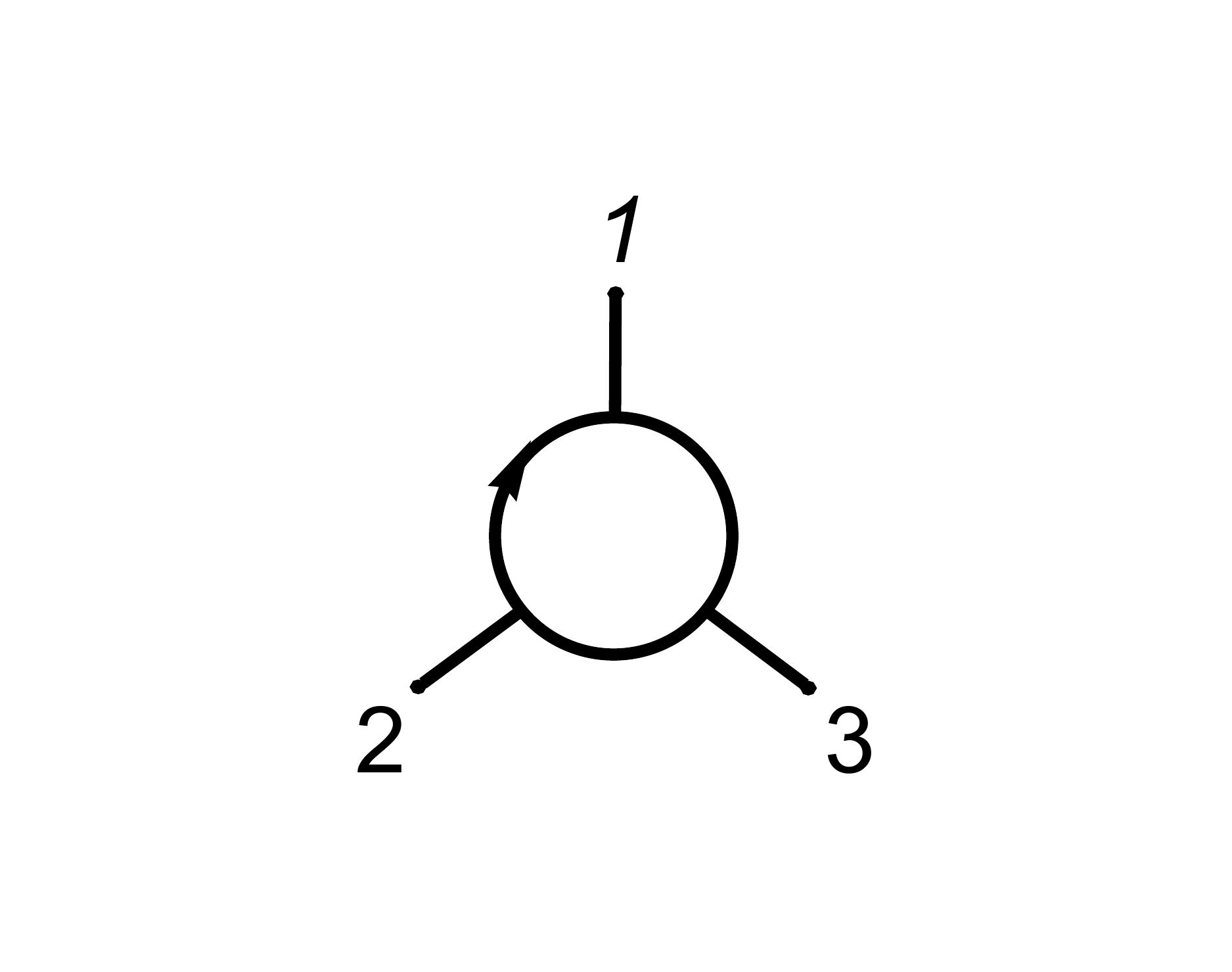}}\,,
\end{eqnarray}
while the rest of terms can be cast in the general form depicted in Fig.  \ref{3-CHY-graph}(a),  and  their computations is totally similar to the one presented in section \ref{Sclassification}

\begin{eqnarray}
&&\frac{1}{2^4} \times s_{a_1 b_1} \times \int d\mu_{3+4}^{\rm tree}
\hspace{-0.4cm}
\parbox[c]{7em}{\includegraphics[scale=0.18]{1-l_3-p_c3.pdf}}
=\frac{1}{s_{23}}\times \frac{1}{(k_{a_2}+k_{b_2})^2 \, (k_{1}+k_{a_1}+k_{b_1})^2}, 
\label{CHY(123.123)-2a}
 \\
&&\frac{1}{2^4} \times s_{a_1 b_1} \times \int d\mu_{3+4}^{\rm tree}
\hspace{-0.4cm}
\parbox[c]{7em}{\includegraphics[scale=0.18]{1-l_3-p_c4.pdf}}
=\frac{1}{s_{13}}\times \frac{1}{(k_{a_2}+k_{b_2})^2 \, (k_{1}+k_3+k_{a_1}+k_{b_1})^2}.
\label{CHY(123.123)-2b}
\end{eqnarray}

Notice that  by performing the integral $\int d\Omega$, i.e. at the forward limit, the momentum conservation condition becomes, $k_1+k_2+k_3=0$ ($s_{12}=s_{13}=s_{23}=0$), and the expressions \eqref{CHY(123.123)-2a}, \eqref{CHY(123.123)-2b} are ill defined. This fact indicates that these type of terms need some regularization\footnote{ Notice that in the linear propagator approach these kind of diagrams are absent \cite{He:2015yua, Baadsgaard:2015hia}, see section \ref{sectionEB}.}.
One way to obtain a well defined result by integrating \eqref{CHY(123.123)-2a} and \eqref{CHY(123.123)-2b} over $\int d\Omega$ is to regularize the forward limit condition, i.e. instead to consider the measure, $d^Dk_{a_2}\,\,d^Dk_{b_2}\,\,\delta^{(D)}(k_{a_2}+k_{a_1}) \,\,\delta^{(D)}(k_{b_2}+k_{b_1})$, we allow for the measure
$d^Dk_{a_2}\,\,d^Dk_{b_2}\,\,\delta^{(D)}(k_{a_2}+k_{a_1}-\frac{\epsilon}{2}) \,\,\delta^{(D)}(k_{b_2}+k_{b_1}-\frac{\epsilon}{2})$, where $\epsilon^\mu$ is an infinitesimal vector ($\epsilon^2\sim 0$) orthogonal to the external vectors\footnote{Let us recall this condition depends of the dimension of the momentum space.}  ($\epsilon\cdot k_i=0$). Using this new measure the momentum conservation condition becomes, $k_1+k_2+k_3=\epsilon$ ($s_{12}=s_{13}=s_{23}=\epsilon^2$),  and now we are able to integrate \eqref{CHY(123.123)-2a} and \eqref{CHY(123.123)-2b}. Considering the leading order term one has
\begin{eqnarray}\label{bubble-legs}
&&\int d\Omega \left[\frac{1}{s_{23}\,(k_{a_2}+k_{b_2})^2 \, (k_{1}+k_{a_1}+k_{b_1})^2}\right] =\frac{1}{s_{23}}\times \frac{1}{\ell ^2 \, (\ell + k_{1})^2} \,\,=
\hspace{-1.0cm}
\parbox[c]{7em}{\includegraphics[scale=0.18]{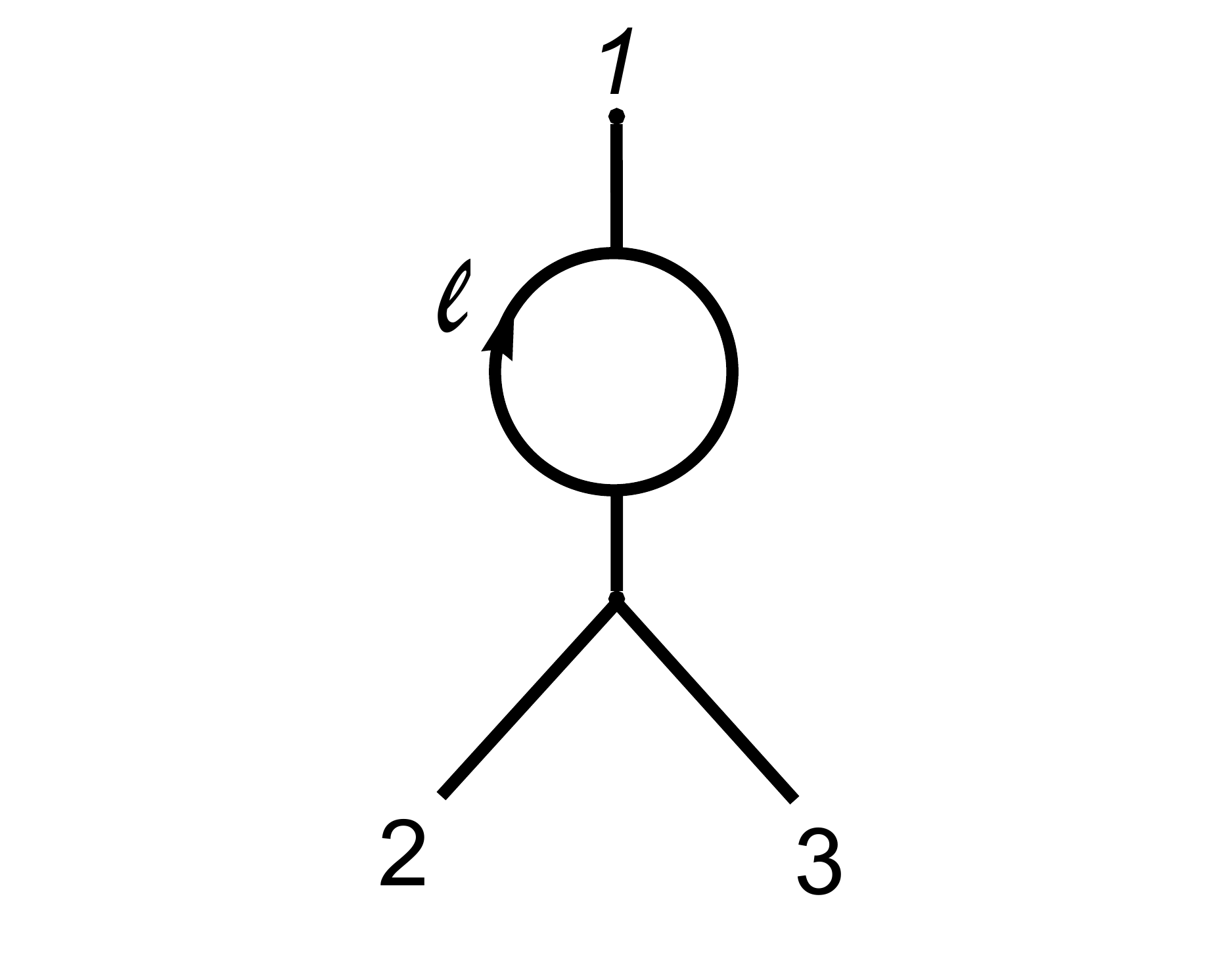}}
, 
 \\
&&\int d\Omega\left[ \frac{1}{ s_{13}\,(k_{a_2}+k_{b_2})^2 \, (k_{1}+k_3+k_{a_1}+k_{b_1})^2}\right]
=\frac{1}{s_{13}} \times \frac{1}{\ell^2 \, (\ell + k_{1}+k_3)^2}\,\,=
\hspace{-1.2cm}
\parbox[c]{7em}{\includegraphics[scale=0.18]{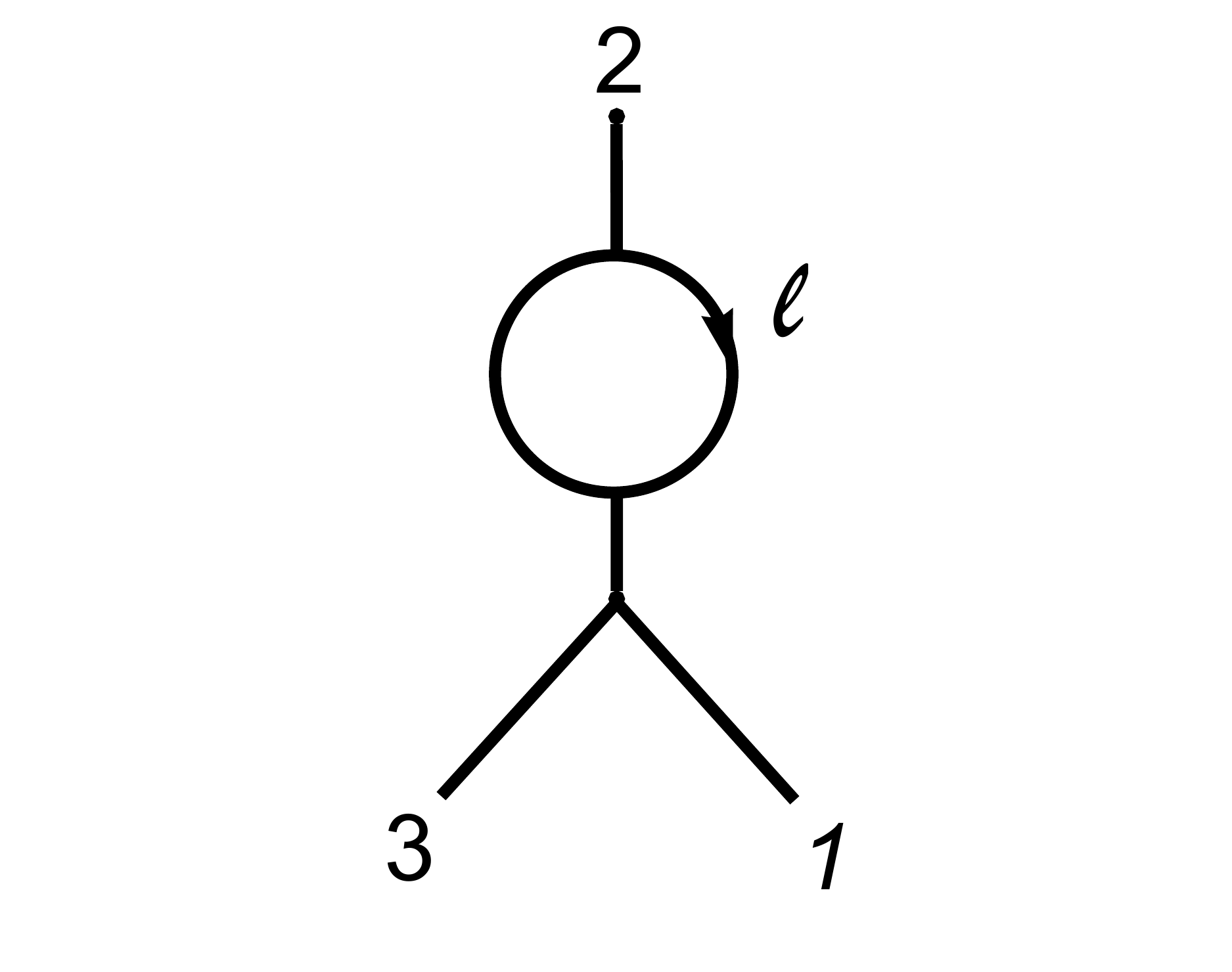}}.\nonumber
\end{eqnarray} 
 Finally, adding all the partial results one gets 
\ba\label{CHY(123.123)-Res}
\mathfrak{M}_3^{\rm 1-loop}[1,2,3|1,2,3] =
\hspace{-1.2cm}
\parbox[c]{9em}{\includegraphics[scale=0.23]{fey-3pts-firstC.pdf}} +
\hspace{-0.7cm}
\parbox[c]{7em}{\includegraphics[scale=0.17]{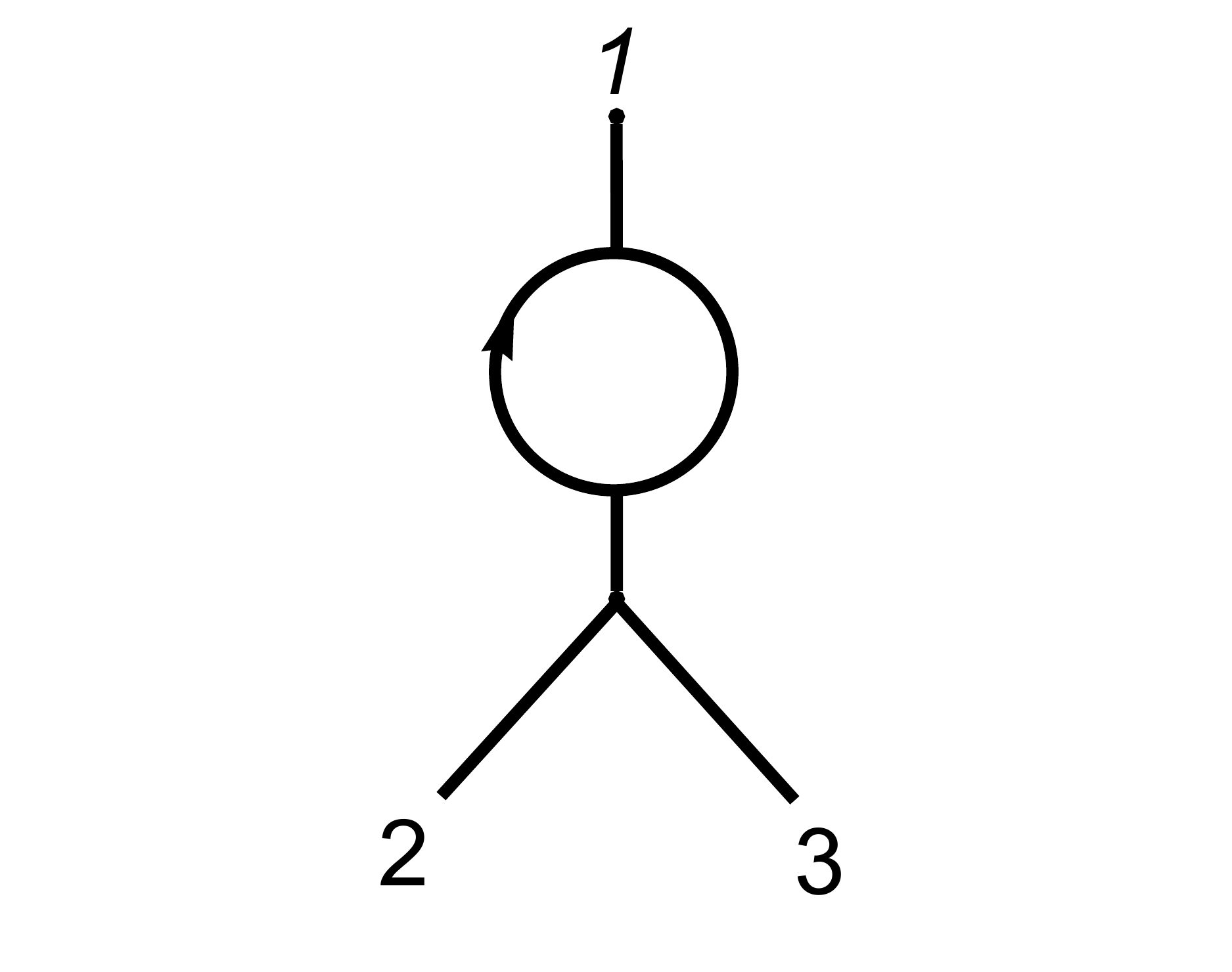}}+
\hspace{-0.6cm}
\parbox[c]{7em}{\includegraphics[scale=0.17]{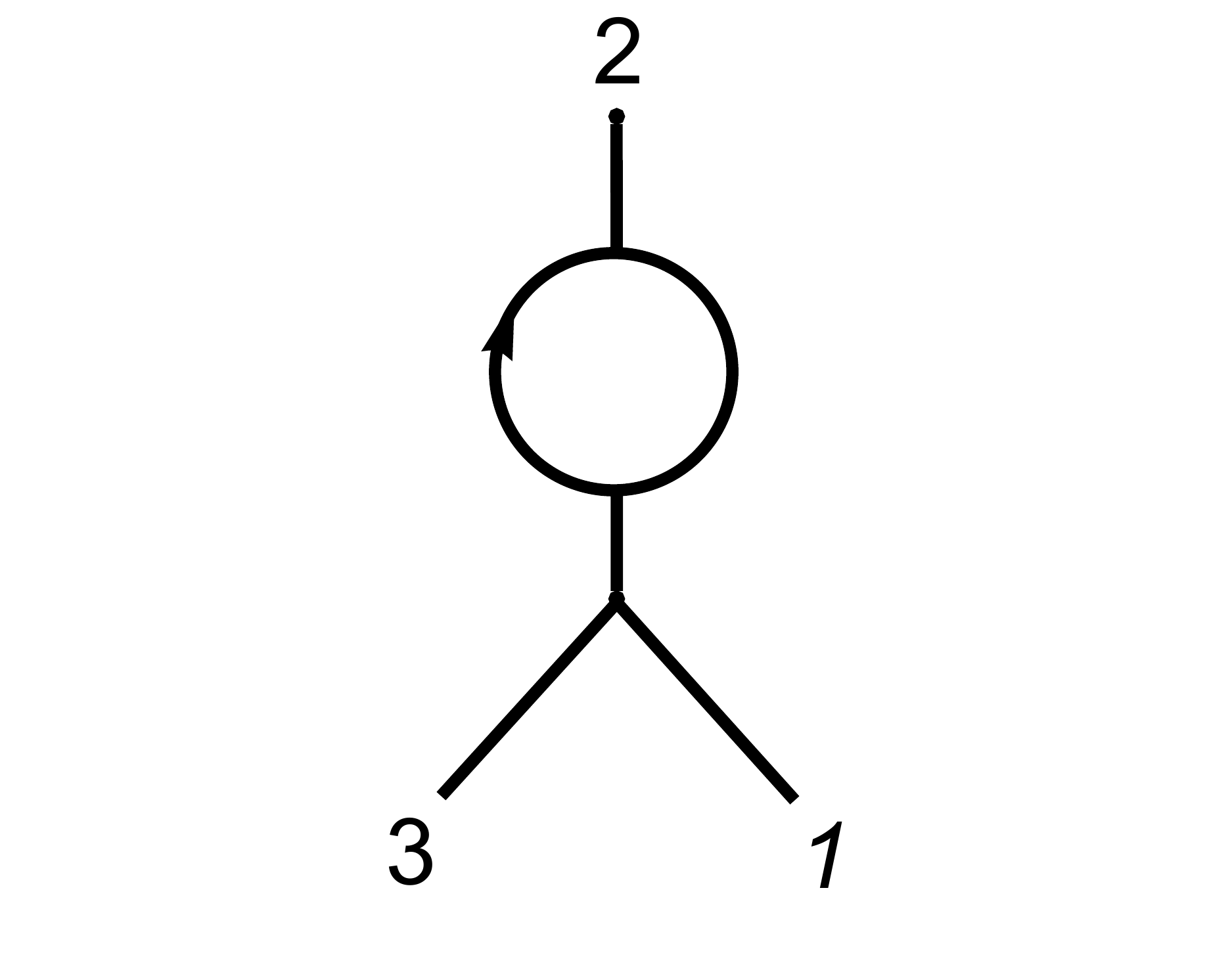}}+
\hspace{-0.7cm}
\parbox[c]{7em}{\includegraphics[scale=0.17]{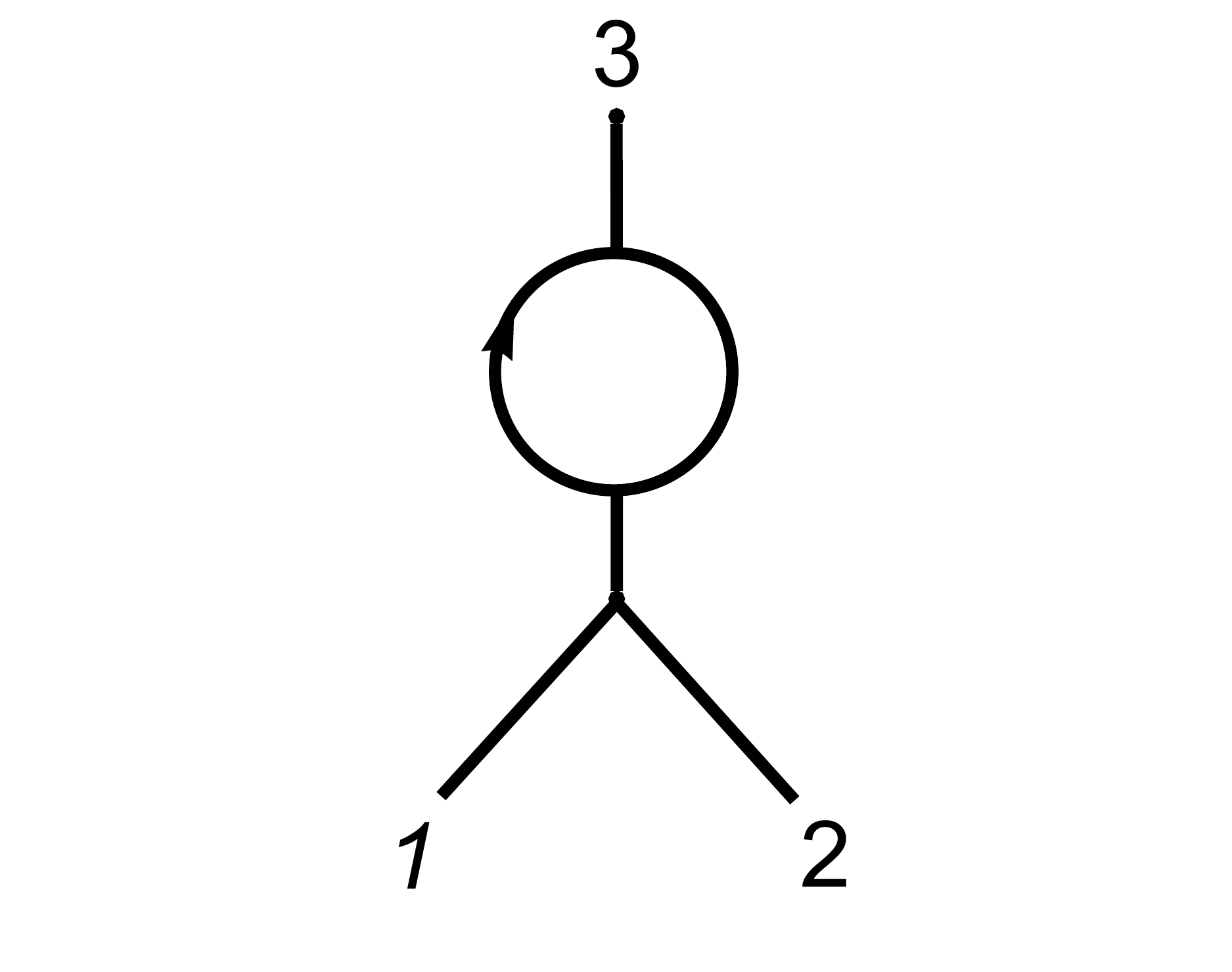}}\,
\ea
which is the expected answer, and has also been checked numerically. In section \ref{sectionEB} we will briefly discuss about the external-leg bubbles contributions in  $\mathfrak{M}_3^{\rm 1-loop}[\ast| \ast]$.

\subsubsection{Obtaining $\mathfrak{M}_3^{\rm 1-loop}[1,2,3|3,2,1] $} 

To compute the next partial amplitude we use the second expression for ${\bf PT}^{\, a_1:a_2}_{\rm 1-loop} [1,2,3]$ in \eqref{PT-3pts_2}. The CHY-integrand for this case is
\ba
\mathbf{I}_{\rm 1-loop}^{\rm CHY}[1,2,3|3,2,1] &=& -\frac{{\rm PT}_{\rm tree}[3,2,1]{\rm PT}_{\rm tree}[3,2,1]}{(a_1,b_1,b_2,a_2)^2}\left[\, \mathbf{D}^{\, a_1:a_2}_{\rm type-0} [3,2,1]^{ 3\o} +  \mathbf{D}^{\, a_1:a_2}_{\rm type-I} [3,2,1]^{ 2\o} \right]\nonumber\\
&&\hspace{4.45cm}\times\left[\s_{32}\, \o^{b_1:b_2}_{3:2} + \s_{21} \, \o^{b_1:b_2}_{2:1} + \s_{13} \o^{b_1:b_2}_{1:3} \right]\nonumber\\
&=&\mathbf{I}^{\rm CHY}_{\rm (I)}[3,2,1]^{3:2} + \mathbf{I}^{\rm CHY}_{\rm (III)}[3,2,1]^{3:2} + \mathbf{I}^{\rm CHY}_{\rm (I)}[3,(2,1)]^{3:2} + \mathbf{I}^{\rm CHY}_{\rm (I)}[(1,3),2]^{3:2}\nonumber\\
&&+ {\rm cyc(3,2,1)}\nonumber\\
&=& \mathbf{I}^{\rm CHY}_{\rm (I)}[3,(2,1)]^{3:2} + \mathbf{I}^{\rm CHY}_{\rm (I)}[(1,3),2]^{3:2} + {\rm cyc(3,2,1)}\, ,
\ea
where in the first line we used the inversion property of the PT factors, and  in the third one the result, $\mathbf{I}^{\rm CHY}_{\rm (III)}[3,2,1]^{3:2} = -\mathbf{I}^{\rm CHY}_{\rm (I)}[3,2,1]^{3:2} + ({\rm zero\,\, after\,\,integration})$, given in \eqref{i3-i1}. Translating to CHY-graphs the amplitude becomes,
\begin{eqnarray}
\mathfrak{M}_3^{\rm 1-loop}[1,2,3|3,2,1]&=&
\frac{(-1)}{2^4} \int d\Omega \times s_{a_1 b_1}   
\nonumber\\
&&
\hspace{-0.25cm}
\times
\int d\mu_{3+4}^{\rm tree}\left\{ 
\hspace{-0.4cm}
\parbox[c]{7em}{\includegraphics[scale=0.18]{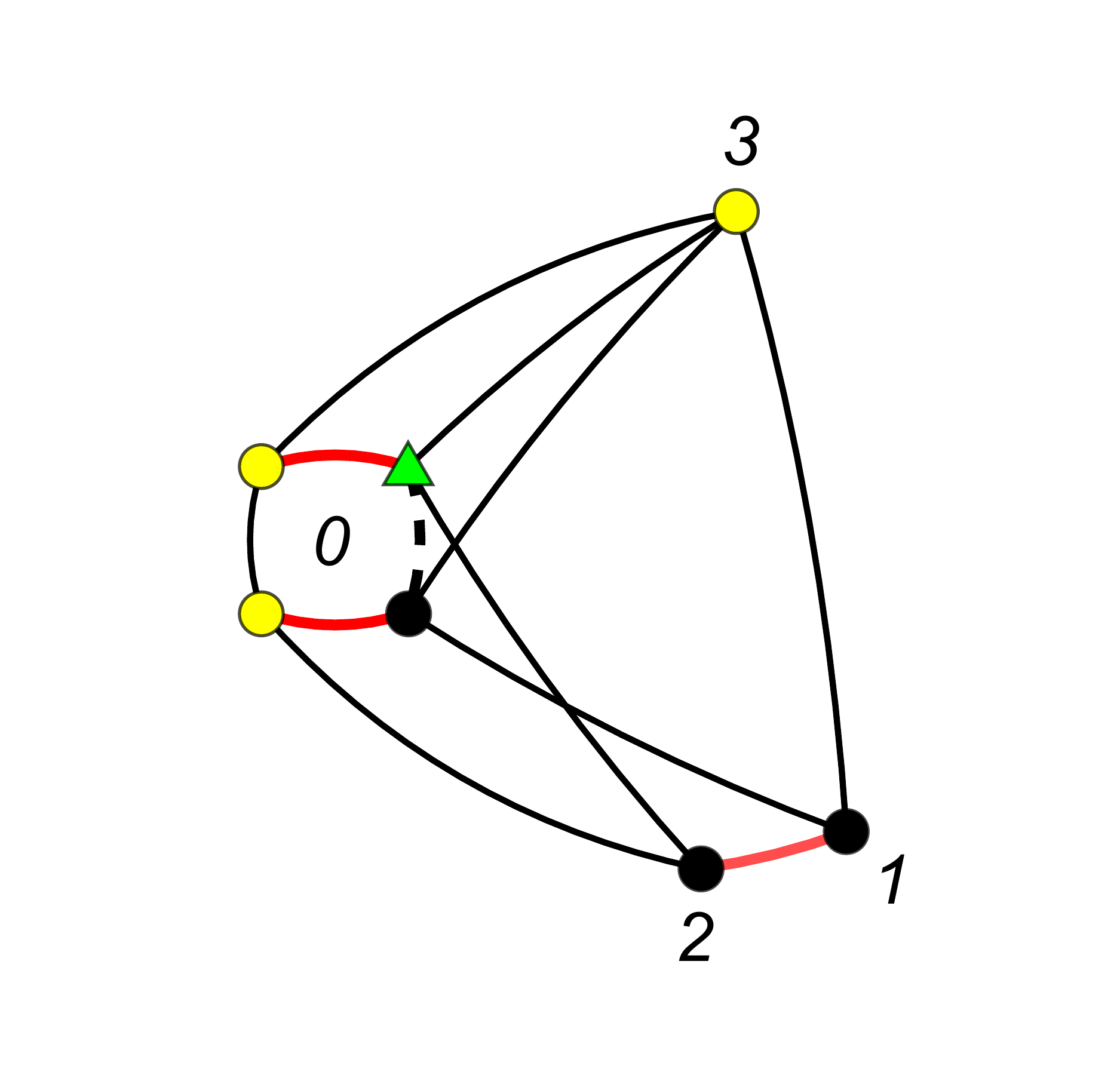}}+
\hspace{-0.4cm}
\parbox[c]{7em}{\includegraphics[scale=0.18]{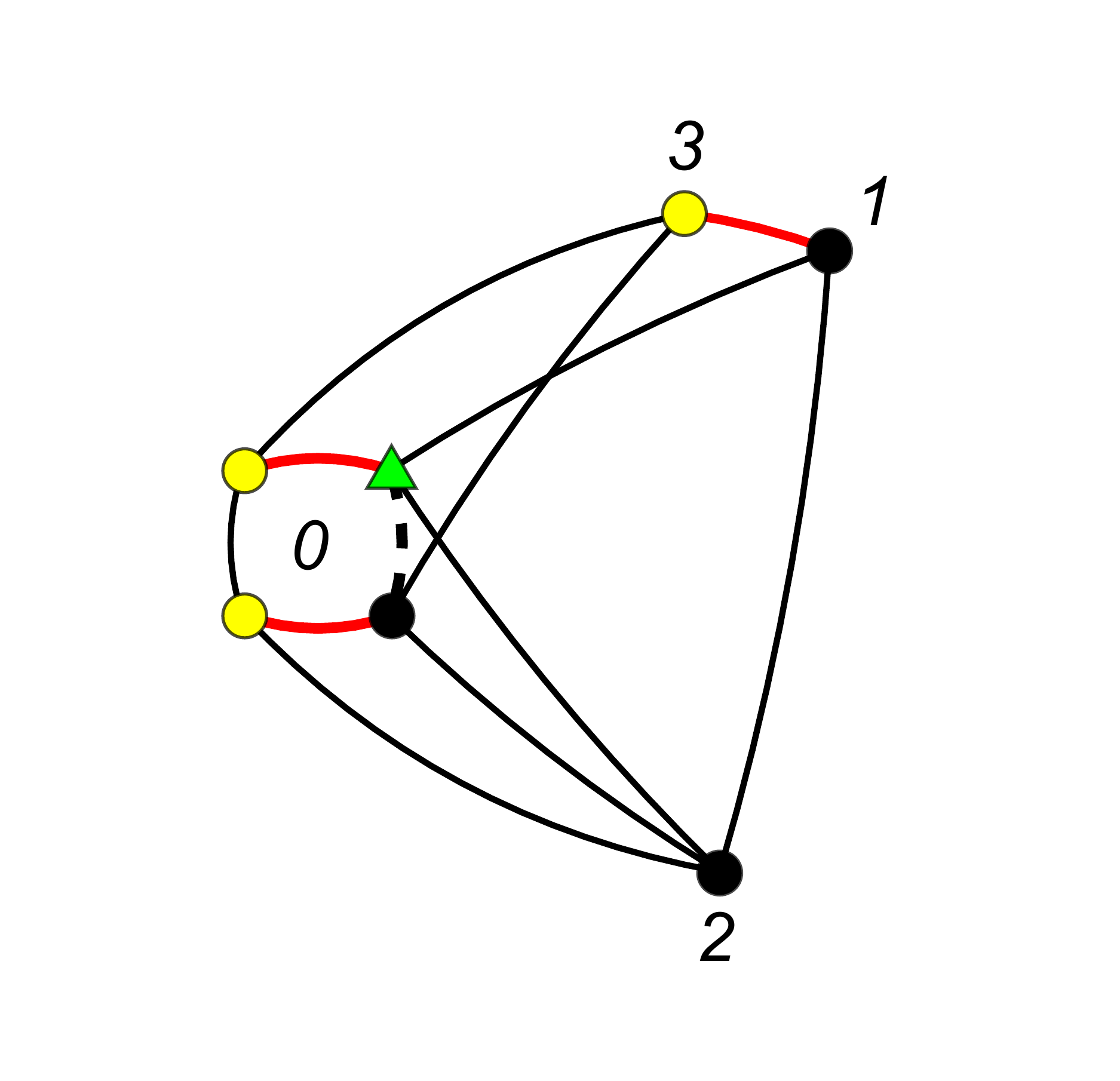}}
+ {\rm cyc(3,2,1)}\right\}.
\label{CHY(123.321)}
\end{eqnarray}
The integrals entering in (\ref{CHY(123.321)}) were already computed in the example above, 
\eqref{CHY(123.123)-2a}--\eqref{bubble-legs}. Collecting then we can write the total result as
\ba\label{CHY(123.321)-Res}
\mathfrak{M}_3^{\rm 1-loop}[1,2,3|3,2,1] =-
\left\{
\hspace{-1.0cm}
\parbox[c]{6em}{\includegraphics[scale=0.17]{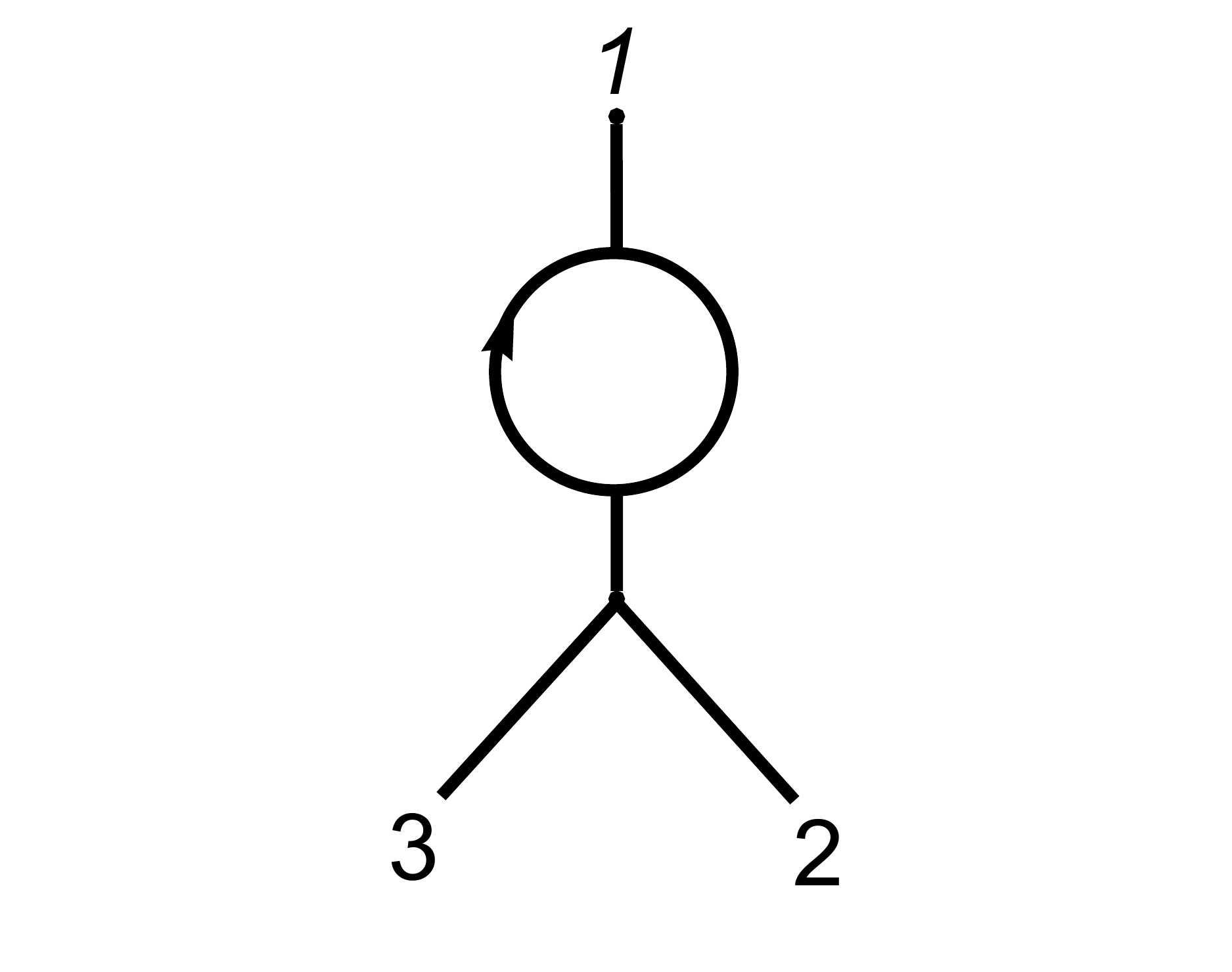}}+
\hspace{-0.6cm}
\parbox[c]{6em}{\includegraphics[scale=0.17]{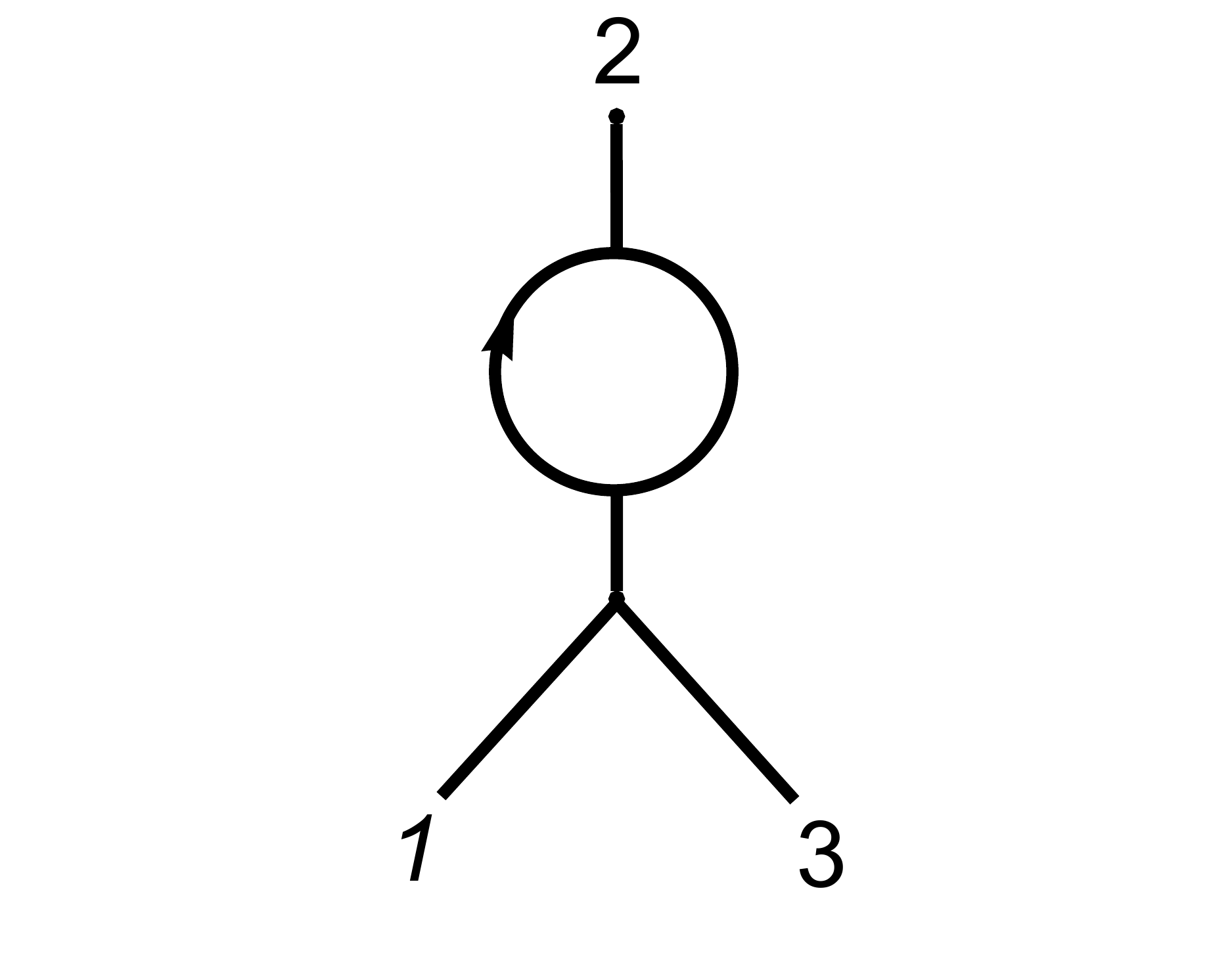}}+
\hspace{-0.7cm}
\parbox[c]{6em}{\includegraphics[scale=0.17]{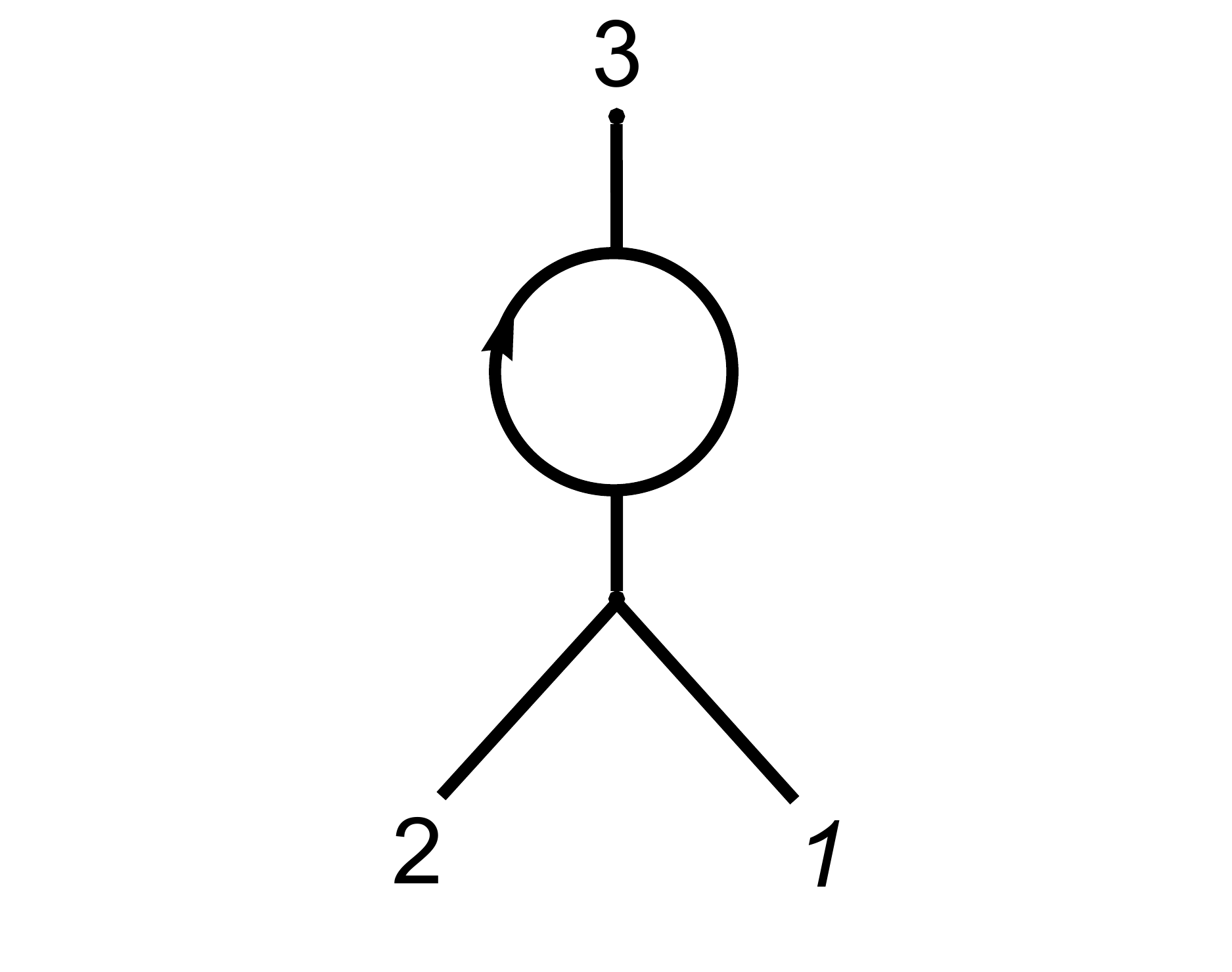}}\right\},
\ea
which has been checked numerically.

Hitherto we have found the Feynman diagrams expansion for the canonical ordering, $\mathfrak{M}_3^{\rm 1-loop}[1,2,3|1,2,3] $ and its opposite  ordering, $\mathfrak{M}_3^{\rm 1-loop}[1,2,3|3,2,1] $, in \eqref{CHY(123.123)-Res},  \eqref{CHY(123.321)-Res} respectively. With those it is now straightforward to verify the relation
\ba\label{Int-1loop-3p}
 (-1)\,\,\mathfrak{M}_3^{\rm 1-loop}[1,2,3|1,2,3]\cap\mathfrak{M}_3^{\rm 1-loop}[3,2,1|3,2,1] = \mathfrak{M}_3^{\rm 1-loop}[1,2,3|3,2,1] \, ,
\ea
which is the one-loop equivalent to \eqref{ord_intersc} \cite{Cachazo:2013iea}. In \cite{He:2015yua}  a general relation at one-loop was conjectured, but in the Q-cut representation, i.e. using the prescription presented in section \ref{oldP}.  Although we have not got a general proof for \eqref{Int-1loop-3p}, we have a {\sl strong numerical evidence} that the conjecture  formulated  in \cite{He:2015yua} can be extended to the  new proposal formulated in this paper as
\ba
\mathfrak{M}_n^{\rm 1-loop}[\pi|\rho] =(-1)^{n_{\pi|\rho}}\,\, \mathfrak{M}_n^{\rm 1-loop}[\pi|\pi]\cap\mathfrak{M}_n^{\rm 1-loop}[\rho|\rho]\,.
\label{inter1lopp}
\ea

\subsection{Four-point}

In this case we will have to deal with three different independent orderings, 

\ba 
[\pi|\rho]\in\{[1234|1234],[1234|4321],[1234|1243]\}\,.
\ea
We will calculate explicitly the first one and rely heavily in the use of (\ref{inter1lopp}) in order to infer
the rest of them.

The CHY-integrand for the partial amplitude $\mathfrak{M}_4^{\rm 1-loop}[1,2,3,4| 1,2,3,4] $ reads
\ba
&&\mathbf{I}_{\rm 1-loop}^{\rm CHY}[1,2,3,4|1,2,3,4] = \frac{{\rm PT}_{\rm tree}[1,2,3,4]{\rm PT}_{\rm tree}[1,2,3,4]}{(a_1,b_1,b_2,a_2)^2}
\,\,\big[\,3\mathbf{D}^{\, a_1:a_2}_{\rm type-0} [1,2,3,4]^{ 4\o} + \, 2\mathbf{D}^{\, a_1:a_2}_{\rm type-I} [1,2,3,4]^{ 3\o}\nonumber\\
&&
+ \mathbf{D}^{\, a_1:a_2}_{\rm type-I} [1,2,3,4]^{ 2\o} + \frac{1}{2}\,\mathbf{D}^{\, a_1:a_2}_{\rm type-II} [1,2,3,4]^{ 2\o}\big]
\times\big[\s_{12}\,\o^{b_1:b_2}_{1:2} + \s_{23}\,\o^{b_1:b_2}_{2:3} + \s_{34}\,\o^{b_1:b_2}_{3:4} + \s_{41}\,\o^{b_1:b_2}_{4:1}\big].\nonumber
\ea
After expanding the terms we obtain several cancellations due to the identity,  $\mathbf{I}^{\rm CHY}_{\rm (III)}[\ast] = -\mathbf{I}^{\rm CHY}_{\rm (I)}[\ast]$, which was proven in \eqref{i3-i1} in the simplest case. After a rather cumbersome calculation the above amplitude can be cast in terms of   CHY-graphs as
\begin{eqnarray}
&&\mathfrak{M}_4^{\rm 1-loop}[1,2,3,4| 1,2,3,4] 
=
\frac{(-1)}{2^5} \int d\Omega \times s_{a_1 b_1} \times\nonumber\\
&&
 \int d\mu^{\rm tree}_{4+4}  
 \left\{
\hspace{-0.4cm}
 \parbox[c]{8em}{\includegraphics[scale=0.17]{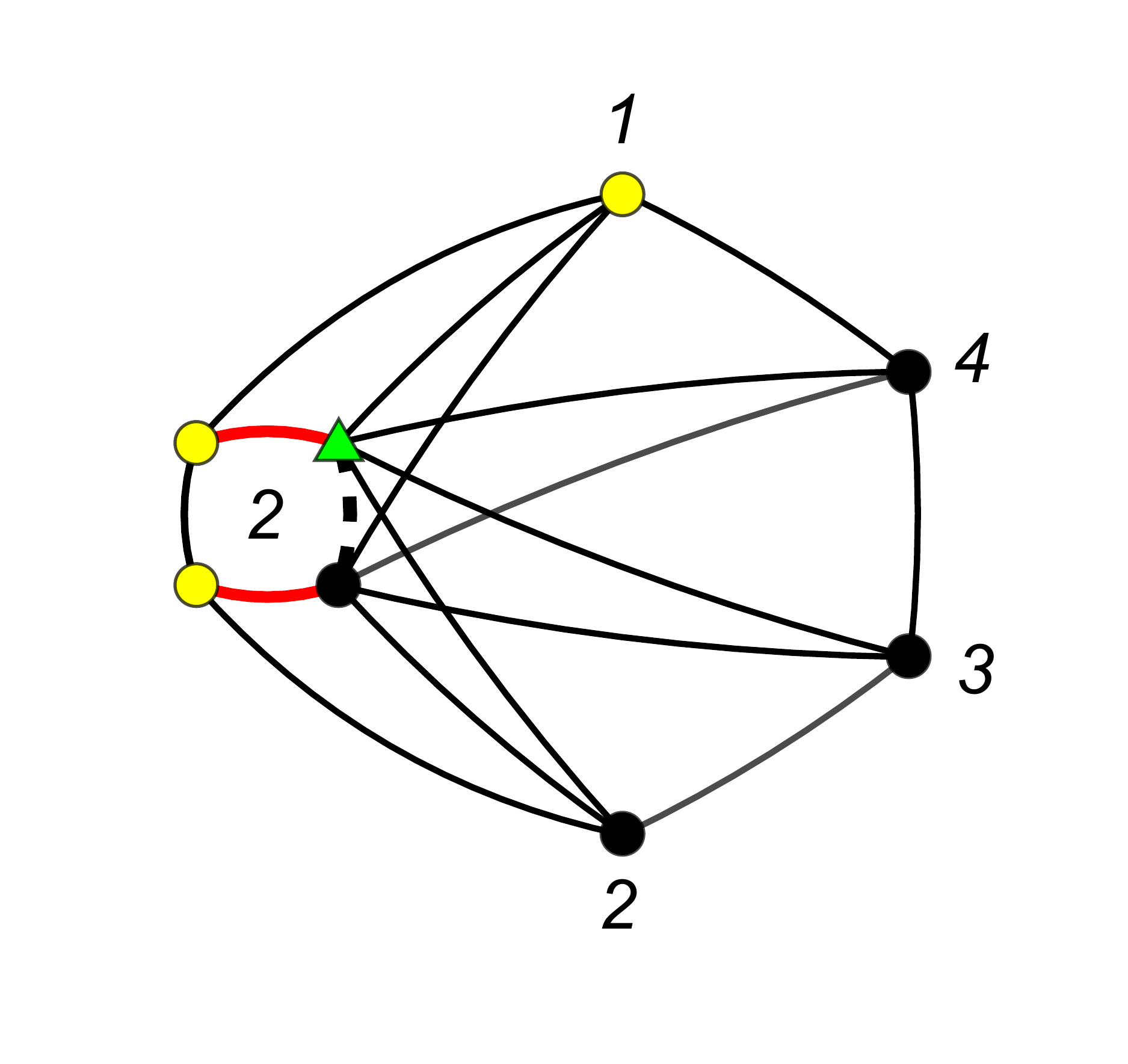}}+
 \hspace{-0.4cm}
 \parbox[c]{7.5em}{\includegraphics[scale=0.16]{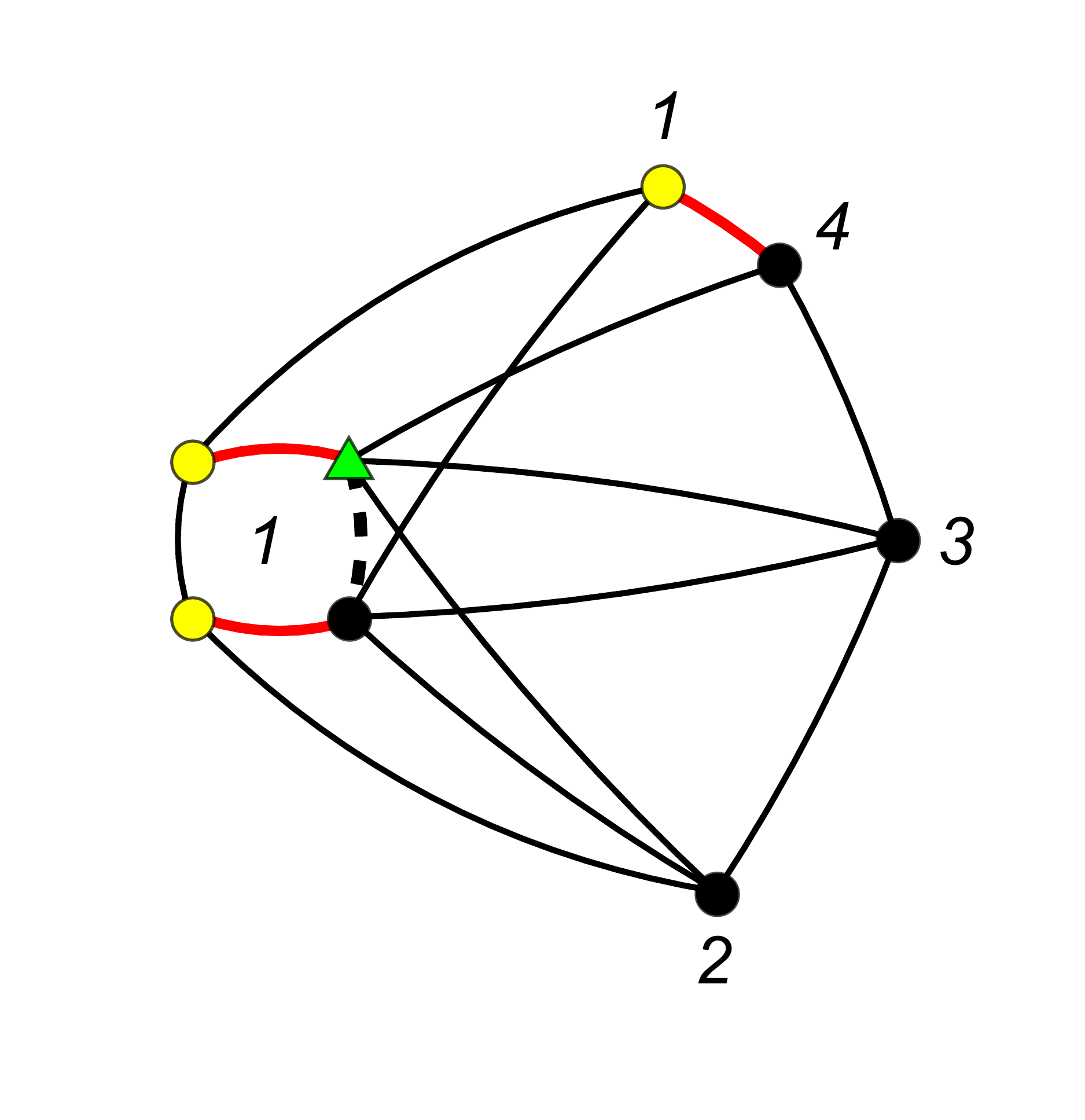}}+
 \hspace{-0.4cm}
 \parbox[c]{8em}{\includegraphics[scale=0.17]{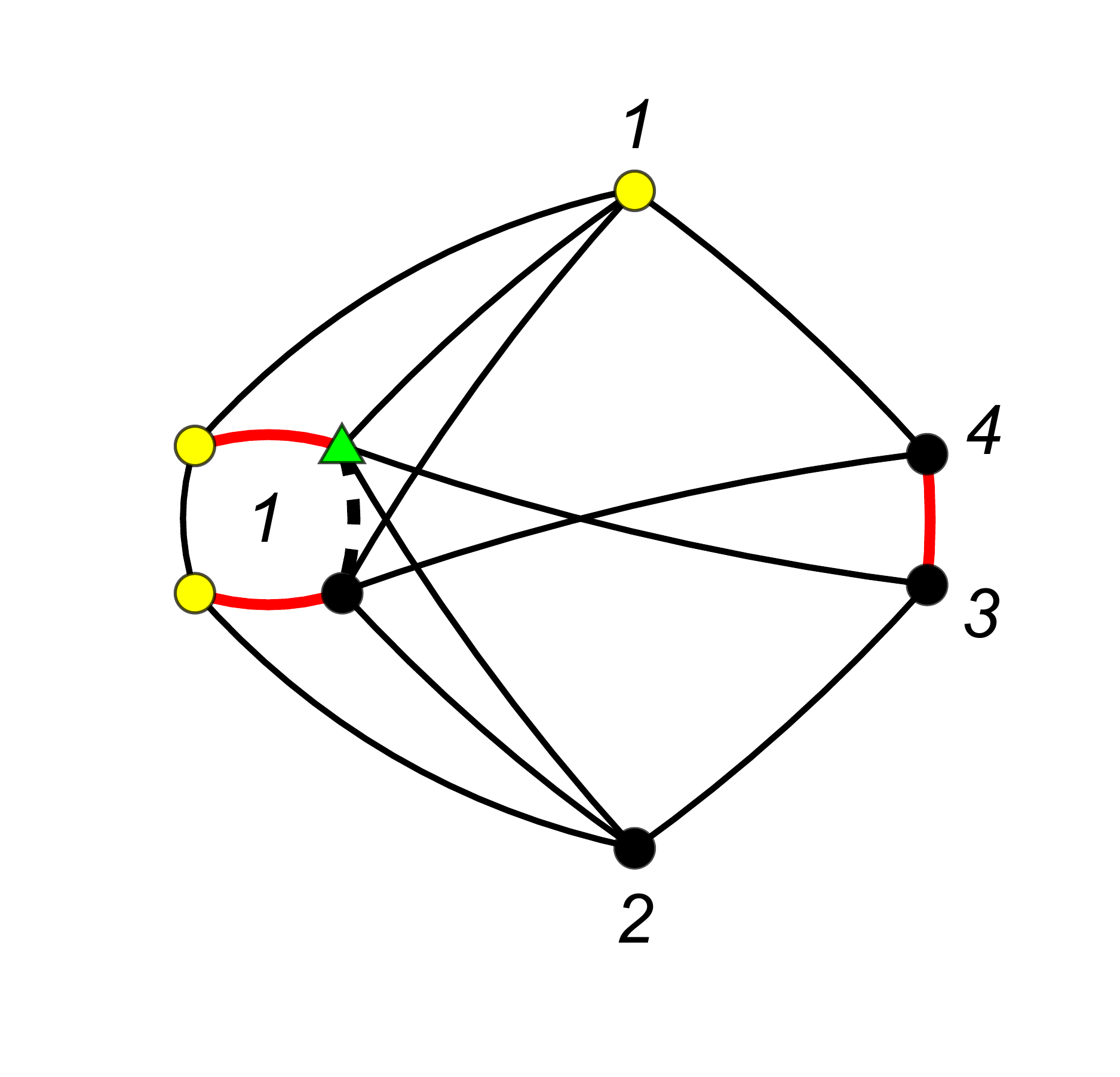}}+
 \hspace{-0.4cm}
\parbox[c]{7.5em}{\includegraphics[scale=0.16]{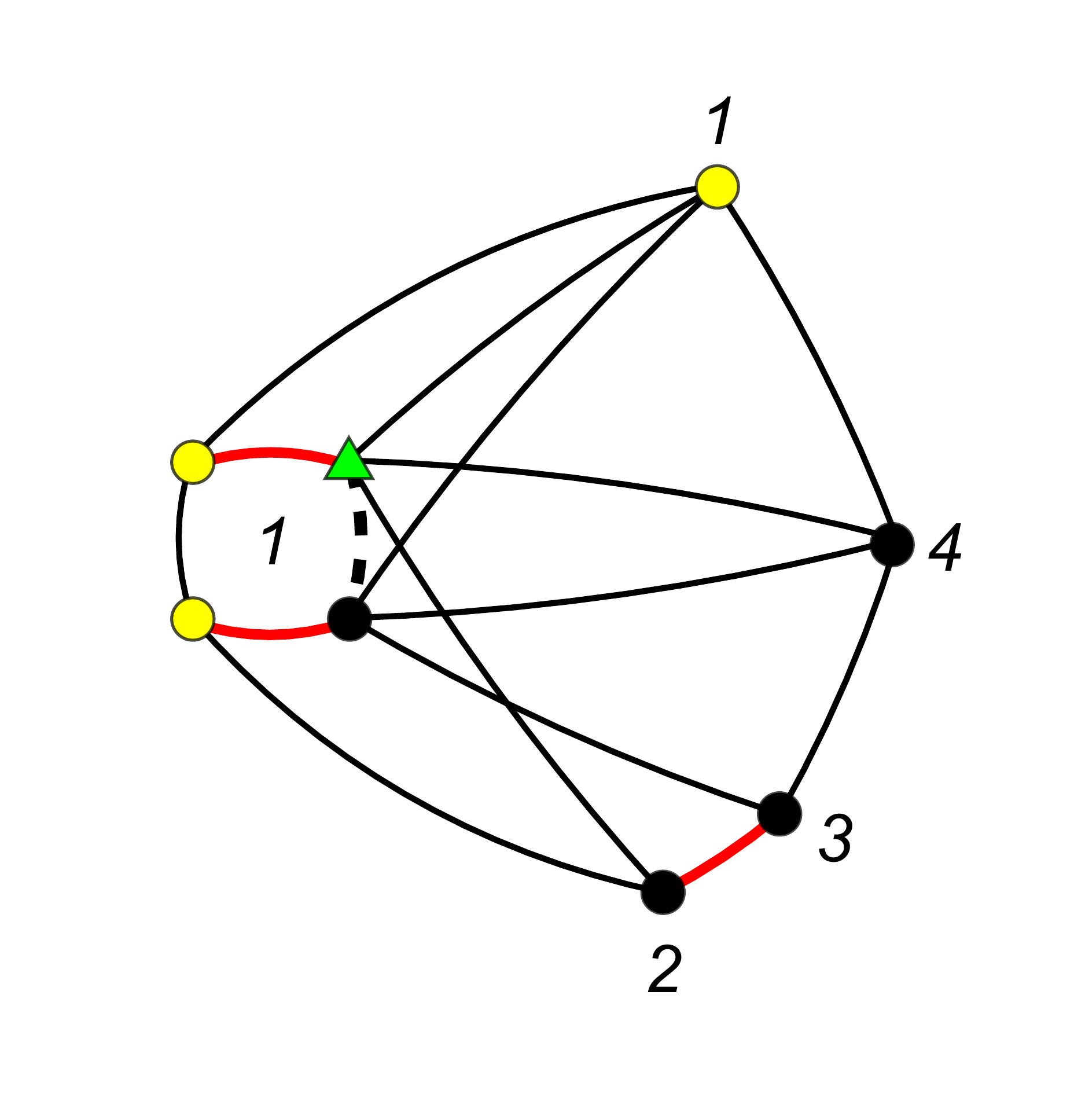}}+ 
 \hspace{-0.4cm}
\parbox[c]{8em}{\includegraphics[scale=0.16]{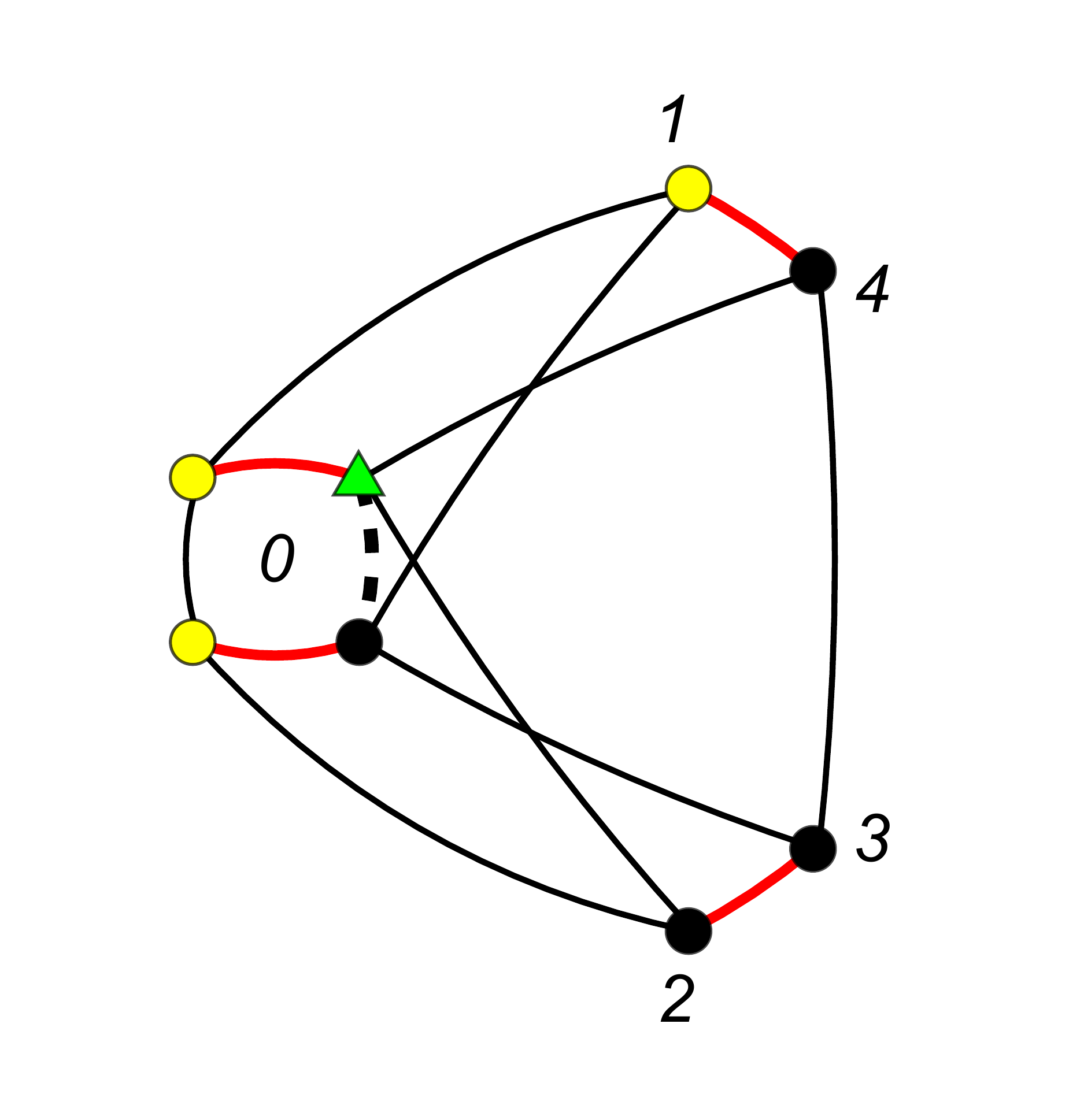}}
 \right.
\nonumber\\
&&\hspace{2cm}
+
\left.
 \hspace{-0.4cm}
\parbox[c]{8em}{\includegraphics[scale=0.16]{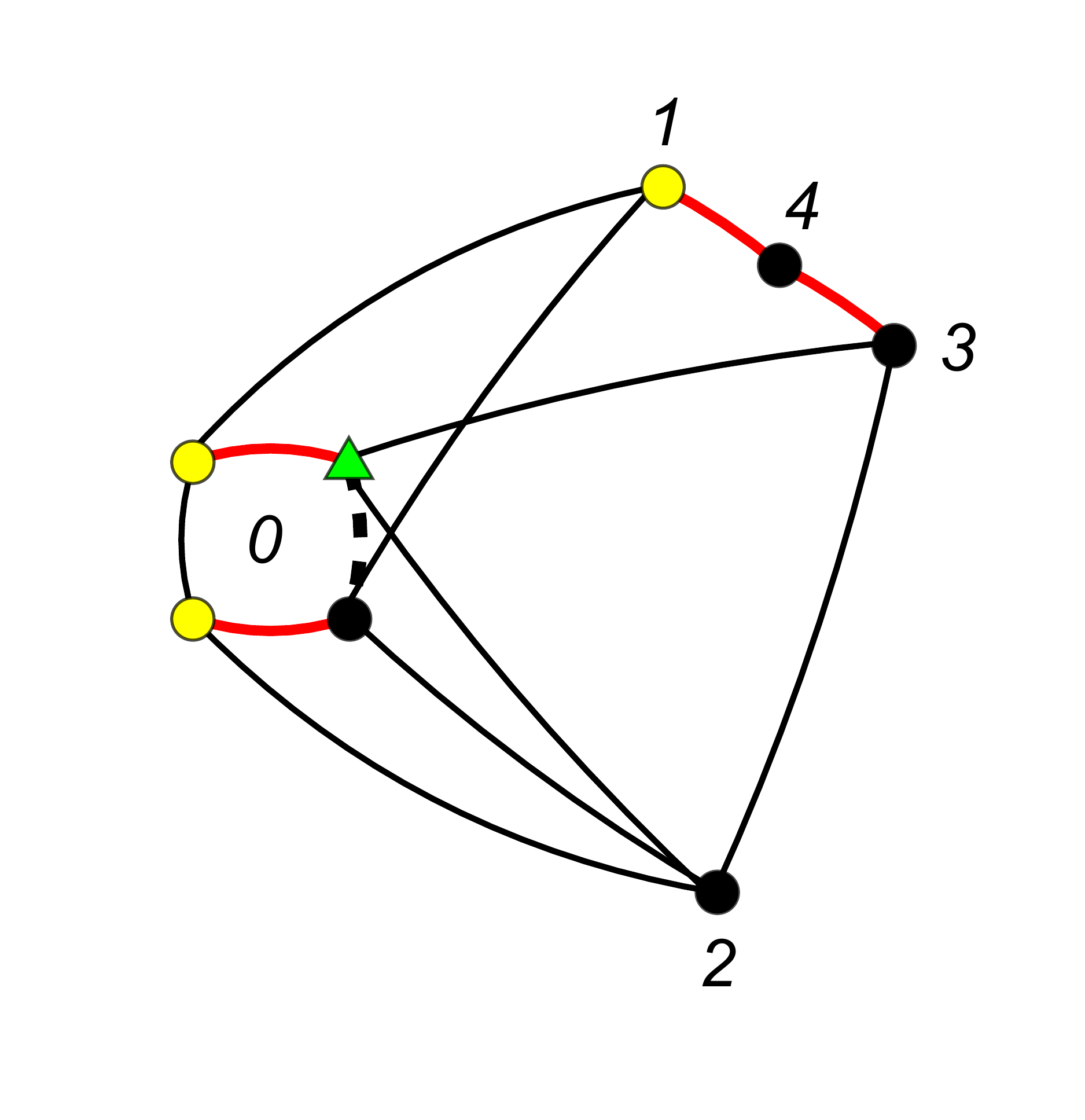}}+ 
 \hspace{-0.4cm}
\parbox[c]{8em}{\includegraphics[scale=0.16]{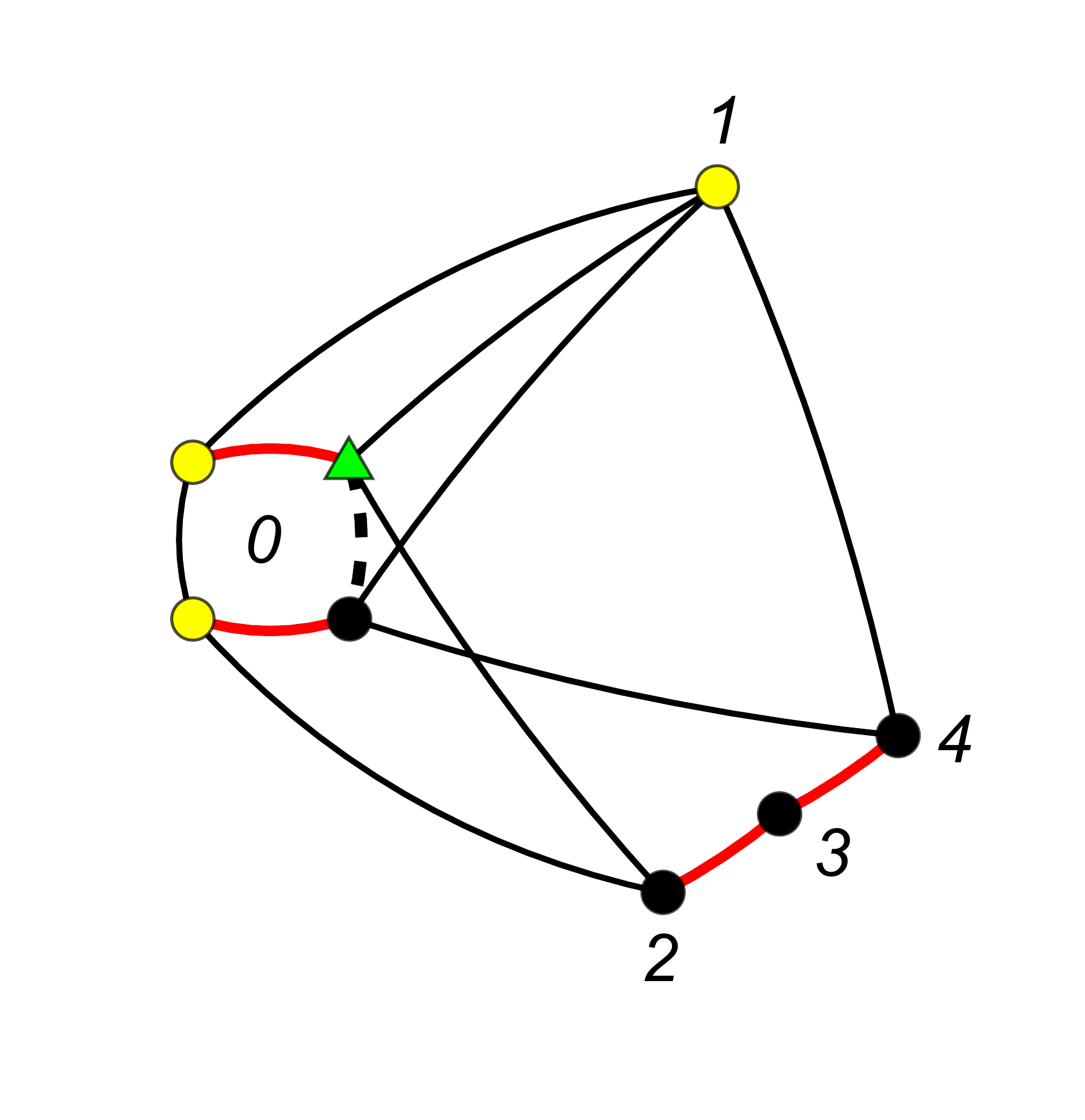}}+
{\rm cyc(1,2,3,4)} \right\}\,.
\label{CHY(1234.1234)}
\end{eqnarray}
All the integrals in (\ref{CHY(1234.1234)}) can be computed using the {\bf propositions 1,2,3} and the technology described in their proofs. The result is, following the same order: one box, four triangles, two bubbles and the external-leg bubbles The latter must be regularized.  In terms of Feynman diagrams one has
\begin{eqnarray}
&&\mathfrak{M}_4^{\rm 1-loop}[1,2,3,4| 1,2,3,4] 
=\,\,-
\hspace{-0.77cm}
 \parbox[c]{8em}{\includegraphics[scale=0.22]{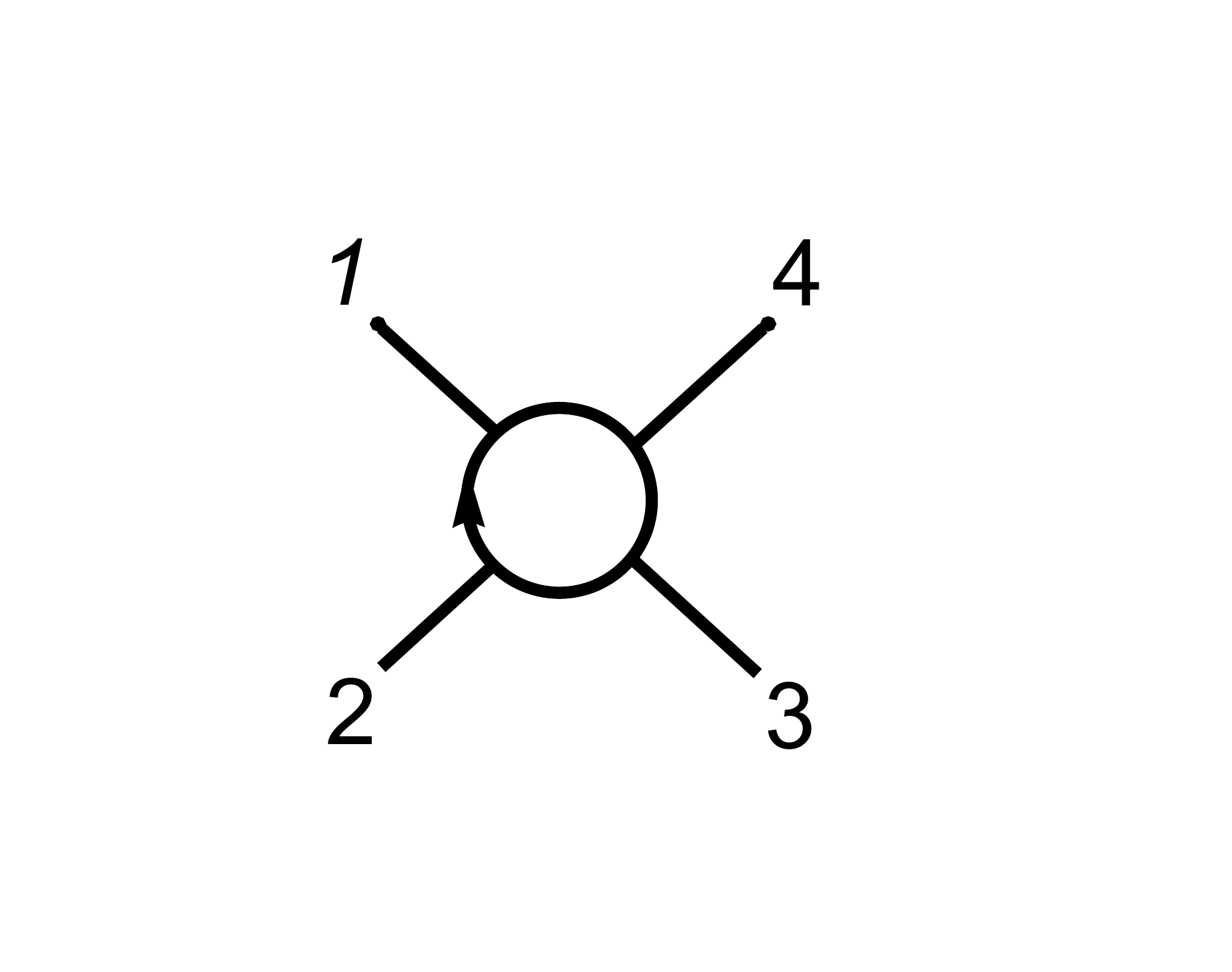}}-
 \,\,\,
 \left(
 \hspace{-0.4cm}
 \parbox[c]{8em}{\includegraphics[scale=0.2]{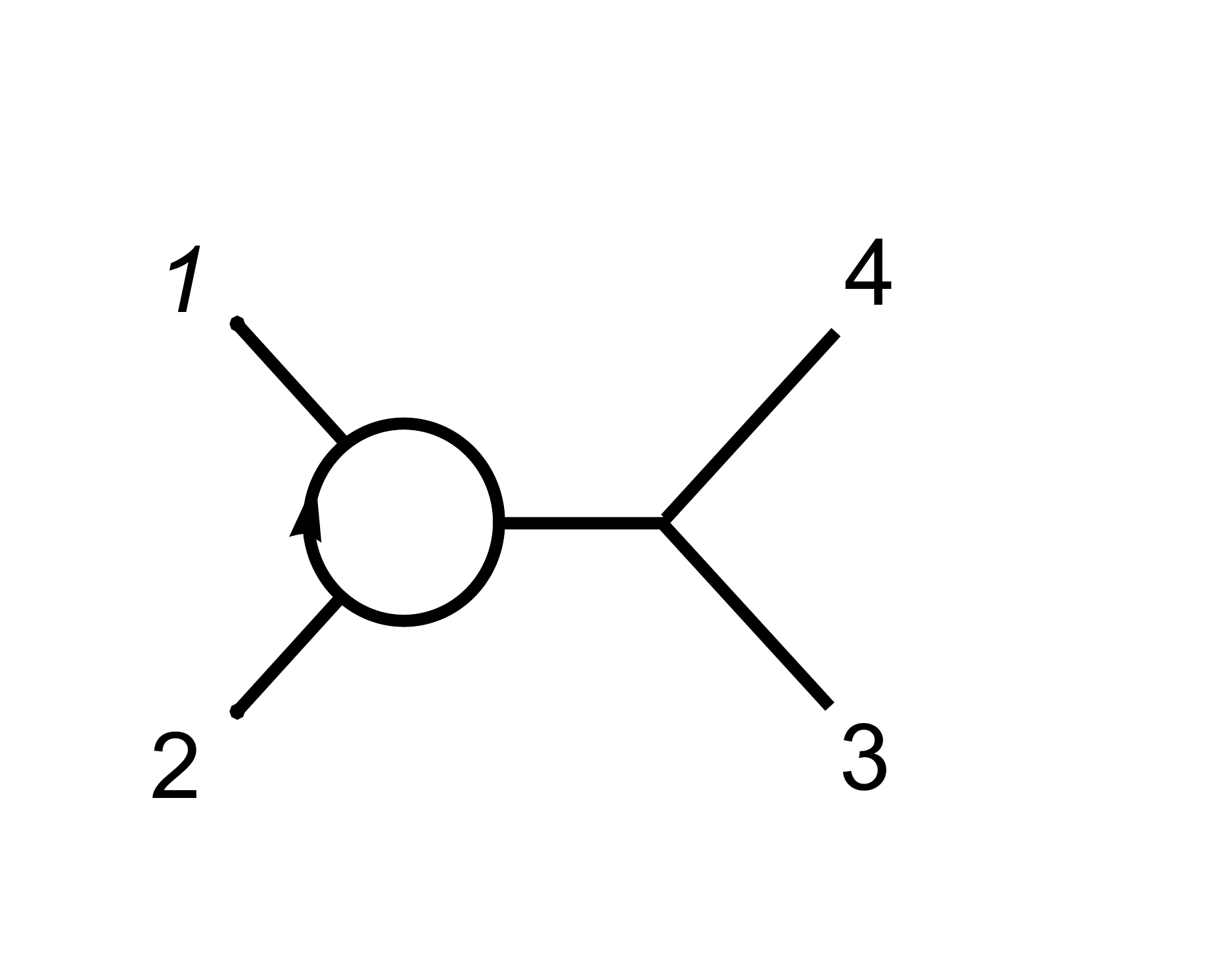}}
+ 
{\rm cyc(1,2,3,4)} \right) \nonumber \\
&&
 -
 \hspace{-0.2cm}
  \parbox[c]{8em}{\includegraphics[scale=0.19]{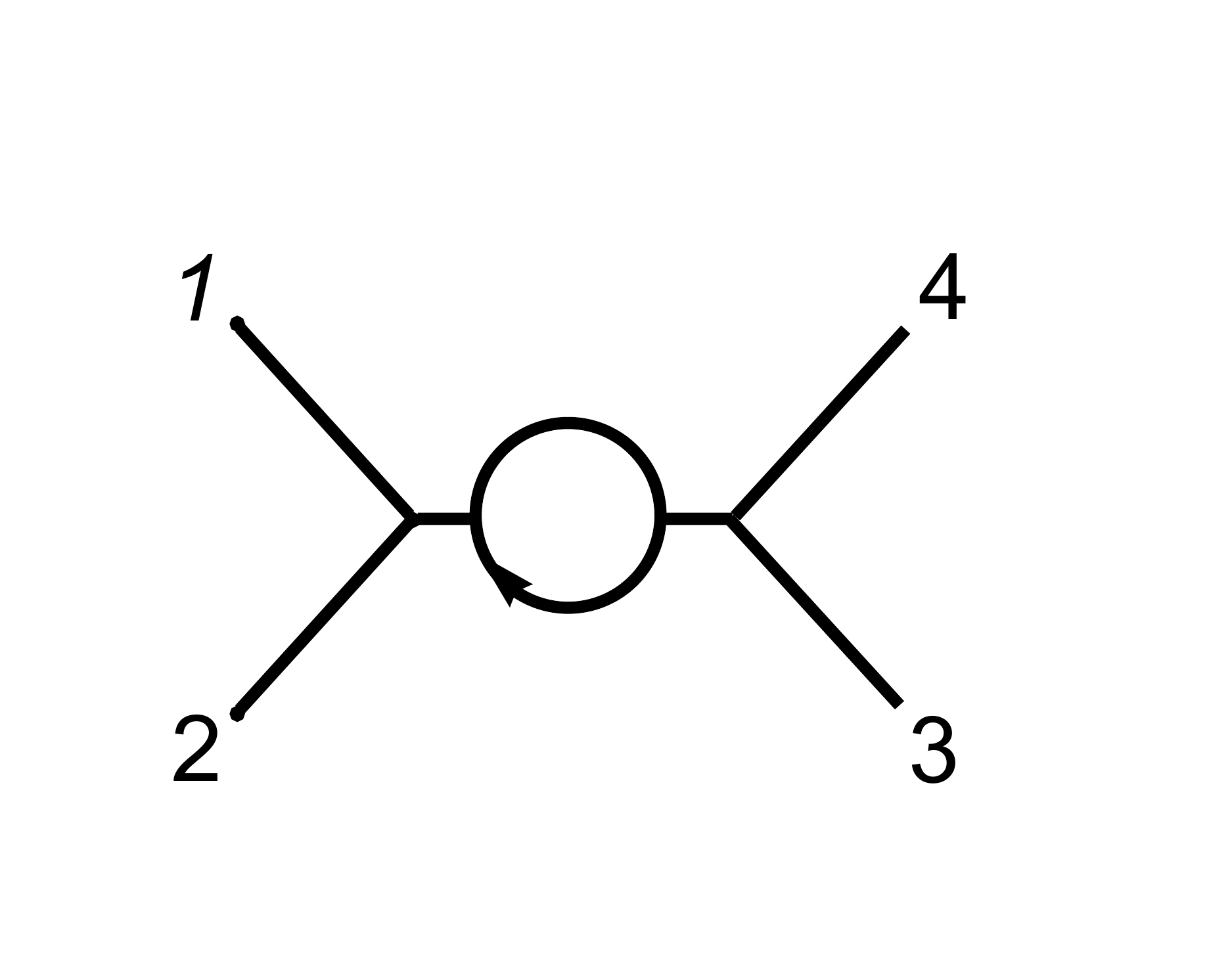}} -
   \hspace{-0.55cm}
 \parbox[c]{7em}{\includegraphics[scale=0.18]{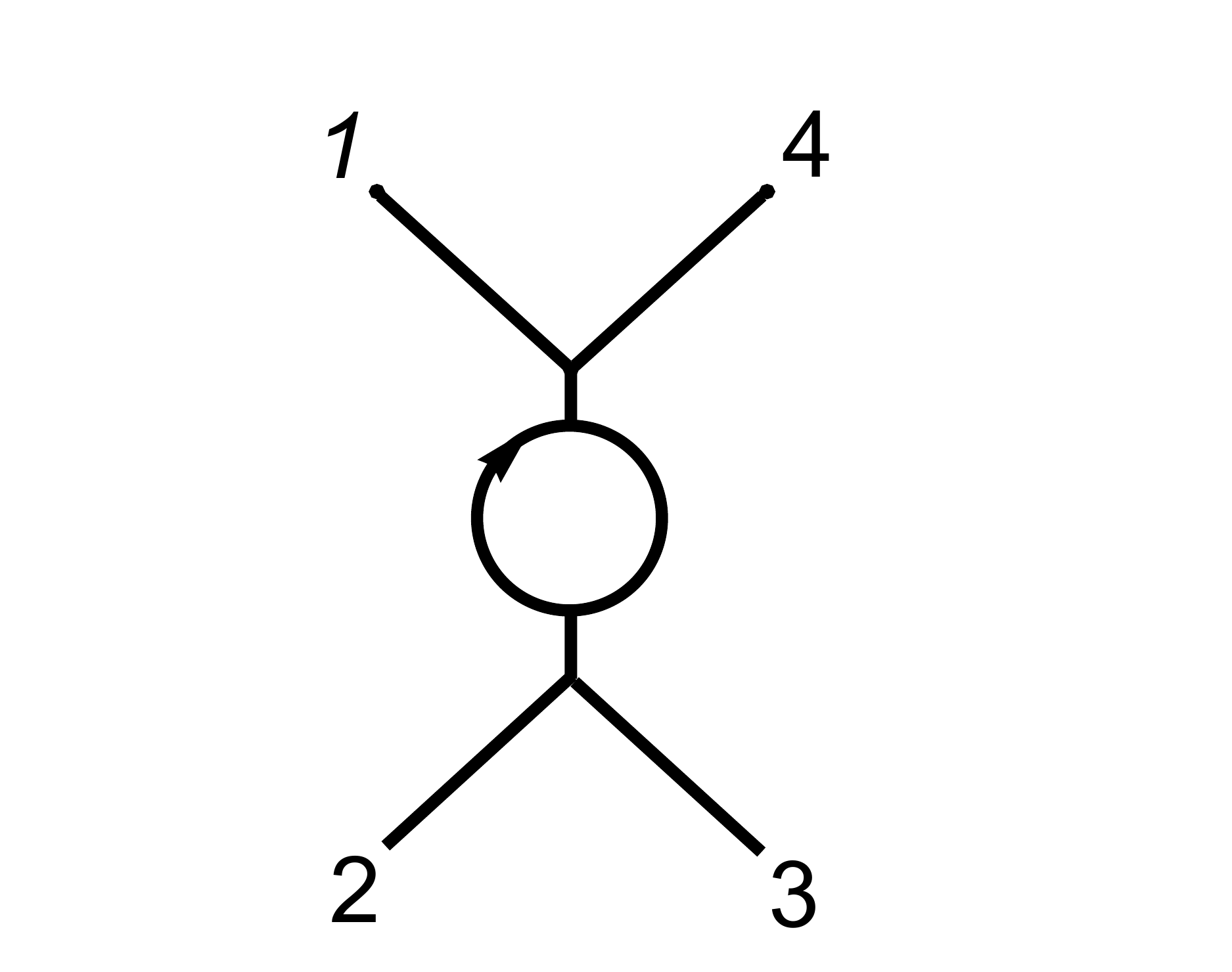}}
-
\,\,
 \left( 
 \hspace{-0.55cm}
\parbox[c]{7em}{\includegraphics[scale=0.2]{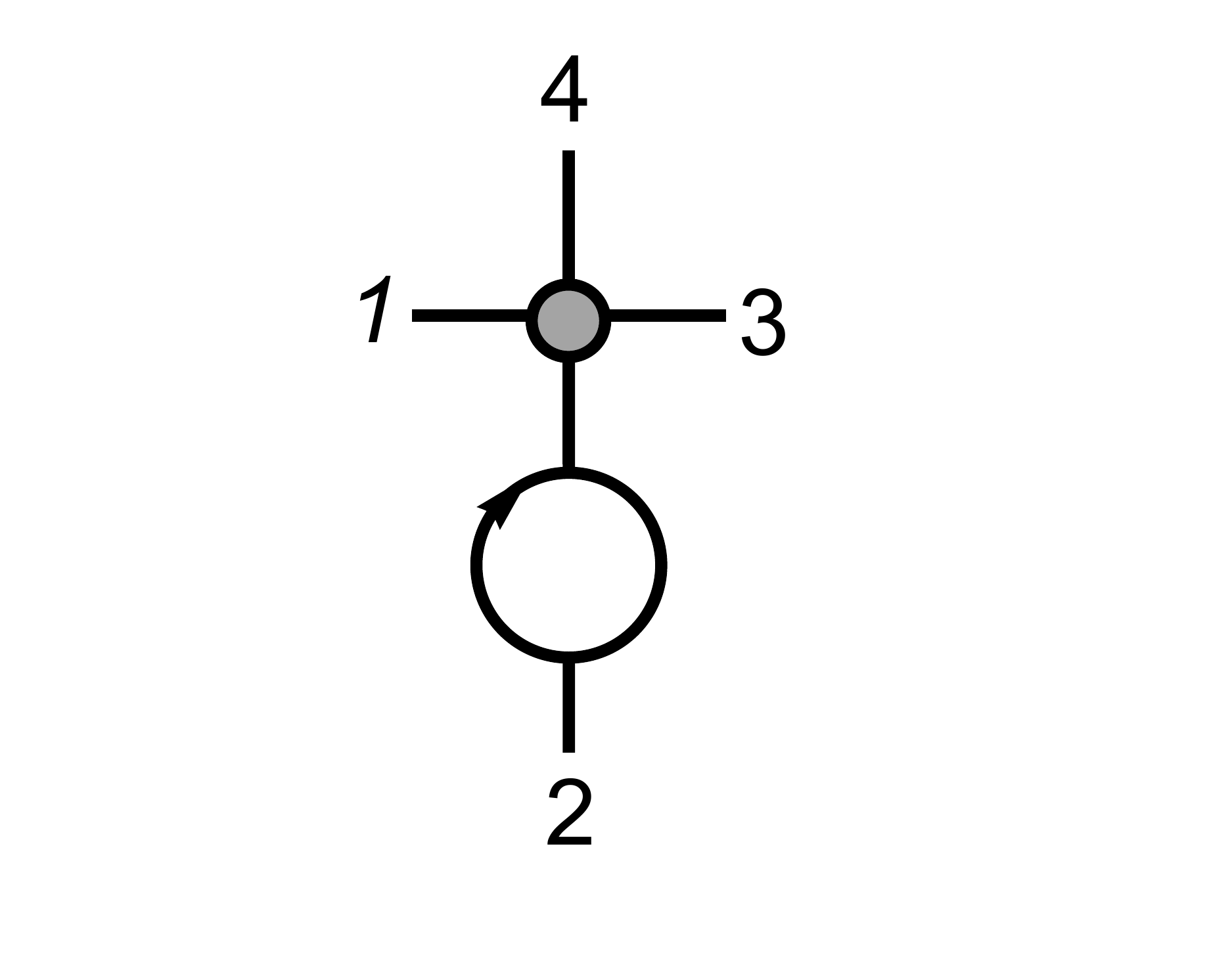}}+ 
{\rm cyc(1,2,3,4)} \right) \,,
\label{FEY(1234.1234)}
\end{eqnarray}
where the grey circle in the third graph inside of the bracket means the sum over all possible trivalent planar diagrams. This result was checked numerically.

In order to calculate analytically the next contributions we are going to use the conjecture in \eqref{inter1lopp}
\begin{eqnarray}
&&\mathfrak{M}_4^{\rm 1-loop}[1,2,3,4| 4,3,2,1] 
=(-1)\,\mathfrak{M}_4^{\rm 1-loop}[1,2,3,4| 1,2,3,4] \cap  \mathfrak{M}_4^{\rm 1-loop}[4,3,2,1| 4,3,2,1] \nonumber\\
&&
=  -
 \hspace{-0.4cm}
  \parbox[c]{8em}{\includegraphics[scale=0.19]{fey-bubble-cano2.pdf}} -
   \hspace{-0.55cm}
 \parbox[c]{7em}{\includegraphics[scale=0.18]{fey-bubble-cano.pdf}}
 -
\,\,
 \left( 
 \hspace{-0.55cm}
\parbox[c]{7em}{\includegraphics[scale=0.2]{fey-bubbleL-cano.pdf}}+ 
{\rm cyc(1,2,3,4)} \right) \, ,
\\
&&\mathfrak{M}_4^{\rm 1-loop}[1,2,3,4| 1,2,4,3] 
=(-1)\,\mathfrak{M}_4^{\rm 1-loop}[1,2,3,4| 1,2,3,4] \cap  \mathfrak{M}_4^{\rm 1-loop}[1,2,4,3| 1,2,4,3]  \nonumber\\
&&
=  -
 \hspace{-0.46cm}
 \parbox[c]{7em}{\includegraphics[scale=0.19]{fey-tri-cano.pdf}}-
   \hspace{-0.55cm}
  \parbox[c]{8em}{\includegraphics[scale=0.19]{fey-bubble-cano2.pdf}}
 -
 \hspace{-0.68cm}
\parbox[c]{5em}{\includegraphics[scale=0.24]{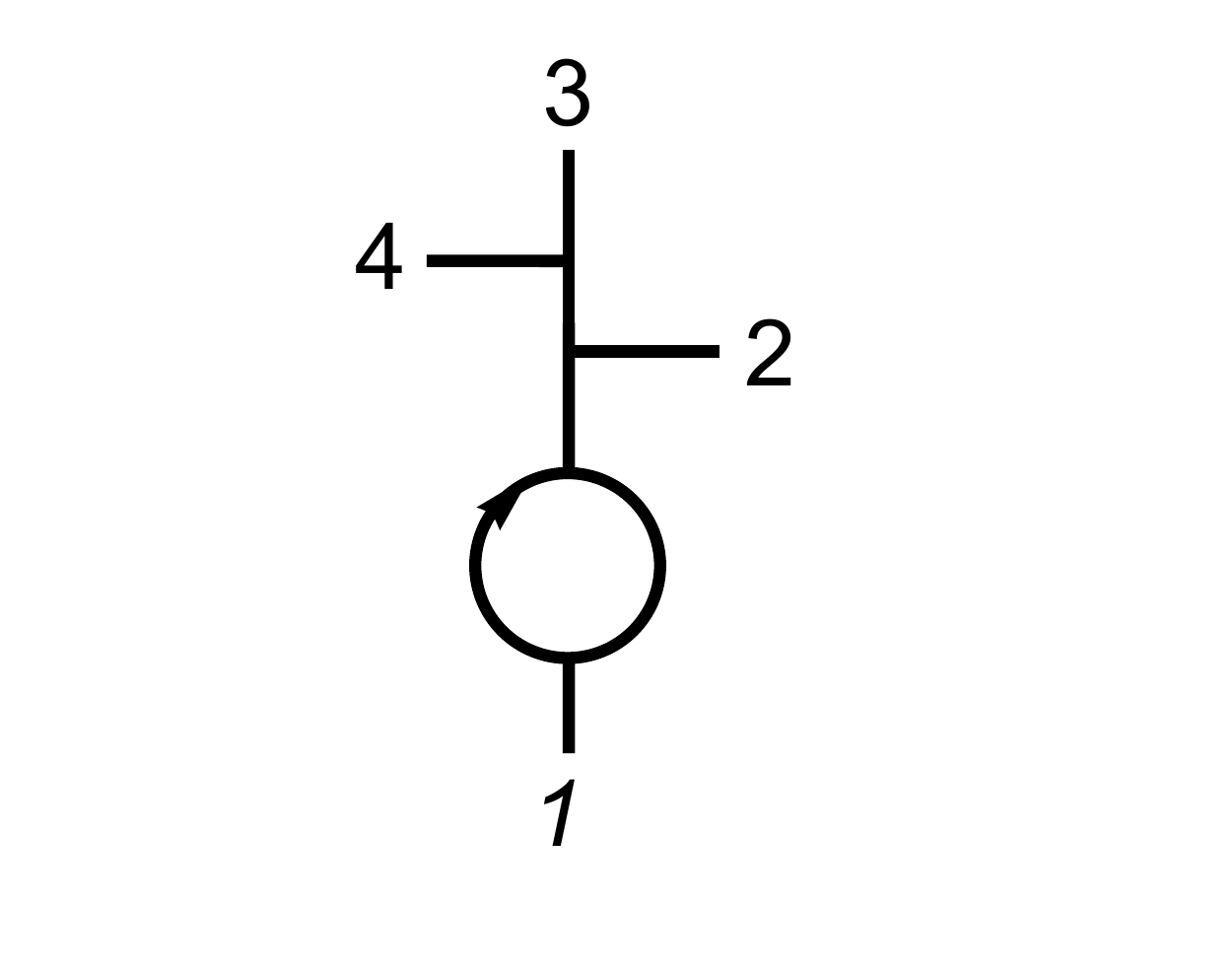}}-
 \hspace{-0.68cm}
\parbox[c]{5em}{\includegraphics[scale=0.24]{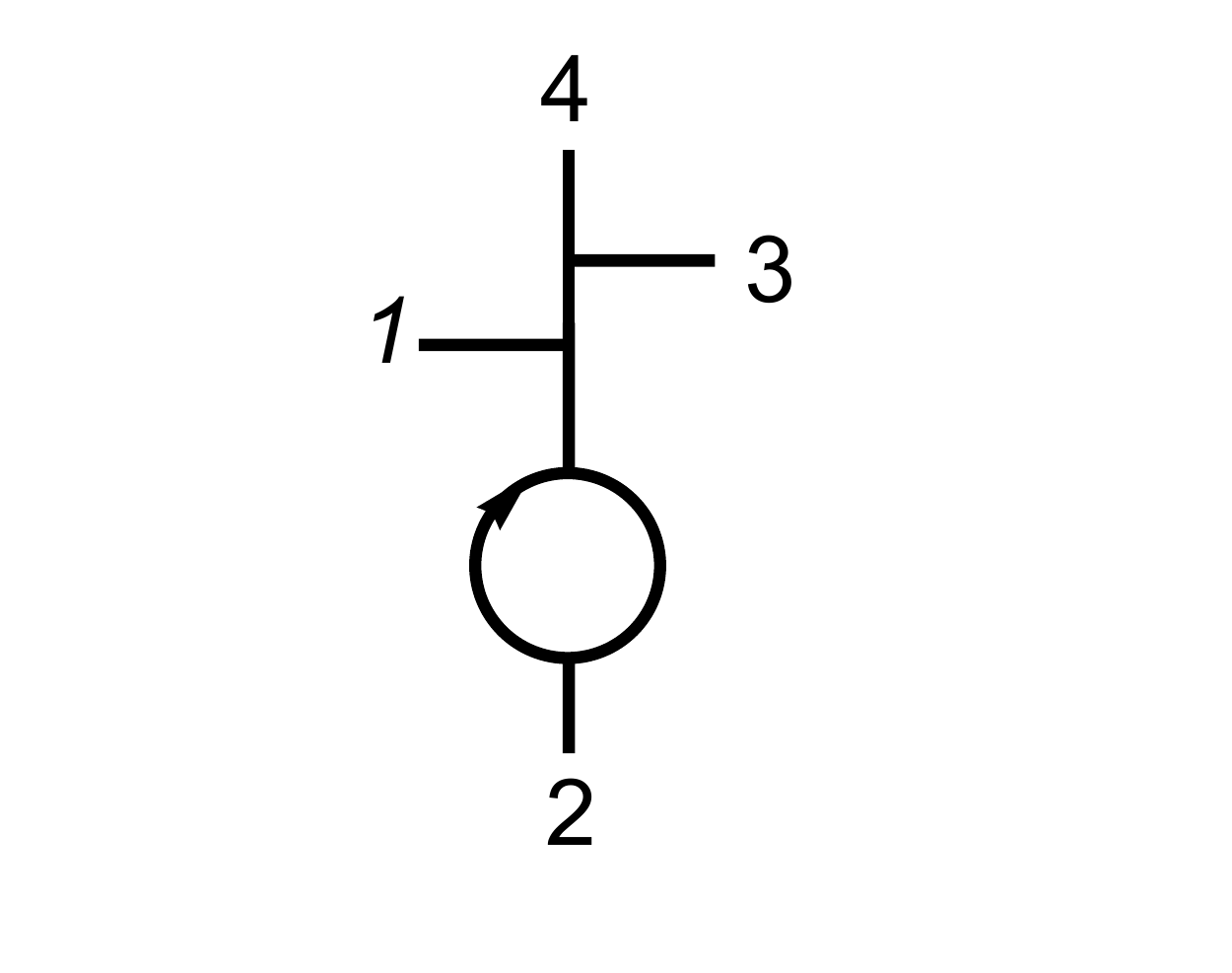}}-
 \hspace{-0.68cm}
\parbox[c]{5em}{\includegraphics[scale=0.24]{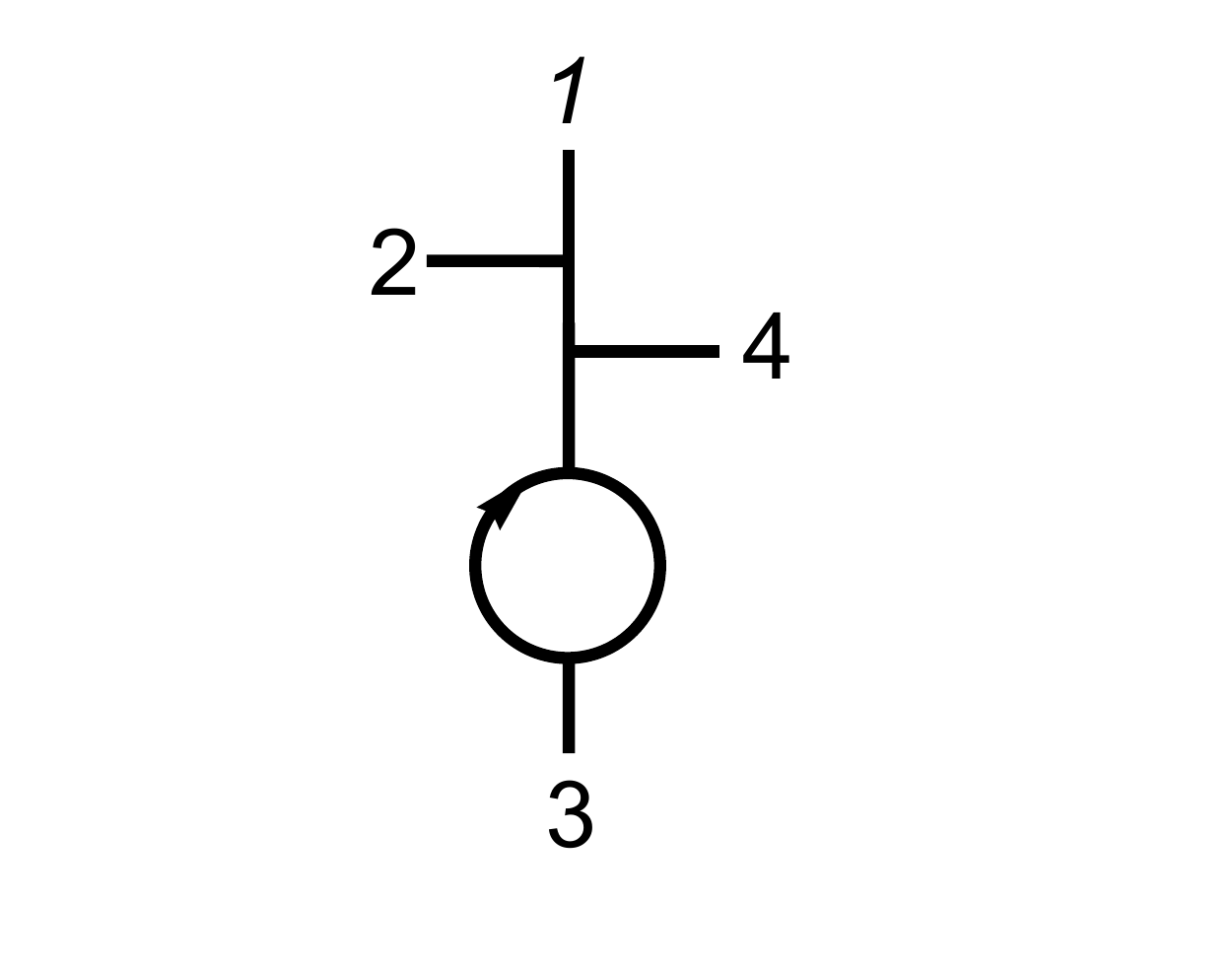}}-
 \hspace{-0.68cm}
\parbox[c]{5em}{\includegraphics[scale=0.24]{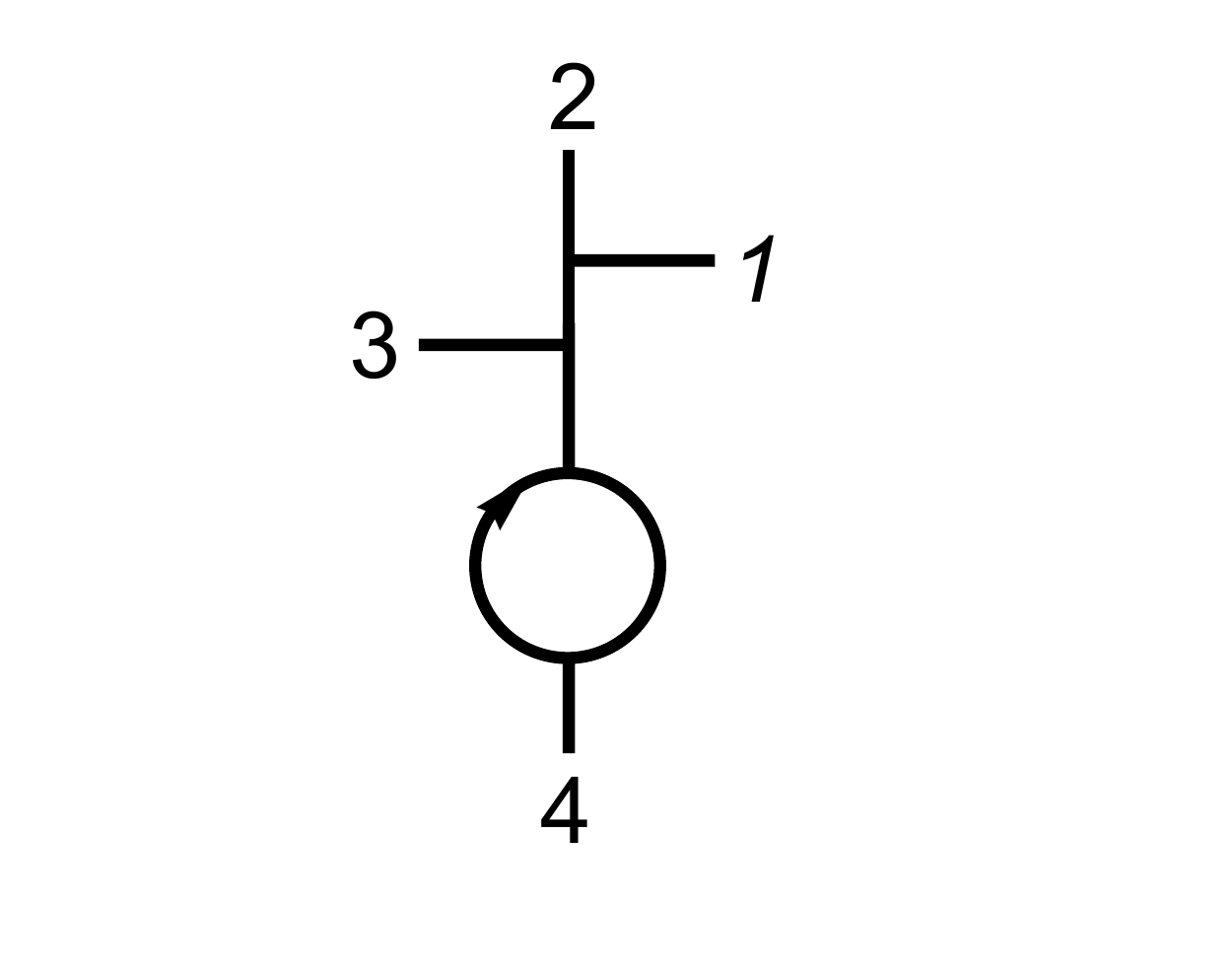}}.
\end{eqnarray}
These results were checked numerically concluding that, in fact, the conjecture in \eqref{inter1lopp} works perfectly. 

With this procedure we can easily go to higher point cases, since we can always construct the 
CHY-graphs for $\mathfrak{M}_n^{\rm 1-loop}[\mathbb{1}|\mathbb{1} ]$, where $\mathbb{1}$ means canonical ordering, and know how to calculate all the CHY-integrals.

\subsection{General Structure of $\mathfrak{M}_N^{\rm 1-loop}[\mathbb{1}_N|\mathbb{1}_N] $}

From the CHY-graphs representation found for $\mathfrak{M}_3^{\rm 1-loop}[\mathbb{1}_3|\mathbb{1}_3 ]$ and $\mathfrak{M}_4^{\rm 1-loop}[\mathbb{1}_4|\mathbb{1}_4 ]$ in \eqref{CHY(123.123)2} and \eqref{CHY(1234.1234)} respectively, where $\mathbb{1}_N$ means the canonical ordering $(1,2,\ldots, N)$, it is direct to obtain the general expression for $\mathfrak{M}_N^{\rm 1-loop}[\mathbb{1}_N|\mathbb{1}_N] $. To be precise, up to global sign, one has 
\begin{eqnarray}\label{M-Ngen}
\hspace{-0.7cm}
\mathfrak{M}_N^{\rm 1-loop}[\mathbb{1}_N|\mathbb{1}_N] = \frac{1}{2^{N+1}} \int d\Omega \,\, s_{a_1 b_1} 
 \int d\mu^{\rm tree}_{N+4}  \left[
 \sum_{a=2}^N \sum_i{\rm Gchy}^{a-2}_{1,\ldots, N}[[i]] +{\rm cyc}(1,\ldots, N)\right]
 \hspace{-0.1cm},
\end{eqnarray}
where the set, ${\rm Gchy}^{a-2}_{1,\ldots, N}$, is defined as
\begin{eqnarray}
{\rm Gchy}^{n-2}_{1,\ldots, N}:=\left\{
{\rm All\,\, possible\,\, CHY-graphs\,\, with\,\, the \,\, form}  \right.
\hspace{-0.2cm}
\parbox[c]{9em}{\includegraphics[scale=0.26]{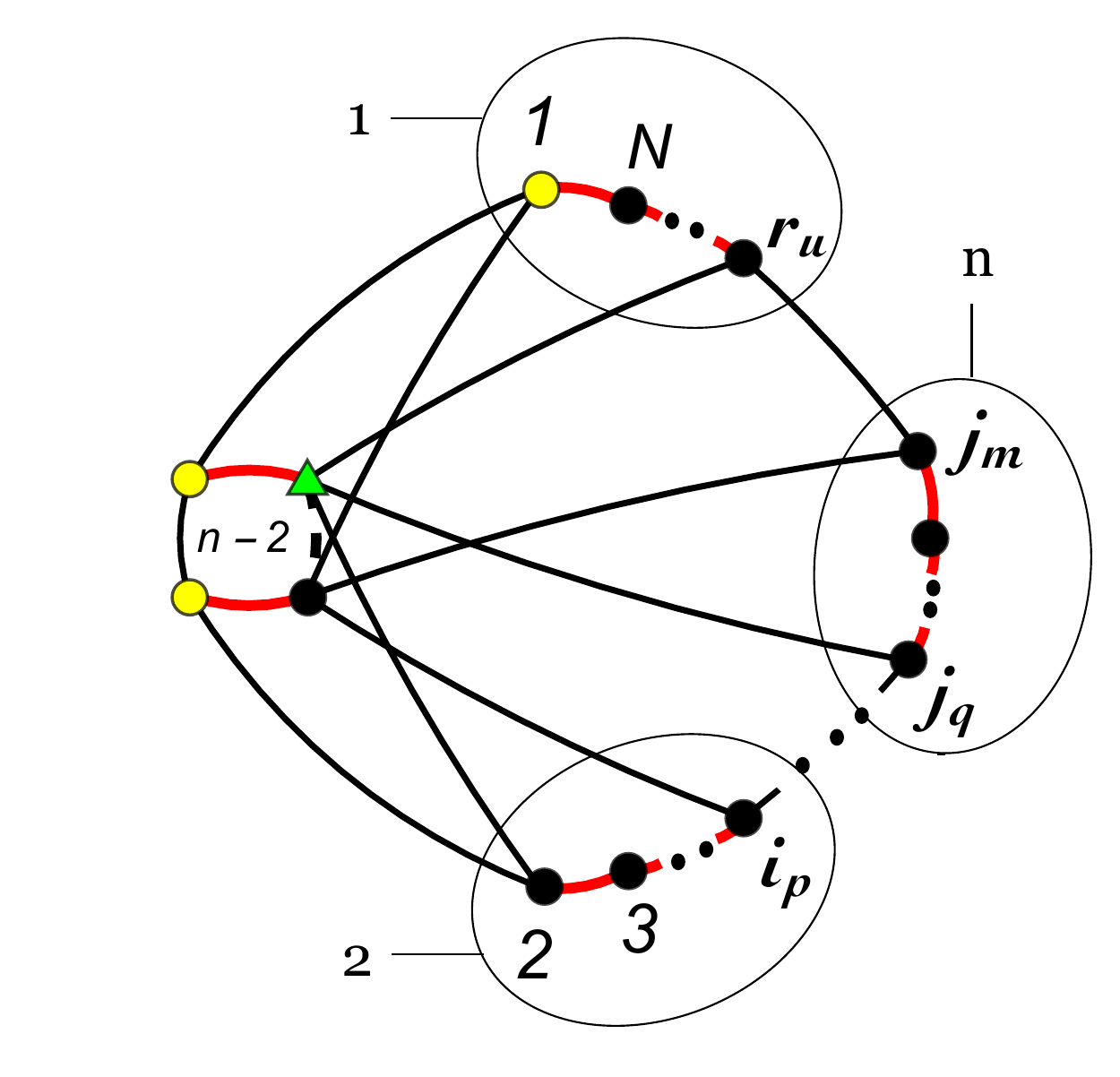}}
\left.
\right\}\,,\label{Gchydefi}
\end{eqnarray}
being ${\rm Gchy}^{n-2}_{1,\ldots, N}[[i]]$ an element in ${\rm Gchy}^{n-2}_{1,\ldots, N}$. For instance 
\begin{eqnarray}
{\rm Gchy}^{1}_{1,2,3,4} 
=\left\{
 \hspace{-0.4cm}
 \parbox[c]{7em}{\includegraphics[scale=0.14]{1-l_4-p_c2U.pdf}},
 \hspace{-0.2cm}
 \parbox[c]{7.5em}{\includegraphics[scale=0.15]{1-l_4-p_c2.pdf}},
 \hspace{-0.2cm}
\parbox[c]{7em}{\includegraphics[scale=0.14]{1-l_4-p_c2D.pdf}} 
\right\}\,.
\end{eqnarray}
Therefore, it now is clear that \eqref{M-Ngen} is in agreement with \eqref{CHY(123.123)2} and \eqref{CHY(1234.1234)}. 

Using the same techniques developed in the proof of {\bf proposition 1}, we can compute the CHY-integral for a generic element in  ${\rm Gchy}^{n-2}_{1,\ldots, N}$. Thus, we assert the following general result at the integrand level 
\begin{eqnarray}\label{MOSTGEN}
\frac{1}{2^{N+1}} \int d\Omega \times s_{a_1 b_1} \times
 \int d\mu^{\rm tree}_{N+4}  
\hspace{-0.4cm}
 \parbox[c]{11em}{\includegraphics[scale=0.34]{mostG-chy.pdf}}
 =
 \hspace{-0.25cm}
  \parbox[c]{12em}{\includegraphics[scale=0.38]{mostG-fey.pdf}}
\end{eqnarray}
where the grey circles  mean the sum over all possible trivalent planar diagrams and the symbol, ``$P_n$'', means the loop circle is  a regular polygon of n edges, where we define $P_2$ as a bubble.

Finally, using the general results given in \eqref{M-Ngen} and  the ``map" in  \eqref{MOSTGEN}, the structure of  $\mathfrak{M}_N^{\rm 1-loop}[\mathbb{1}_N|\mathbb{1}_N]$ becomes simple. In \cite{wp}, we also analyse the general case for the integrands that contribute to $\mathfrak{M}_N^{\rm 1-loop}[\pi|\rho]$.

\section{External-Leg Bubbles}\label{sectionEB}

As it is known, in the linear propagators formalism the external-leg bubbles vanish. To illustrate that, let us  consider the following Feynman integrand
\vspace{-0.8cm}
\begin{eqnarray}\label{bubbleS}
 \parbox[c]{9.6em}{\includegraphics[scale=0.34]{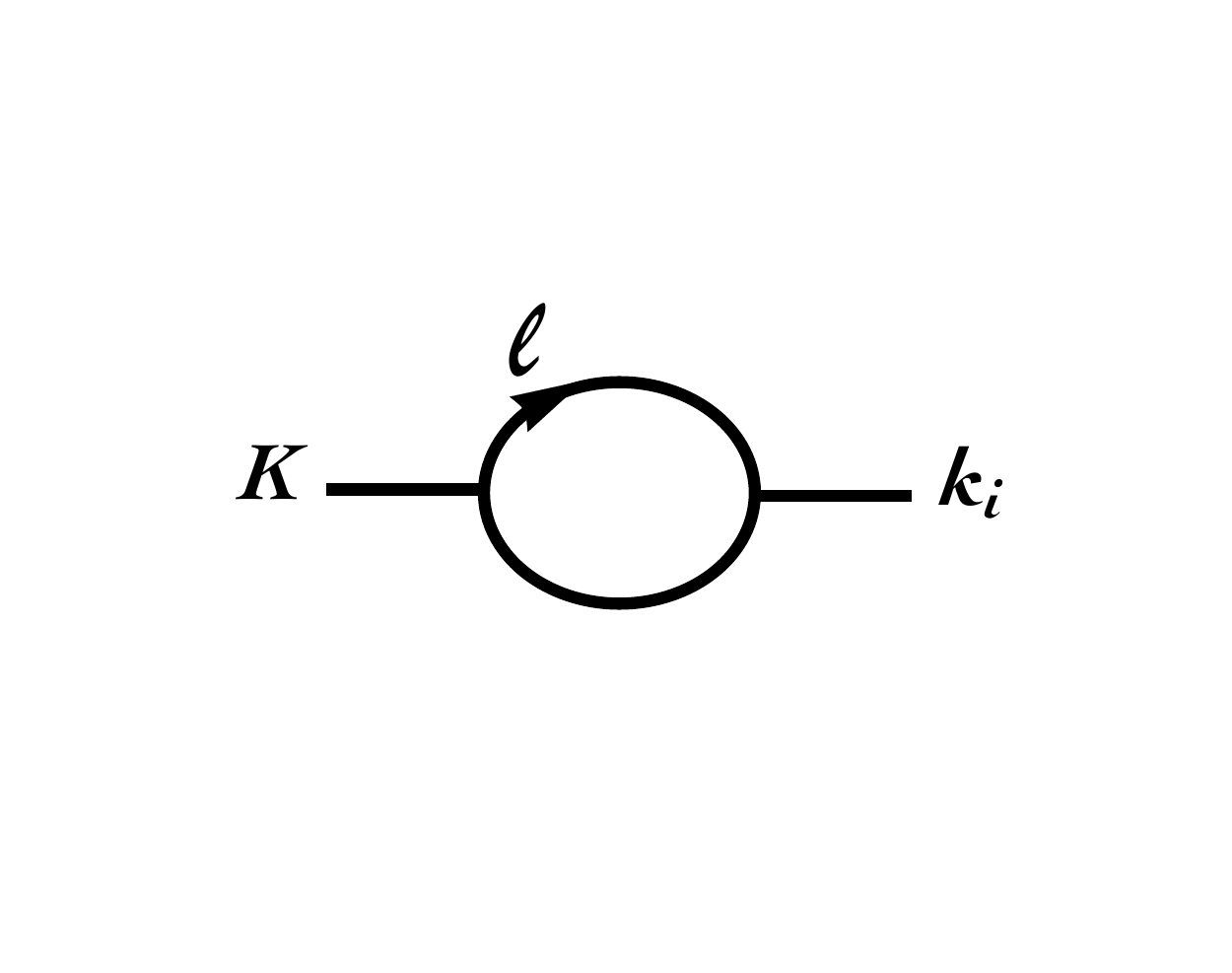}}
 =\,\,\frac{1}{\ell^2 (\ell+k_i)^2}=\,\,\frac{1}{\ell^2 (2\,\ell\cdot k_i + k_i^2)}-\frac{1}{(\ell+k_i)^2 (2\,\ell\cdot k_i + k_i^2)},
\end{eqnarray}
where we have applied the partial fraction identity, $\frac{1}{A\,B}=\frac{1}{A\,(B-A)}+\frac{1}{B\,(A-B)}$. Performing a shifting over the loop momentum variable in the second term in \eqref{bubbleS}, i.e. $\ell\rightarrow \tilde\ell=\ell+k_i$, one obtains\footnote{We are assuming that the integration measure over $\ell$ is invariant under this transformation.} 
\begin{equation}\label{PFI}
\frac{1}{\ell^2 (\ell+k_i)^2}=
\frac{1}{\ell^2 (2\,\ell\cdot k_i + k_i^2)}-\frac{1}{\ell^2 (2\ell\cdot k_i - k_i^2)}.
\end{equation}
Since we are interested on the external leg bubbles\footnote{On this example we are considering that the propagator,$\frac{1}{K^2}$, is regularized.}, we assume that $k_i$ is an external massless on-shell particle and therefore \eqref{PFI} vanishes. This is the reason,  in principle, why in the linear propagator formalism the external-leg bubbles do not appear \cite{Baadsgaard:2015hia}.  In \cite{He:2015yua}
it was also argued that the external-leg bubbles contribution must be regularized.
In the proposal we present in this work, the external-leg bubbles appear in a natural way and been also singular their contribution need to be regularize. 
Notice that from the map \eqref{MOSTGEN}, it is straightforward to identify the external-leg bubbles contributions in $\mathfrak{M}_N^{\rm 1-loop}[\mathbb{1}_N|\mathbb{1}_N]$. This contribution is given by the expression
\begin{eqnarray*}
\hspace{-0.4cm}
\mathfrak{M}_{{\rm singular}}^{\rm 1-loop}[\mathbb{1}_N|\mathbb{1}_N] = \frac{1}{2^{N+1}}\int d\Omega \times  s_{a_1 b_1}
 \int d\mu^{\rm tree}_{N+4}  \left[ \sum_{j=1}^2{\rm Gchy}^{\rm 0-ELB}_{1,\ldots, N}[[j]] +{\rm cyc}(1,\ldots, N)\right]
 \hspace{-0.1cm} ,
\end{eqnarray*}
where ${\rm Gchy}^{\rm 0-ELB}_{1,\ldots, N}$ is defined as
\begin{eqnarray}
{\rm Gchy}^{\rm 0-ELB}_{1,\ldots, N} 
:=\left\{
 \hspace{-0.3cm}
 \parbox[c]{7em}{\includegraphics[scale=0.21]{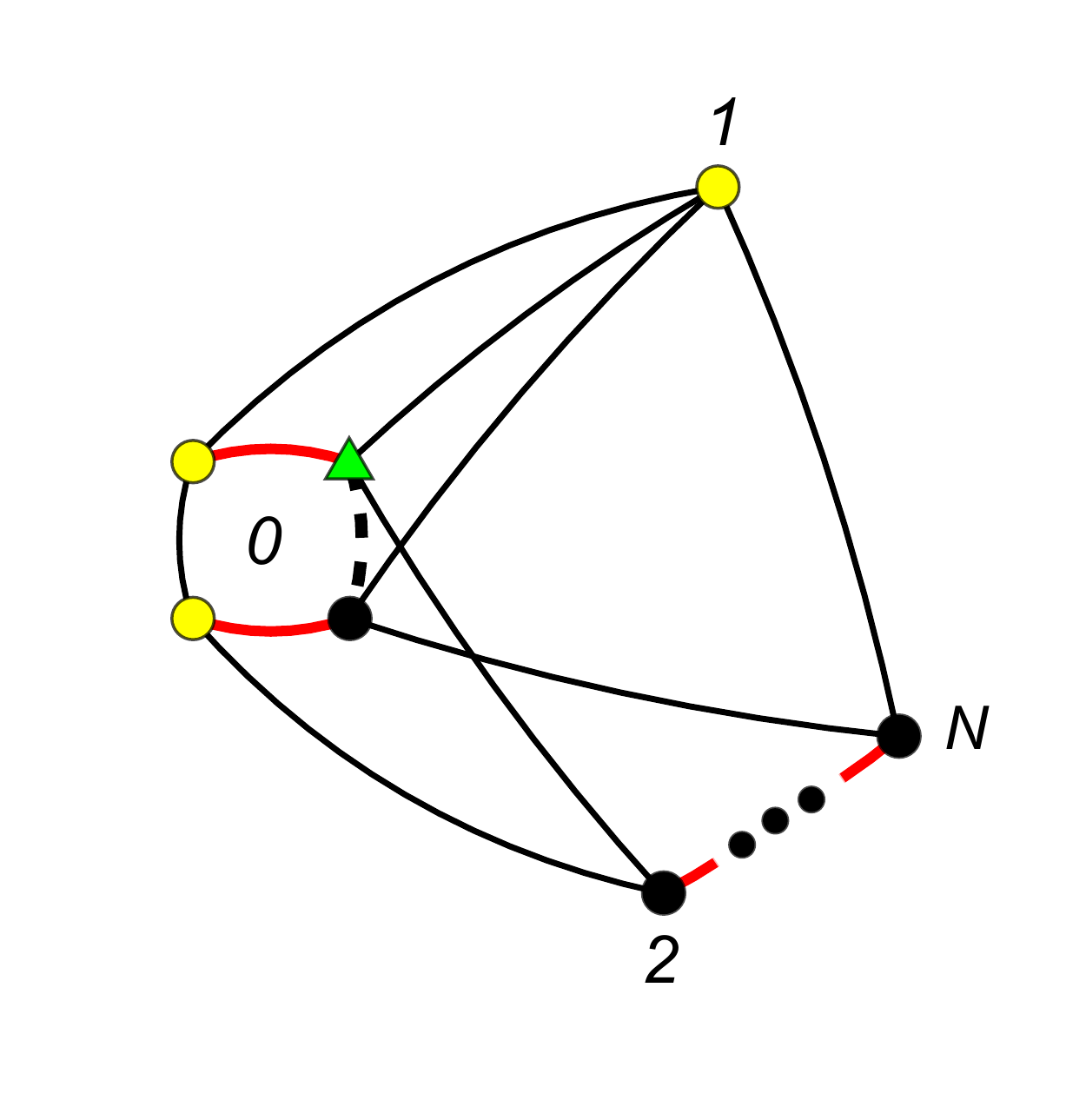}},
 \hspace{-0.2cm}
 \parbox[c]{6.6em}{\includegraphics[scale=0.21]{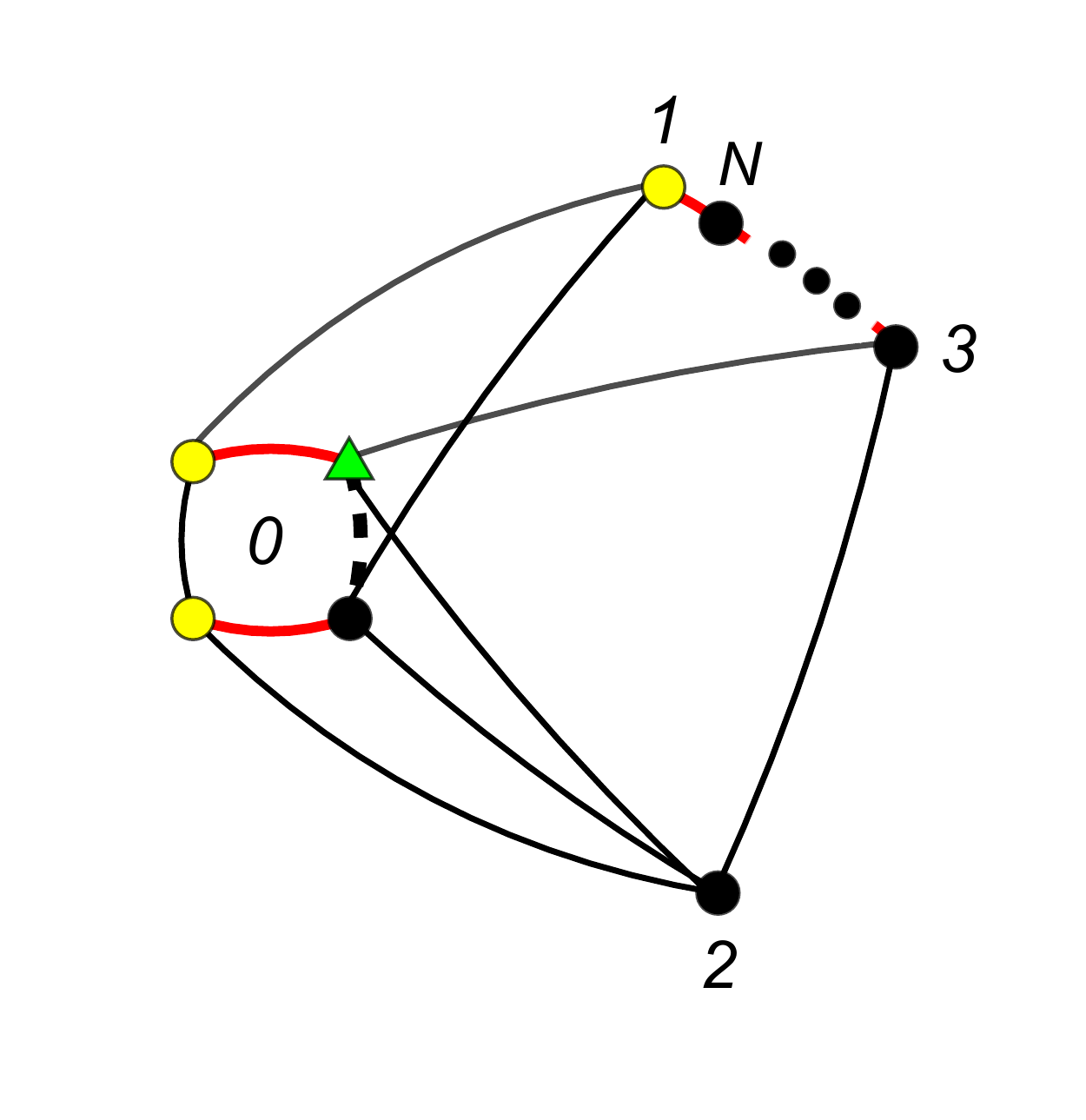}}
\,\,\,
\right\}\,.
\end{eqnarray}
Clearly,  ${\rm Gchy}^{\rm 0-ELB}_{1,\ldots, N} \subset {\rm Gchy}^0_{1,\ldots, N}$,  and this is the generator of the external-leg bubbles (${\rm ELB}$). 

Finally, we believe perhaps it would be interesting to find a regularization method in the CHY coordinates. 


\section{Feynman $i\epsilon$ prescription}\label{sectionepsilon}

In order to obtain the full form of the traditional Feynman propagators, we show that our proposal is able to reproduce the Feynman $i\epsilon$. This term can be included at each propagator in a pragmatic and simple way, by dimensional reduction in the momenta of the auxiliary punctures, i.e. $(k_{a_1},k_{a_2},k_{b_1},k_{b_2})$. 

Let us consider that the momenta of the auxiliary punctures has one more dimension than the rest of the kinematic  data, namely,
\begin{eqnarray}
&& K_{a_{1,2}}^M:=(k_{a_{1,2}}^\mu, k_{a_{1,2}}^{D+1}),\quad K_{b_{1,2}}^M:=(k_{b_{1,2}}^\mu, k_{b_{1,2}}^{D+1}), \quad\mu =1,\ldots , D\,,\\
&& k_{\a}^M:=(k_{\a}^\mu, 0), \hspace{1.6cm }  \ell^M:= (\ell^\mu , 0) ,\hspace{1.63cm} \a=1,\ldots n\, , \nonumber
\end{eqnarray} 
where $D$ is the number of physical dimensions and $n$ is the number of external particles. The forward limit measure, $d\Omega$, is modified in the following way 
\begin{equation}
d\Omega:= d^D(k_{a_1}+k_{b_1})\delta^{(D)}(k_{a_1}+k_{b_1}-\ell) d^{D+1}K_{a_2} d^{D+1}K_{b_2} \delta^{(D+1)}(K_{a_2}+K_{a_1}) \delta^{(D+1)}(K_{b_2}+K_{b_1}).\nonumber
\end{equation}
After performing all  integrals, the propagators become
\begin{eqnarray}
\frac{1}{(K_{a_1}+K_{b_1}+P)^2} = \frac{1}{(k_{a_1}+k_{b_1}+p)^2+2\,k^{D+1}_{a_1} \, k^{D+1}_{b_1}}= \frac{1}{(\ell+p)^2+2\,k^{D+1}_{a_1} \, k^{D+1}_{b_1}},
\end{eqnarray}
where $P^M=(p^\mu,0)$ is a momentum vector given by the sum of external momenta, i.e $P^M=k^M_{\a_1}+\cdots k^M_{\a_i} = (k^\mu_{\a_1}+\cdots + k^\mu_{\a_i},0):=(p^\mu,0)$, therefore $(K_{a_1}+K_{b_1})\cdot P = (k_{a_1}+k_{b_1})\cdot p $.  Finally, with the identification, $2\,k^{D+1}_{a_1} \, k^{D+1}_{b_1}$ = $i\epsilon$,  we obtain the Feynman propagators in full form.

\section{Discussions}\label{discussions}

Along the line of reasoning introduced in \cite{Gomez:2017lhy}, we have proposed a reformulation for the one-loop Parke-Taylor factors  given  in \cite{He:2015yua, Baadsgaard:2015hia,Geyer:2015jch}. 
Exploiting the algebraic (Schouten-like) identity between the $\s_{ij}$'s in the PT factors, we were able to show that they can be expanded in such a way that no tadpoles integrands will appear later, so for bi-adjoint $\Phi^3$ theory this cancellation is not necessarily related to the anti-symmetry of the structure constant in the cubic vertex. The construction also allowed us to build and classify systematically all the contributing CHY-integrals to the one-loop $n-$point case.

These new PT factors were used to calculate the partial amplitudes for the bi-adjoint $\Phi^3$ theory at one-loop. It can be seen that the prescription presents advantages over the previous ones:

\begin{itemize}

\item In this approach the CHY-integrals are supported over $n+4$ massless scattering equations only, where $n$ is the total number of external particles. Thus, all the known techniques to compute these type of integrals can be used  \cite{Gomez:2016cqb,Chen:2017edo,  Cardona:2016gon,Huang:2016zzb,Cardona:2015ouc,Kalousios:2015fya,Mafra:2016ltu,Cachazo:2013iea,Baadsgaard:2015voa,Cachazo:2015nwa,Cardona:2015eba}.

\item It gives directly the quadratic Feynman propagators, unlike the already known prescriptions, which must  apply the partial fractions identity, namely, their results are written in the $Q-$cut language \cite{Baadsgaard:2015twa}.   

\item The corresponding CHY-graphs are well suited to be easily solved using the $\L$-algorithm allowing to calculate the integrands  for higher points cases in a similar fashion to Feynman diagrams. In addition, with the technology developed in this work, we are able to compute the CHY-integrals directly in the forward limit, up to external-leg bubble configurations. The reason is because the singular solutions of the scattering equations do not contribute. 
\end{itemize}

In addition, it is straightforward to note that if one takes all the permutations instead of the cyclic ones, i.e,
\begin{align}
{\cal I}^{a_1:a_2}_{L} = \sum_{\a\in {\rm S_{n-1}}}
\frac{1}{\s_{\a_1\a_2}\s_{\a_2\a_3}\cdots \s_{\a_{n-1}\a_n}} \o^{a_1:a_2}_{\a_n:\a_1}
&=(-1)^n\,
{\rm PT}_{\rm tree}[1,\ldots, n]\,\mathbf{D}^{\, a_1:a_2}_{\rm type-0} [n,\ldots, 1]^{ n\o} \nonumber\\
&= \, \o^{a_1:a_2}_{1:1}\,  \o^{a_1:a_2}_{2:2}\, \cdots \, \o^{a_1:a_2}_{n:n} \\
&=\, \o^{a_1:a_2}_{1:2}\,  \o^{a_1:a_2}_{2:3}\, \cdots \, \o^{a_1:a_2}_{n:1}\nonumber, 
\end{align}
where $\a_1:=1$ and $\rm S_{n-1}$ is the set of all permutations of $\{2,3,\ldots, n\}$,  and by integrating it with, $ {\cal I}^{b_1:b_2}_{R} = {\bf PT}^{\, b_1:b_2}_{\rm 1-loop} [1,2,\ldots, n]$, one obtains the $n-$gon with the canonical ordering, which is a consequence of {\bf proposition 1}.   Therefore, we can say that the calculations in \cite{Gomez:2017lhy} are a particular case of our prescription.

At this time, most of computations performed at loop level using the CHY prescription have been obtained in the Q-cut language \cite{Chen:2016fgi,Feng:2016nrf,Baadsgaard:2015hia,Cachazo:2015aol,He:2015yua,Gomez:2016cqb,Cardona:2016wcr,Cardona:2016bpi,Geyer:2015jch,Geyer:2015bja,Geyer:2016wjx,Casali:2014hfa,Adamo:2015hoa}, i.e. linear propagators in the loop momenta. Additionally, it is very well known that the PT factors is one of the most important ingredient to define several types of theories in the CHY approach, most notably  Yang-Mills and Einstein gravity (from the Kawai--Lewellen--Tye (KLT) relations  point of view). 

We are confident that by using the new formulation of the PT factors at one-loop, as it has been proposed in this work, we will be  able to  extend our ideas beyond the bi-adjoint $\Phi^3$ case, in particular for the  Yang-Mills theory (see appendix \ref{chy-integrands} and eq. \eqref{YMprescription}), 
Recently, many works about the Bern-Carrasco-Johansson (BCJ) duality and KLT kernel in the CHY context have been published   \cite{Bern:2008qj,BjerrumBohr:2010hn,He:2016mzd,He:2017spx,Bjerrum-Bohr:2016axv,Bjerrum-Bohr:2016juj,Huang:2017ydz,Teng:2017tbo}. In particular, at one-loop all found results have been written in the ${\cal Q}-$cuts representation \cite{He:2016mzd,He:2017spx}.  Thus, following the lines of the new proposal developed here, it would be very interesting to obtain result in terms of the conventional propagators,  $(\ell + K)^{-2}$,  and to compare with the technologies presented in 
\cite{He:2016mzd,He:2017spx,Stieberger:2016lng,Hohenegger:2017kqy,Tourkine:2016bak}.

Moreover, extensions to higher loops are being developed \cite{wp}. In addition, it would be fascinating to found the origin of this new prescription or its relationship with the Ambitwistor string theory 
\cite{Mason:2013sva,Adamo:2013tsa,Casali:2015vta,Ohmori:2015sha}.

\acknowledgments

H.G. would like to thank to E. Bjerrum-Bohr, J. Bourjaily, and
P. Damgaard for discussions.  H.G. is very grateful to the Niels Bohr Institute - University of Copenhagen for hospitality and partial financial support during this work.  We thank to S. Mizera and P. Damgaard for useful comments. 
The work of  H.G.  is supported by USC grant DGI-COCEIN-No 935-621115-N22. 
P.T. is partially supported by MINECO grant FPA2016-76005-C2-1-P.


\appendix

\section{CHY-integrands at One-loop}\label{chy-integrands}

In this appendix we will give a simple way to construct CHY-integrands that have a particular dependence on the loop momenta after integration, i.e. they come as couples, $(k_{a_1}+k_{b_1})$ and $(k_{a_2}+k_{b_2})$. 

As it was mentioned previously, the proposal given in this paper follows the idea presented in \cite{Gomez:2017lhy}. The main idea that motivated the one-loop calculation in \cite{Gomez:2017lhy} is that the CHY-integral
\begin{align}
&\frac{1}{2^{n+1}}\int d^Dk_{a_2}\,d^Dk_{b_2} \delta^{(D)}(k_{a_2} + k_{a_1}) \delta^{(D)}(k_{b_2} + k_{b_1})  \,\,\times \label{chy-UC}\\
&
\int d\mu_{n+4}^{\rm tree} 
   \left[ \frac{\o^{a_1:a_2}_{1:2}\,  \o^{a_1:a_2}_{2:3}\, \cdots \, \o^{a_1:a_2}_{n:1}}{(a_1,b_1,b_2,a_2)} \right]  \times  \left[ \frac{\o^{b_1:b_2}_{1:2}\,  \o^{b_1:b_2}_{2:3}\, \cdots \, \o^{b_1:b_2}_{n:1}}{(a_1,b_1,b_2,a_2)} \right] =
   \hspace{-0.5cm}
\parbox[c]{10.5em}{\includegraphics[scale=0.35]{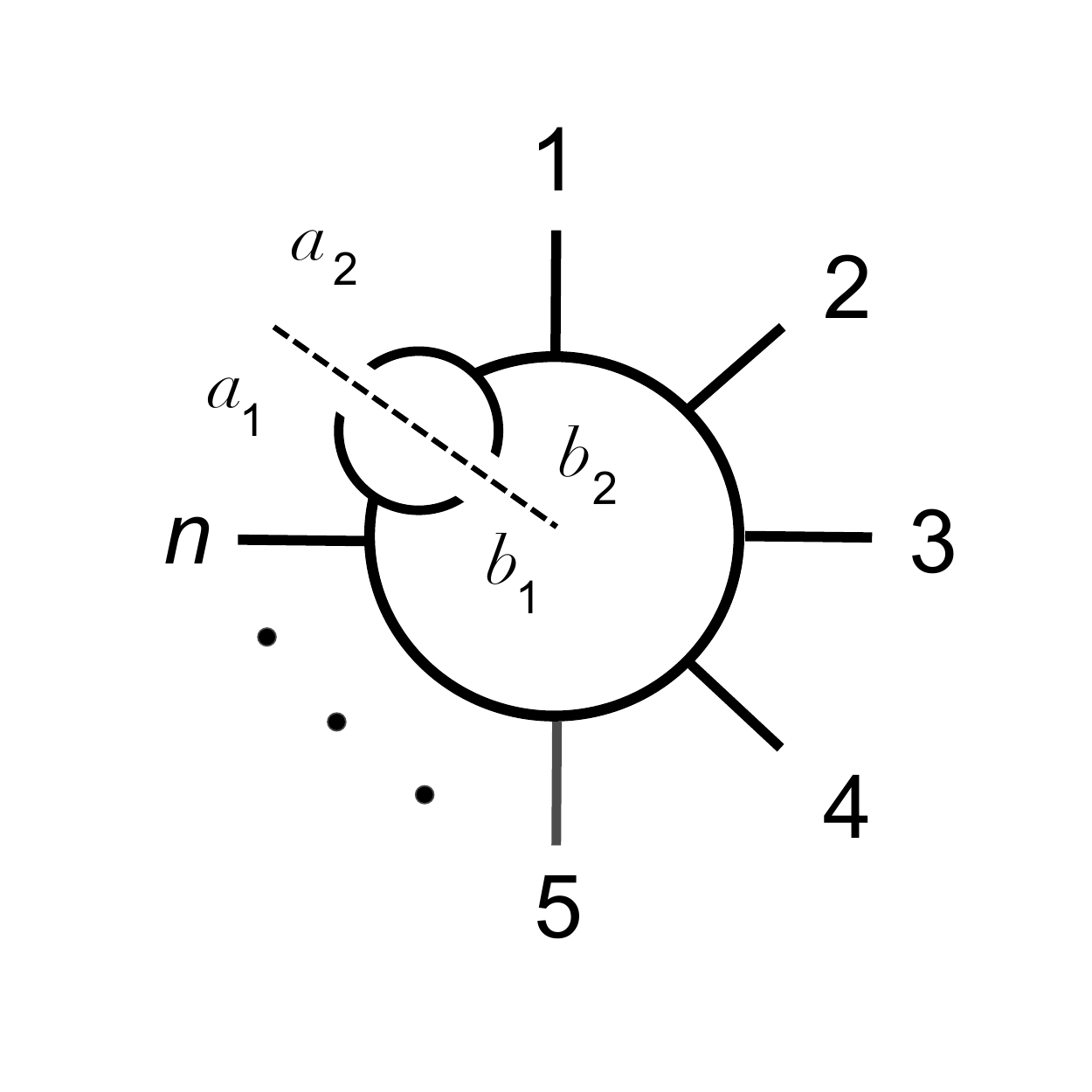}} + {\rm per}  (1,...,n) 
\nonumber
\end{align}
is just the unitary cut of the two-loop diagram \cite{Gomez:2016cqb}, at the integrand level, as it has been represented above.  Therefore, in order to obtain a one-loop integrand, we multiply by the factor, $(k_{a_1}+k_{b_1})^2$, and we make the identification, $k_{a_1}+k_{b_1} = \ell$. This process is performed by the integral, $\int d^D(k_{a_1}+k_{b_1}) \, \delta^{(D)}(k_{a_1} + k_{b_1}-\ell)$. This would be a simple explanation why the measure, $d\Omega \times s_{a_1b_1}$, is introduced in that particular way.  Note that the \eqref{chy-UC}  CHY-integrand is a generalization for the one found in \cite{Cardona:2016bpi,Geyer:2015jch} 
\begin{align}\label{linear-1loop}
&\int d\mu_{n+2}^{\rm 1-loop} 
   \left[ \frac{\o^{\ell^+:\ell^-}_{1:2}\,  \o^{\ell^+:\ell^-}_{2:3}\, \cdots \, \o^{\ell^+:\ell^-}_{n:1}}{(\ell^+,\ell^-)} \right]  \times  \left[ \frac{\o^{\ell^+:\ell^-}_{1:2}\,  \o^{\ell^+:\ell^-}_{2:3}\, \cdots \, \o^{\ell^+:\ell^-}_{n:1}}{(\ell^+,\ell^-)} \right] \\
   &
   ^{\underrightarrow{\quad {\rm Linear\,\, propagators}  \quad}}
   \hspace{-0.2cm}
\parbox[c]{8.7em}{\includegraphics[scale=0.2]{feyn-p.pdf}} + \,\,{\rm per}  (1,...,n) ,
\nonumber
\end{align}
which is the one that reproduces only linear propagators.

Generalizing the \eqref{chy-UC} idea, our proposal is
\begin{align}
\mathfrak{I}_n :=
\frac{1}{2^{n+1}}\int d\Omega \, s_{a_1b_1}\, \int d\mu_{n+4}^{\rm tree} 
   \left[ \frac{{\cal I}_{L}^{\, a_1:a_2} (\s)}{(a_1,b_1,b_2,a_2)} \right]  \times  \left[ \frac{{\cal I}_{R}^{\, b_1:b_2} (\s)}{(a_1,b_1,b_2,a_2)} \right].  \nonumber
\end{align}
So, a natural question is: What must be the form of the integrands, $\{{\cal I}_{L}^{\, a_1:a_2},{\cal I}_{R}^{\, b_1:b_2}  \}$, in order to obtain a function of the couples, $(k_{a_1}+k_{b_1})$ and $(k_{a_2}+k_{b_2})$ ?  Before giving an answer of this question, it is useful to remind  our one-loop Parke-Taylor factors construction.
 
In \cite{He:2015yua,Baadsgaard:2015hia}, the planar one-loop Parke-Taylor factors for linear propagators were presented, and they can be written like
\begin{align}
 {\rm PT}_{\rm 1-loop} [\pi]:=\frac{1}{(\ell^+,\ell^-)} 
 \sum_{\a\in {\rm cyc (\pi)}}
\frac{1}{\s_{\a_1\a_2}\s_{\a_2\a_3}\cdots \s_{\a_{n-1}\a_n}} \o^{\ell^+:\ell^-}_{\a_n:\a_1}.
\end{align}
Following the previous proposal and the generalization of \eqref{linear-1loop} given in \eqref{chy-UC}, in this paper we proposed the  planar  Parke-Taylor factors at one-loop  for quadratic propagators as
\begin{align}\label{Nproposal}
 &\frac{{\cal I}_{L}^{\, a_1:a_2} (\s)}{(a_1,b_1,b_2,a_2)} =\frac{ {\bf PT}^{a_1:a_2}_{\rm 1-loop} [\pi]}{(a_1,b_1,b_2,a_2)}:=\frac{1}{(a_1,b_1,b_2,a_2)} 
 \sum_{\a\in {\rm cyc (\pi)}}
\frac{1}{\s_{\a_1\a_2}\s_{\a_2\a_3}\cdots \s_{\a_{n-1}\a_n}} \o^{a_1:a_2}_{\a_n:\a_1} \, ,\nonumber\\
&\frac{{\cal I}_{R}^{\, b_1:b_2} (\s)}{(a_1,b_1,b_2,a_2)} =\frac{ {\bf PT}^{b_1:b_2}_{\rm 1-loop} [\rho]}{(a_1,b_1,b_2,a_2)}:=\frac{1}{(a_1,b_1,b_2,a_2)} 
 \sum_{\b\in {\rm cyc (\rho)}}
\frac{1}{\s_{\b_1\b_2}\s_{\b_2\b_3}\cdots \s_{\b_{n-1}\b_n}} \o^{b_1:b_2}_{\b_n:\b_1}\, .
\end{align}
It is not obvious that by using these integrands we will obtain a functional dependence of the loop momenta like $(k_{a_1}+k_{b_1})$ and $(k_{a_2}+k_{b_2})$, nevertheless, it is not difficult to show that this turn out to be the case.  First of all, as in \eqref{PTn-PTn}, it is straightforward to check that the integrands in \eqref{Nproposal} can be written as 
 \begin{align}\label{Nproposal2}
 &\frac{ {\bf PT}^{a_1:a_2}_{\rm 1-loop} [\pi]}{(a_1,b_1,b_2,a_2)}=
 \sum_{\a\in {\rm cyc (\pi)}}
{\rm PT_{tree}}[\a_1,\a_2,\ldots,\a_n,a_1,b_1,b_2,a_2]
 \, ,\nonumber\\
&\frac{ {\bf PT}^{b_1:b_2}_{\rm 1-loop} [\rho]}{(a_1,b_1,b_2,a_2)}= \sum_{\b\in {\rm cyc (\rho)}}
{\rm PT_{tree}}[\b_1,\b_2,\ldots,\b_n,b_1,a_1,a_2,b_2]\, .
\end{align}
Each term in the \eqref{Nproposal2} sums is called a {\it partial planar one-loop Parke-Taylor factor}, and we denote them as
 \begin{align}\label{Npartial}
 {\rm PT}^{\rm 1-loop}_{L} [\a]:=&
{\rm PT_{tree}}[\a_1,\ldots,\a_n,a_1,b_1,b_2,a_2]=\frac{1}{(a_1,b_1,b_2,a_2)}\left[\frac{(a_1,a_2)}{(\a_1,\dots,\a_n,a_1,a_2)}
\right]
 \, ,\nonumber\\
 {\rm PT}^{\rm 1-loop}_{R} [\b]:=&
{\rm PT_{tree}}[\b_1,\ldots,\b_n,b_1,a_1,a_2,b_2]=
\frac{1}{(a_1,b_1,b_2,a_2)}\left[\frac{(b_1,b_2)}{(\a_1,\dots,\a_n,b_1,b_2)}
\right]
\, .
\end{align}

Next, by taking the CHY-integral of the  product of these two factors, $ {\rm PT}^{\rm 1-loop}_{L}$ and  $ {\rm PT}^{\rm 1-loop}_{R}$,  for two generic orderings\footnote{Note that this is a general case of the example shown in \eqref{PT3-PT3}.} $\a$ and $\b$, it can be represented as
\begin{align}\label{PTL-PTR}
&\int d\mu_{n+4}^{\rm tree} \,\,
    {\rm PT}^{\rm 1-loop}_{L} [\a_1,\ldots,\a_n]  \times   {\rm PT}^{\rm 1-loop}_{R} [\b_1,\ldots,\b_n] 
\\    
&    =
   \hspace{-0.5cm}
\parbox[c]{10em}{\includegraphics[scale=0.32]{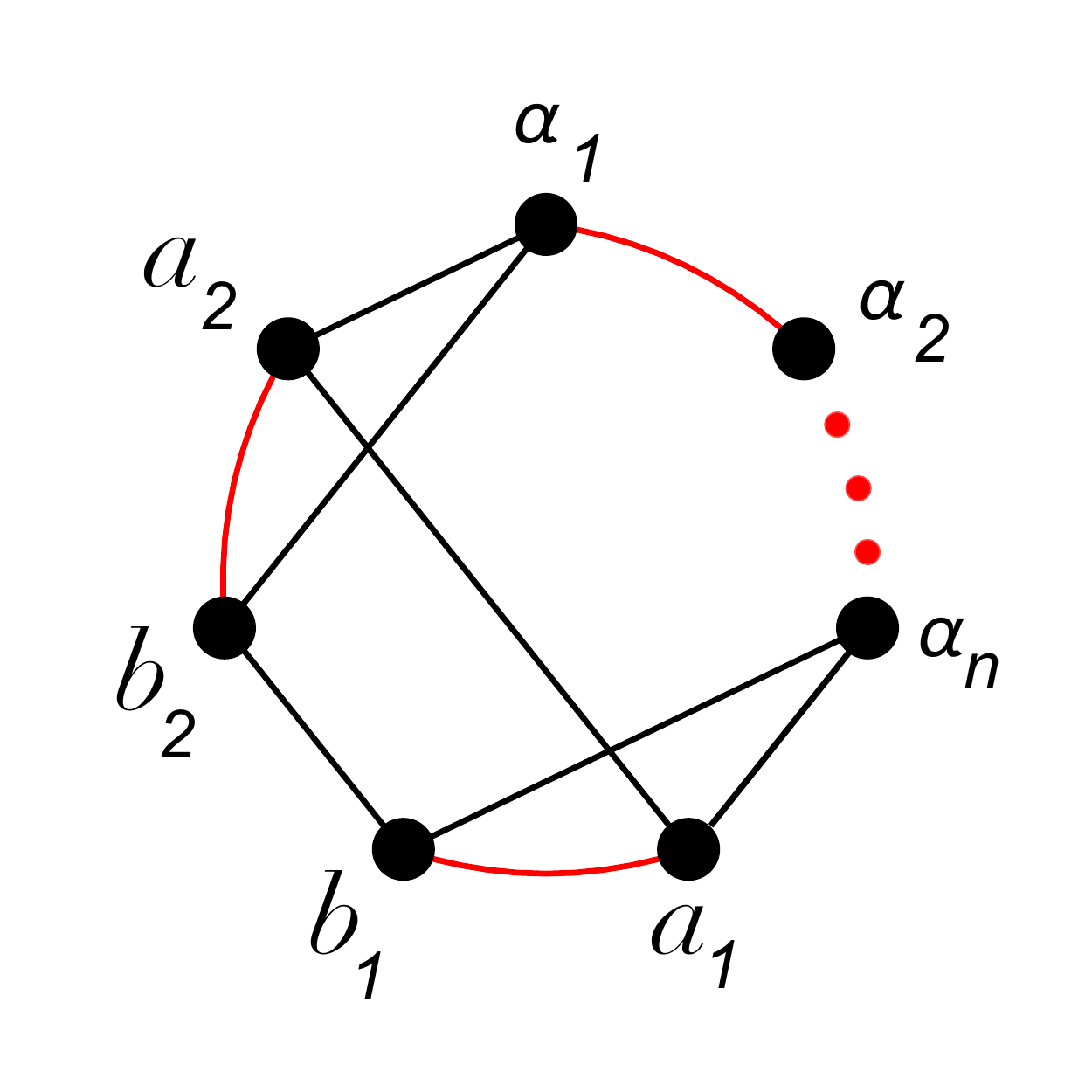}}  \bigcap
   \hspace{-0.3cm}
\parbox[c]{10em}{\includegraphics[scale=0.32]{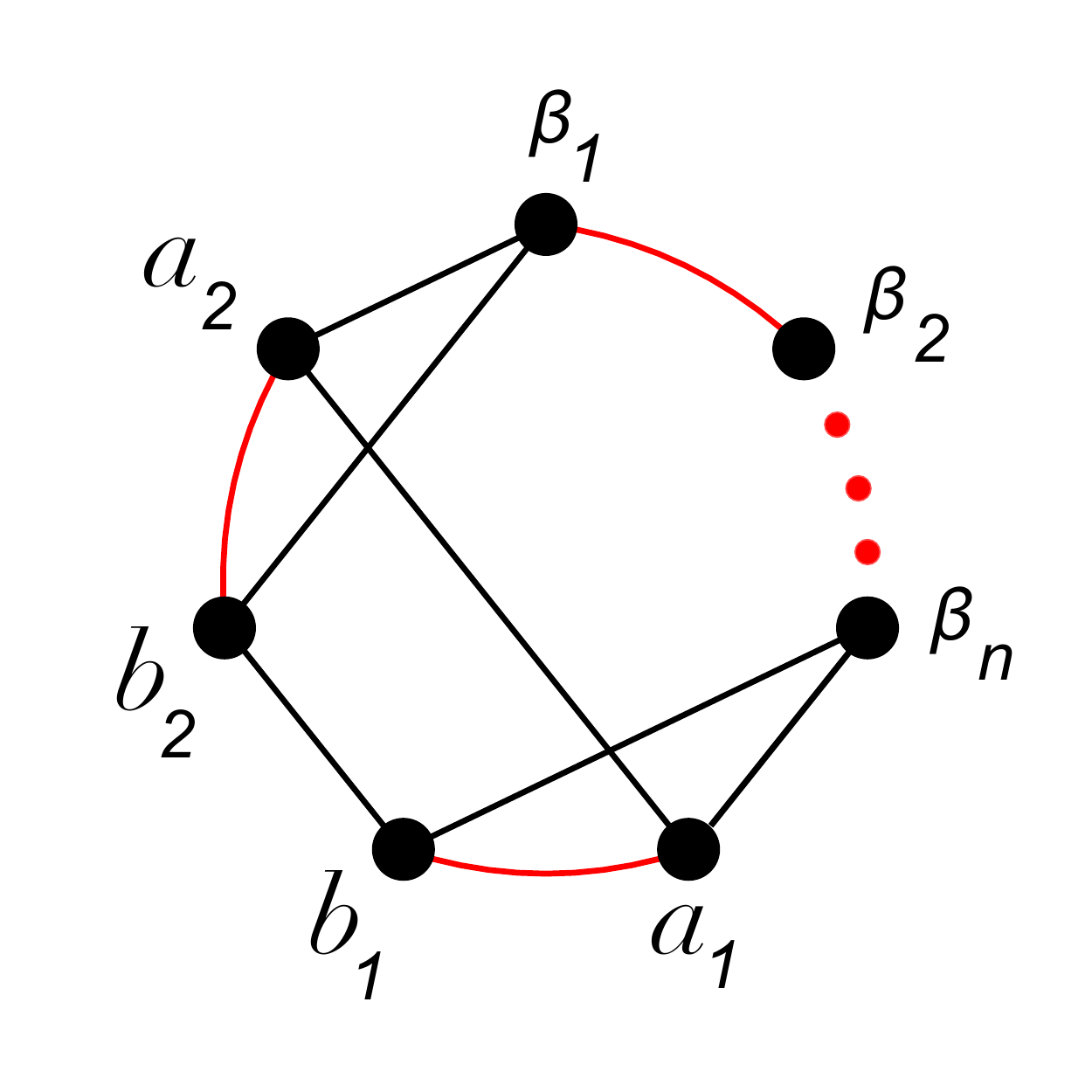}}=
   \hspace{-0.5cm}
\parbox[c]{10em}{\includegraphics[scale=0.32]{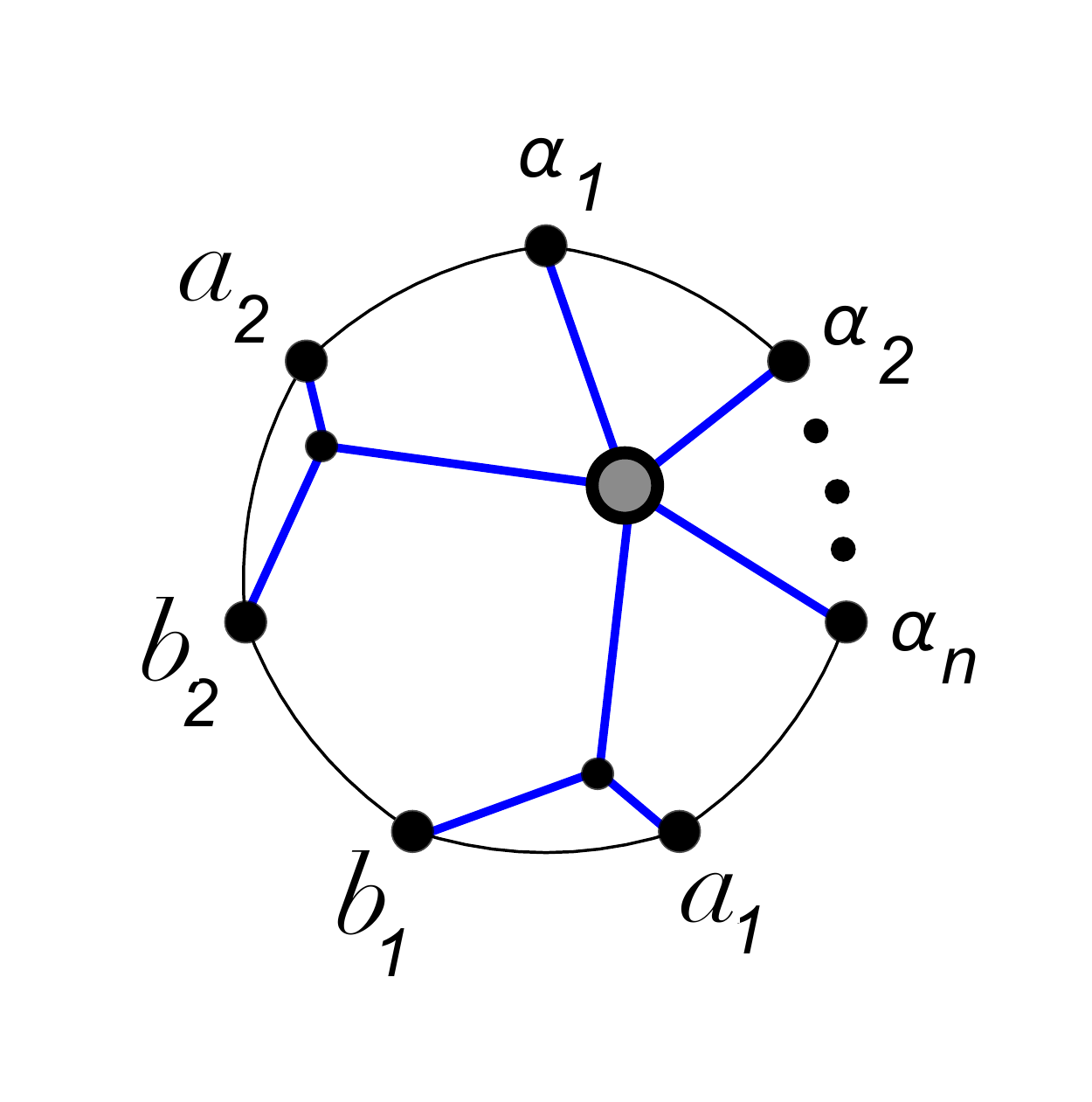}}  \bigcap
   \hspace{-0.3cm}
\parbox[c]{10em}{\includegraphics[scale=0.32]{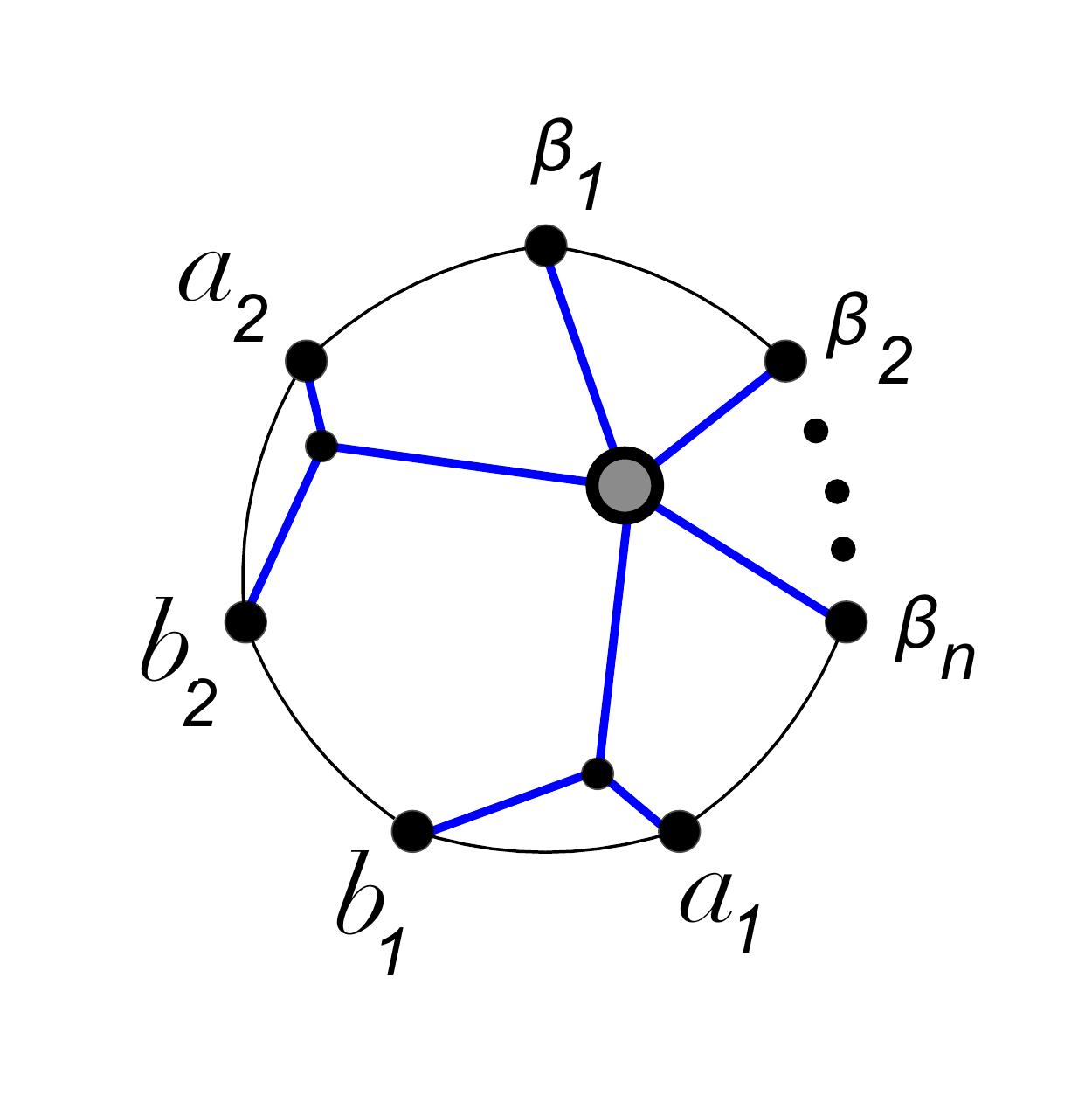}}\, ,
\nonumber
\end{align}
where we have used the intersection property \cite{Cachazo:2013iea} from section \ref{biadjoint}, and the grey circles  mean the sum over all possible trivalent planar diagrams. Clearly, in \eqref{PTL-PTR} we have shown that the CHY-integral,  $ \int d\mu_{n+4}^{\rm tree} \, {\rm PT}^{\rm 1-loop}_{L} [\a]  \times   {\rm PT}^{\rm 1-loop}_{R} [\b] $,  is in fact a function of the two off-shell momenta which come from the combinations of four on-shell momenta: $(k_{a_1}+k_{b_1})$ and $(k_{a_2}+k_{b_2})$.  This implies that the whole construction developed in this paper is well defined, i.e. the three types of CHY-integrands:
\begin{eqnarray}
\frac{{\cal I}_{L}^{\, a_1:a_2} (\s)}{(a_1,b_1,b_2,a_2)}\times \frac{{\cal I}_{R}^{\, b_1:b_2} (\s)}{(a_1,b_1,b_2,a_2)}
&=&{\rm PT}^{\rm 1-loop}_{L} [\a]\times {\rm PT}^{\rm 1-loop}_{R} [\b]\, ,\\
\frac{{\cal I}_{L}^{\, a_1:a_2} (\s)}{(a_1,b_1,b_2,a_2)}\times \frac{{\cal I}_{R}^{\, b_1:b_2} (\s)}{(a_1,b_1,b_2,a_2)}
&=&\frac{ {\bf PT}^{a_1:a_2}_{\rm 1-loop} [\pi]}{(a_1,b_1,b_2,a_2)}\times {\rm PT}^{\rm 1-loop}_{R} [\b]\, ,\\
\frac{{\cal I}_{L}^{\, a_1:a_2} (\s)}{(a_1,b_1,b_2,a_2)}\times \frac{{\cal I}_{R}^{\, b_1:b_2} (\s)}{(a_1,b_1,b_2,a_2)}
&=&\frac{ {\bf PT}^{a_1:a_2}_{\rm 1-loop} [\pi]}{(a_1,b_1,b_2,a_2)}\times\frac{ {\bf PT}^{b_1:b_2}_{\rm 1-loop} [\rho]}{(a_1,b_1,b_2,a_2)} \, ,
\end{eqnarray}
give a functional dependence of the loop momenta like $(k_{a_1}+k_{b_1})$ and $(k_{a_2}+k_{b_2})$. In addition, from the identities
\begin{eqnarray}
\sum_{\a\in S_{n-1}} {\rm PT}^{\rm 1-loop}_{L} [\a_1,\ldots,\a_{n-1},n] &=& \frac{\o^{a_1:a_2}_{1:2}\,  \o^{a_1:a_2}_{2:3}\, \cdots \, \o^{a_1:a_2}_{n:1}}{(a_1,b_1,b_2,a_2)} \, ,\\
\sum_{\b\in S_{n-1}} {\rm PT}^{\rm 1-loop}_{R} [\b_1,\ldots,\b_{n-1},n] &=& \frac{\o^{b_1:b_2}_{1:2}\,  \o^{b_1:b_2}_{2:3}\, \cdots \, \o^{b_1:b_2}_{n:1}}{(a_1,b_1,b_2,a_2)} \, ,
\end{eqnarray}
where $S_{n-1}$ is the group of $(n-1)$-permutations,  the CHY-integrand in \eqref{chy-UC},
\begin{eqnarray}
\frac{{\cal I}_{L}^{\, a_1:a_2} (\s)}{(a_1,b_1,b_2,a_2)}\times \frac{{\cal I}_{R}^{\, b_1:b_2} (\s)}{(a_1,b_1,b_2,a_2)}
=  \left[ \frac{\o^{a_1:a_2}_{1:2}\,  \o^{a_1:a_2}_{2:3}\, \cdots \, \o^{a_1:a_2}_{n:1}}{(a_1,b_1,b_2,a_2)} \right]  \times  \left[ \frac{\o^{b_1:b_2}_{1:2}\,  \o^{b_1:b_2}_{2:3}\, \cdots \, \o^{b_1:b_2}_{n:1}}{(a_1,b_1,b_2,a_2)} \right]\, , \nonumber
\end{eqnarray}
gives also a functional dependence of  $(k_{a_1}+k_{b_1})$ and $(k_{a_2}+k_{b_2})$. 

Therefore,  we have found an answer for the question formulated previously,  to obtain a CHY-integral that is able to give a functional dependence of the momenta, $(k_{a_1}+k_{b_1})$ and $(k_{a_2}+k_{b_2})$,  the integrands, $\frac{{\cal I}_{L}^{\, a_1:a_2} (\s)}{(a_1,b_1,b_2,a_2)}$ and $ \frac{{\cal I}_{R}^{\, b_1:b_2} (\s)}{(a_1,b_1,b_2,a_2)}$, must be a linear combination of the factors,  ${\rm PT}^{\rm 1-loop}_{L}$ and ${\rm PT}^{\rm 1-loop}_{R}$, respectively. 

Finally, from the ideas given in this appendix, in order to reproduce the planar contribution at one-loop (with quadratic propagators) for Yang-Mills theory, we  propose the following prescription \cite{wp}
\begin{eqnarray}\label{YMprescription}
&&A^{\rm YM}_{\rm 1-loop}(1,2,\ldots, n) = \int \frac{d^D\ell}{(2\pi)^D}\times \int \frac{d\Omega\,\,s_{a_1b_1}}{2^{n+1}}\times \int d\mu_{n+4}^{\rm tree} \,\,\frac{ {\bf PT}^{a_1:a_2}_{\rm 1-loop} [1,2,\ldots,n]}{(a_1,b_1,b_2,a_2)}\nonumber \\
&&
\hspace{3.2cm}
\times
\sum_{\rho\in S_n}\, {\bf n}_{a_2, b_2 |\rho_1\cdots \rho_n | b_1,a_1} \times  {\rm PT}^{\rm 1-loop}_{R} [\rho_1,\ldots, \rho_n],
\end{eqnarray}
where  ${\bf n}_{a_2, b_2 |\rho_1\cdots \rho_n | b_1,a_1}$ are the BCJ numerators, which must be found (see \cite{He:2016mzd,He:2017spx} for linear representation).  Additionally,  some progress towards the construction of the {\it partial non-planar one-loop Parke-Taylor factors} is in development. In \cite{wp}, we have shown that those non-planar  Parke-Taylor factors can be written as a linear combination of ${\rm PT}^{\rm 1-loop}_{L}$ (or ${\rm PT}^{\rm 1-loop}_{R}$), as it is claimed in this appendix.

\section{Proof of the one-loop Parke-Taylor factor expansion}\label{Aproof}

The path we found to write \eqref{ptdef} as a sum of terms with a minimum number of two $\o^{a_1:a_2}_{i:j}$'s was not a straightforward one, but it allowed us to see that the space for the one-loop CHY diagrams with a fixed number of external points is bigger than the one of Feynman diagrams. We have seen that the diagrams we have encounter so far can be written in terms of the diagrams we computed in section \ref{classification}, even diagrams that cannot be solved using the $\L$-algorithm, which happens to be the case for the original Parke-Taylor factor. Our proof will be supported only in the use of the {\it Schouten-like} identity
\ba 
\mathbb{1}&=&\frac{\s_{ac}\s_{bd}-\s_{ad}\s_{bc}}{\s_{ab}\s_{cd}}\,,
\label{schouten}
\ea
which give us a cross-ratio to relate the diagrams algebraically, without the use of the scattering equations.

Our starting point is the expression \eqref{ptdef} for the one-loop Parke-Taylor factor with ordering $\pi$, it can be rewritten as follows 
\ba
 {\bf PT}^{\, a_1:a_2}_{\rm 1-loop} [\pi]:= {\rm PT}_{\rm tree}[\pi]\sum_{\a\in {\rm cyc (\pi)}}
\s_{\a_n\a_1}\o^{a_1:a_2}_{\a_n:\a_1}\,.
\ea
Since the proof works the same way for any ordering, we can take $\pi=(12...n)$, then

\ba
{\bf PT}^{\, a_1:a_2}_{\rm 1-loop} [1,2,...,n]= 
{\rm PT}_{\rm tree}[1,2,...,n]\left(\s_{12}\o^{a_1:a_2}_{1:2} + \s_{23}\o^{a_1:a_2}_{2:3} + ... + \s_{n1}\o^{a_1:a_2}_{n:1}\right)\,.
\ea

One thing to attempt would be to solve the diagram corresponding to the first term and then take cyclic permutations of the result, but it leads to singular cuts, so we cannot apply the $\L$-algorithm. Since the ${\rm PT}_{\rm tree}$ is a global factor, we can perform our analysis just on the first term $\s\o$ and then apply the cyclic permutations to get the whole ${\bf PT}_{\rm 1-loop}$. Writing it explicitly 

\ba
\s_{12}\o^{a_1:a_2}_{1:2}=\frac{\s_{12}\s_{a_1a_2}}{\s_{1a_1}\s_{2a_2}}\,.
\label{singomega}
\ea

We want to find terms with the higher order of $\o$'s, since we are missing $2n-2$ factors in the denominator, we use the cross-ratios \eqref{schouten} to obtain them. Now we have \eqref{singomega} times ``1'' 

\ba
\frac{\s_{12}\s_{a_1a_2}}{\s_{1a_1}\s_{2a_2}}\left(\frac{\s_{23}\s_{a_1a_2}+\s_{2a_2}\s_{3a_1}}{\s_{2a_1}\s_{3a_2}}\right)\left(\frac{\s_{34}\s_{a_1a_2}+\s_{3a_2}\s_{4a_1}}{\s_{3a_1}\s_{4a_2}}\right)\cdots\left(\frac{\s_{n1}\s_{a_1a_2}+\s_{na_2}\s_{1a_1}}{\s_{na_1}\s_{1a_2}}\right)\,.
\ea

Expanding all the products and performing the sum over cyclic permutations will give us more than just the term with n $\o$'s, it will give all the correct terms\footnote{By correct we mean that each one of these terms will give the integrand for a contributing Feynman diagram.} down to $\o^2$, but we will still have terms linear in $\o$. Actually, those linear terms belong to the inverse ordering. Schematically, the expansion now looks like this
\ba
{\bf PT}^{\, a_1:a_2}_{\rm 1-loop} [1,2,...,n]&=& 
{\rm PT}_{\rm tree}[1,2,...,n]\bigg[n\times(\s\s...\s)^{(n)}(\o\o...\o)^{(n)}\nonumber\\
&&\hspace{2.6cm} + (n-1)\times\big\lbrace(\s\s...\s\o)^{(n)}(\o\o...\o)^{(n-2)}+...\big\rbrace+...\nonumber\\
&&\hspace{2.6cm}+2\times(\s\o)^{(2)}(\s\o)^{(2)}\bigg]^{\mathbb{1}_n} - {\bf PT}^{\, a_1:a_2}_{\rm 1-loop} [n,n-1,...,1]\nonumber\\
\label{1loopPTlin}
\ea
where the round brackets mean closed cycles, and the super indices on them mean the number of factors inside. All the terms inside the square brackets belong to the (12...,n) ordering, so we put a super index $\mathbb{1}$ on them. A somehow unexpected result, is that we can write ${\bf PT}^{\, a_1:a_2}_{\rm 1-loop} [n,n-1,...,1]$ as a sum of all the terms on the square brackets (i.e. from its inverse order), but with all the coefficients equal to 1. The expression we have, again schematically, is

\ba
{\bf PT}^{\, a_1:a_2}_{\rm 1-loop} [n,n-1,...,1]&=& 
{\rm PT}_{\rm tree}[1,2,...,n]\bigg[(\s\s...\s)^{(n)}(\o\o...\o)^{(n)}\nonumber\\
&&\hspace{2.6cm} + \big\lbrace(\s\s...\s\o)^{(n)}(\o\o...\o)^{(n-2)}+...\big\rbrace+...\nonumber\\
&&\hspace{2.6cm}+(\s\o)^{(2)}(\s\o)^{(2)}\bigg]^{\mathbb{1}_n}\,.
\label{PTrel}
\ea

To prove the previous relation we apply the inverse procedure with the {\it Schouten like} identity, we dismantle the numerator by mixing the $\s_{ij}$'s with the $\s_{a_1a_2}$'s, factors that cancel out the denominators will appear and we will arrive to the ${\rm PT}^{\, a_1:a_2}_{\rm 1-loop} [n,n-1,...,1]$. 

Replacing \eqref{PTrel} in \eqref{1loopPTlin} we will have an expression with no linear terms in $\o$. Its coefficients are the only ones modified

\ba
{\rm PT}^{\, a_1:a_2}_{\rm 1-loop} [1,2,...,n]&=& 
{\rm PT}_{\rm tree}[1,2,...,n]\bigg[(n-1)\times(\s\s...\s)^{(n)}(\o\o...\o)^{(n)}\nonumber\\
&&\hspace{2.6cm} + (n-2)\times\big\lbrace(\s\s...\s\o)^{(n)}(\o\o...\o)^{(n-2)}+...\big\rbrace+...\nonumber\\
&&\hspace{2.6cm}+1\times(\s\o)^{(2)}(\s\o)^{(2)}\bigg]^{\mathbb{1}_n}\,.
\label{1loopPT}
\ea

Now this one-loop Parke-Taylor factor will enter into CHY integrands that can be easily solved using the $\L$-algorithm, these give also the correct contributions for the bi-adjoint $\Phi^3$ scalar theory.


\section{From quadratic to Linear propagators in the CHY-graphs}\label{Aconjeture}


The computational techniques developed in this work can be applied to the linear propagators approach as well. 

Schematically, the CHY-graphs that lead to linear propagators can be obtained from the ones related to quadratic ones just by replacing the box loop by a ``line loop" in the CHY-graphs, meaning
\begin{eqnarray}
 \parbox[c]{7em}{\includegraphics[scale=0.35]{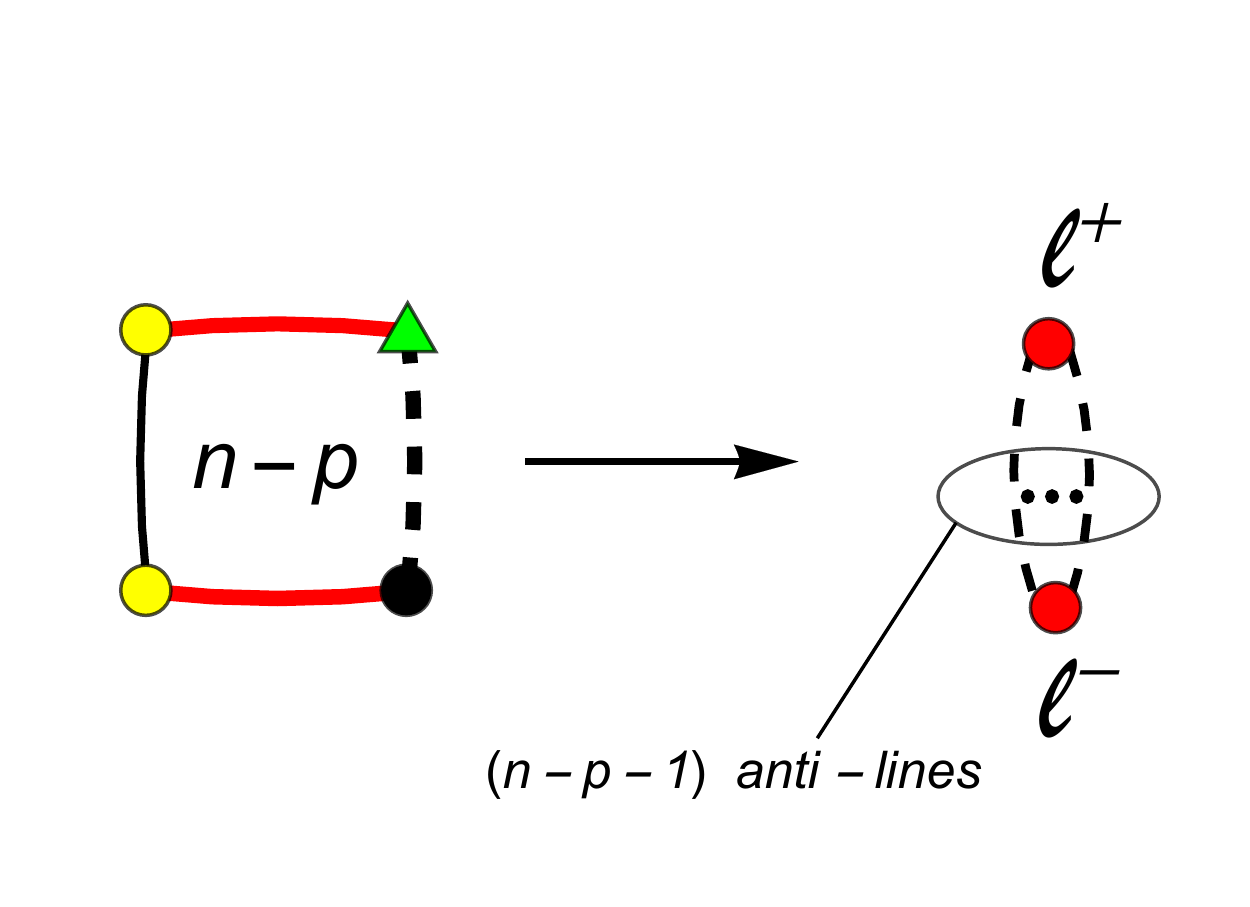}}
\end{eqnarray}
where, $\ell^+=-\ell^-:=\ell$, and $\ell $ is the off-shell loop momentum, $\ell^2\neq0$. For example, the general CHY-graph given in \eqref{G-graph11} becomes
\begin{eqnarray}\label{from-Qu-Li}
\parbox[c]{5em}{\includegraphics[scale=0.21]{G-graph11.pdf}}
\parbox[c]{7em}{\includegraphics[scale=0.35]{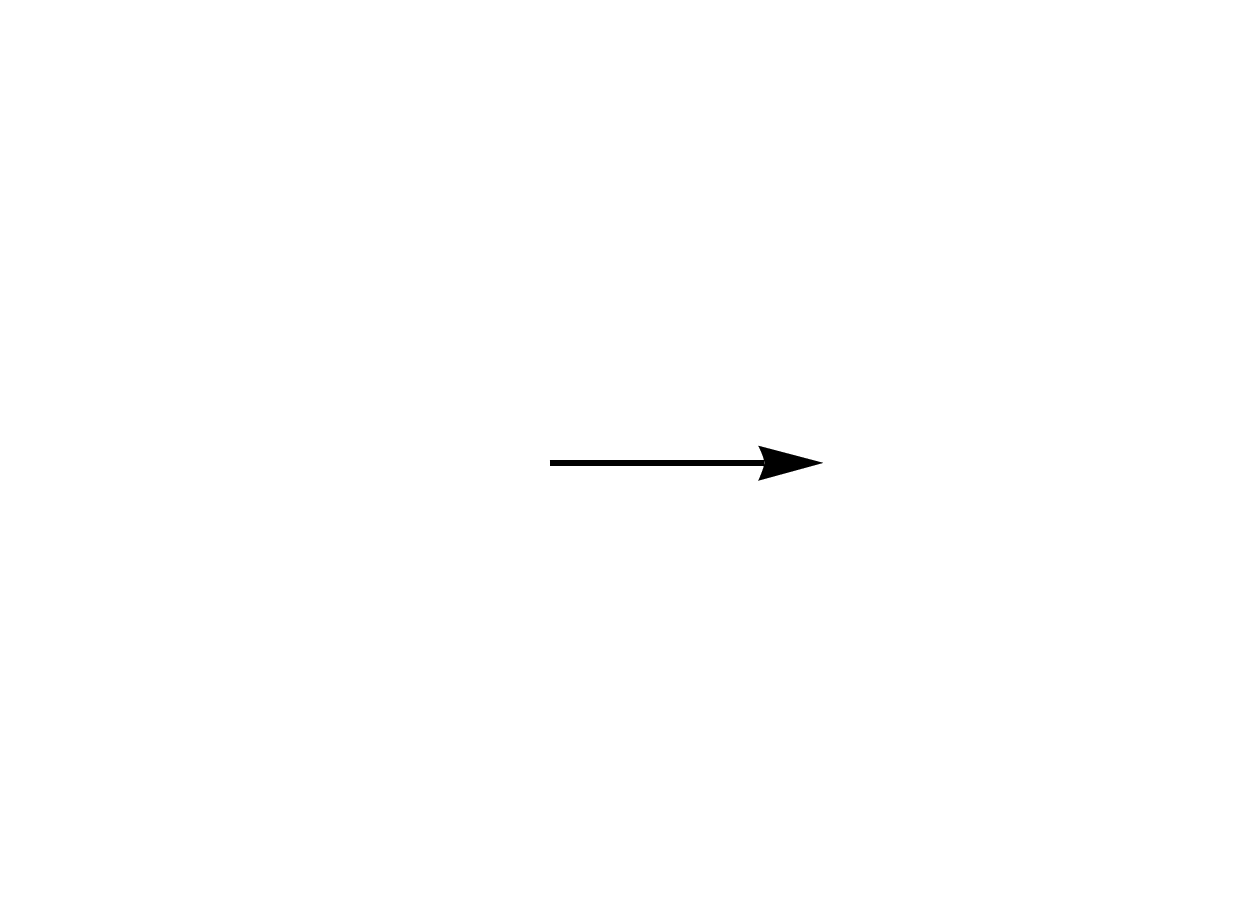}}
 \parbox[c]{7em}{\includegraphics[scale=0.3]{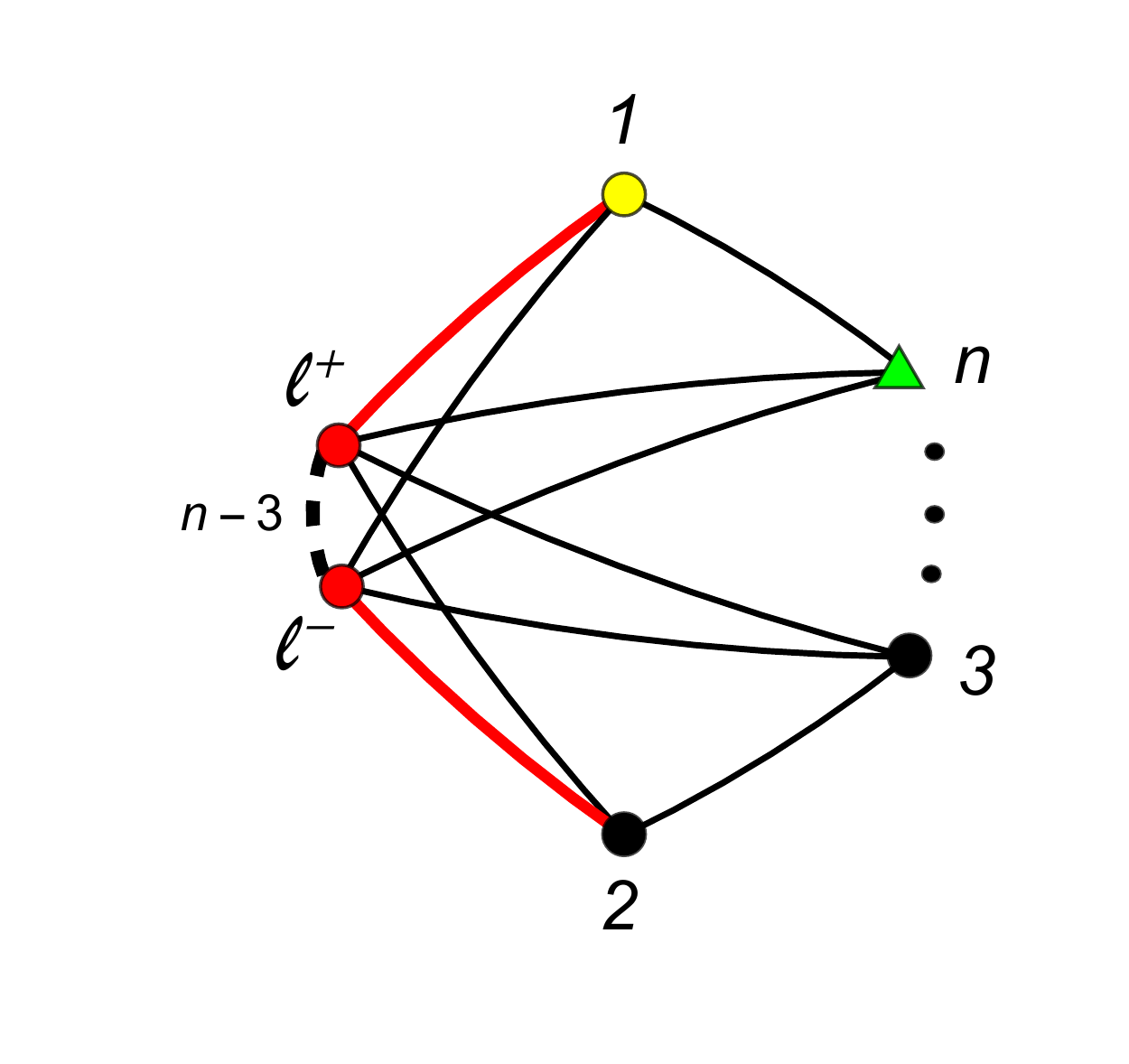}}.
\end{eqnarray}
Here we apply the nearest neighbour gauge fixing to compute the CHY-graph on the left hand side of \eqref{from-Qu-Li}, which was described in detail in the proof of {\bf proposition 1}, where it was applied to calculate the CHY-graph on the right hand side. Following exactly the same steps we obtain the following result
\begin{eqnarray}\label{finally}
&&\int d\mu_{n+2}^{\rm 1-loop}
\hspace{-0.3cm}
 \parbox[c]{7em}{\includegraphics[scale=0.3]{gen-linear.pdf}} \\
 =
&& \frac{1}{[(\ell+k_1)^2-\ell^2][(\ell+k_1+k_n)^2-\ell^2][(\ell+k_1+k_n+k_{n-1})^2-\ell^2]\cdots[(\ell+\sum_{i,i\neq 2}^nk_i)^2-\ell^2]}\nonumber,
\end{eqnarray}
which is linear in $\ell$ and where the measure $d\mu_{n+2}^{\rm 1-loop}$ was defined  in \eqref{1LL-measure}.

Finally, if the result found in \eqref{finally} is summed over all possible permutations, it provides a proof to  the conjecture proposed in \cite{Cardona:2016bpi},  equation (7.18).

\bibliographystyle{JHEP}
\bibliography{mybib}
\end{document}